\documentclass[11pt]{article}
\usepackage[margin=2cm]{geometry}
\usepackage{titlesec}
\usepackage{url}
\usepackage{graphicx}
\usepackage{amssymb}
\usepackage{natbib}
\usepackage{epstopdf}

\usepackage{multicol} 

\titlespacing\section{0pt}{10pt plus 4pt minus 0pt}{2pt plus 2pt minus 0pt}
\titlespacing\subsection{0pt}{10pt plus 4pt minus 0pt}{2pt plus 2pt minus 0pt}

\newcommand{\reffig}[1]{Fig.~\ref{#1}}
\newcommand{\reftab}[1]{Tab.~\ref{#1}}
\newcommand{\refwp}[1]{\S\ref{sec:#1}}

\def\ergcms{erg~cm$^{-2}$~s$^{-1}$}

\def\cm2{cm$^{-2}$}
\usepackage{color}
\definecolor{red}{rgb}{0.7,0,0}
\definecolor{blue}{rgb}{0,0,0.7}

\begin{document}






\begin{center}
\noindent {\Huge\bf The Deep and Transient Universe:\\ New Challenges and Opportunities} \vspace{1mm}

\vspace{4\baselineskip} 


\noindent {\huge\bf Scientific prospects of the \textit{SVOM} mission} \vspace{1.5mm}

\vspace{2\baselineskip}

\noindent {\LARGE\bf J. Wei, B. Cordier, et al. \\
\noindent {\Large\bf(Version of 05-10-2016, for full list of contributors see overleaf)} \vspace{1.5mm}}
\end{center}



\vspace{1\baselineskip}

\begin{figure}[hptb]
   \begin{center}
	\includegraphics[scale=0.25]{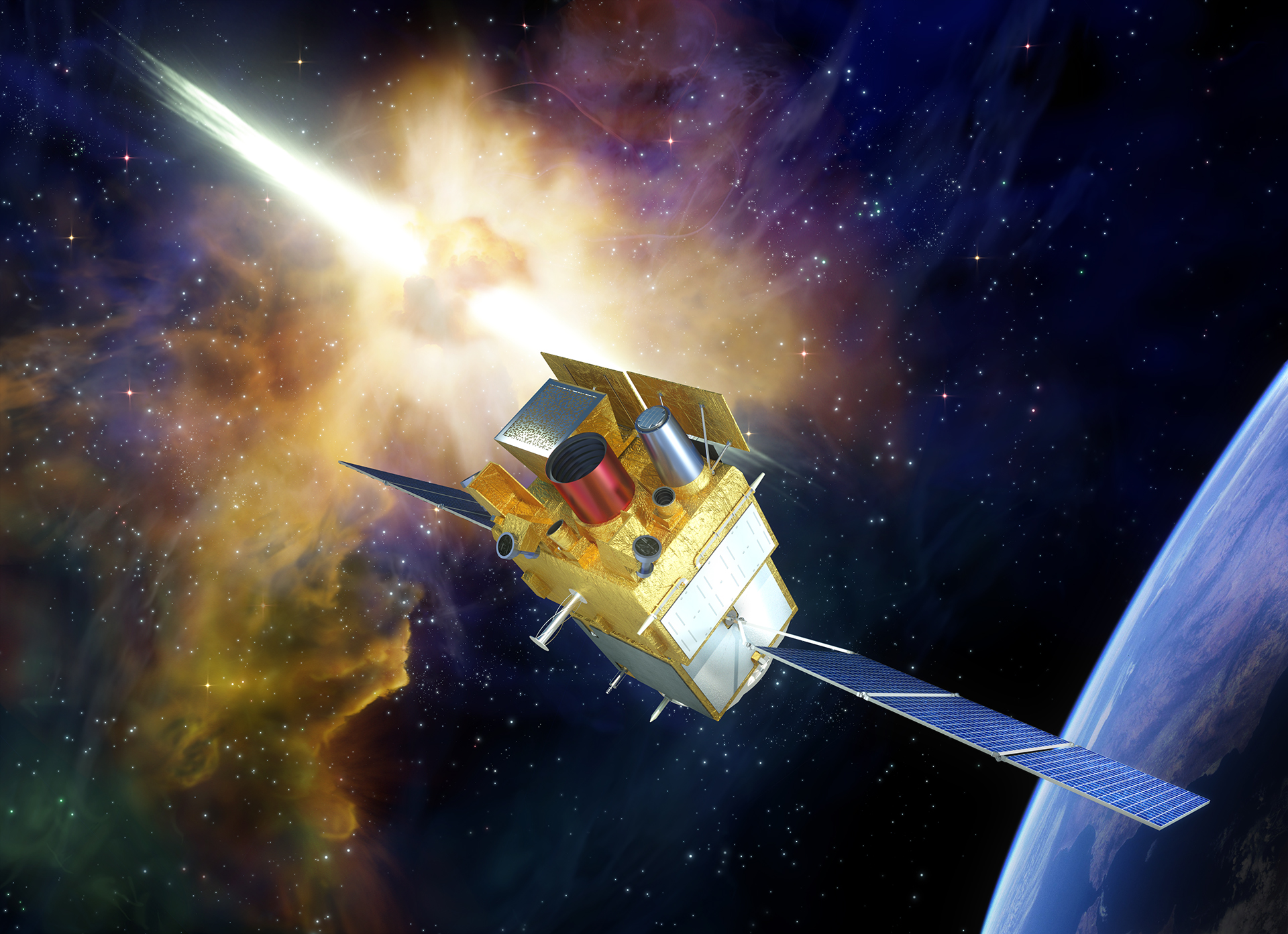}
	\end{center}
\end{figure}

\begin{center}
\Large Frontispiece : Artist view of the \textit{SVOM} satellite
\end{center}

\vspace{3\baselineskip}




\begin{center}
\noindent {\Large\bf Principal Investigators}
\end{center}

\noindent {J. Wei, National Astronomical Observatories/Chinese Academy of Science}\\
20A Datun Road, Beijing, 100012, China \vspace{1.5mm}\\ 
B. Cordier, CEA Saclay, DSM/IRFU/Service d'Astrophysique\\
91191 Gif-sur-Yvette, France\\ 
email: wjy@nao.cas.cn \& bertrand.cordier@cea.fr\\

\begin{center}
\noindent {\Large\bf Contributors}
\end{center}

\begin{multicols}{2}
\noindent {S. Antier, CEA, Saclay, France} \\
P. Antilogus, LPNHE, Paris, France\\
J.-L. Atteia, IRAP, Toulouse, France\\
A. Bajat, IRAP, Toulouse, France\\
S. Basa, LAM, Marseille, France\\
V. Beckmann, APC, Paris, France\\
M.G. Bernardini, LUPM, Monpellier, France\\
S. Boissier, LAM, Marseille, France\\
L. Bouchet, IRAP, Toulouse, France\\
V. Burwitz, MPI, Garching, Germany\\
A. Claret, CEA, Saclay, France\\
Z.-G. Dai, NJU, Nanjing, China\\
F. Daigne, IAP, Paris, France\\
J. Deng, NAOC, Beijing, China\\
D. Dornic, CPPM, Marseille, France\\
H. Feng, DEP, Tsinghua University, Beijing, China\\
T. Foglizzo, CEA, Saclay, France\\
H. Gao, BNU, Beijing, China\\
N. Gehrels, NASA, Greenbelt, USA\\
O. Godet, IRAP, Toulouse, France\\
A. Goldwurm, APC, Paris, France\\
F. Gonzalez, CNES, Toulouse, France\\
L. Gosset, CEA, Saclay, France\\
D. G\"otz, CEA, Saclay, France\\
C. Gouiffes, CEA, Saclay, France\\
F. Grise, Obs. astron. de Strasbourg, France\\
A. Gros, CEA, Saclay, France\\
J. Guilet, MPA, Garching, Germany\\
X. Han, NAOC, Beijing, China\\
M. Huang, NAOC, Beijing, China\\
Y.-F. Huang, NJU, Nanjing, China\\
M. Jouret, CNES, Toulouse, France\\
A. Klotz, IRAP, Toulouse, France\\
O. La Marle, CNES, Toulouse, France\\
C. Lachaud, APC, Paris, France\\
E. Le Floch, CEA, Saclay, France\\
W. Lee, UNAM, Mexico, Mexico\\
N. Leroy, LAL, Orsay, France\\
L.-X. Li, KIAA, PKU, Beijing, China\\
\smallbreak
\noindent
S. C. Li, SECM, Shanghai, China\\
\noindent
Z. Li, Dept. of Astronomy, PKU, Beijing, China\\
E.-W. Liang, GXU-NAOC, Nanning, China\\
H. Lyu, Guangxi University, Nanning, China\\
K. Mercier, CNES, Toulouse, France\\
G. Migliori, CEA, Saclay, France\\
R. Mochkovitch, IAP, Paris, France\\
P. O’Brien, Dept. of Astronomy, UL, Leicester, UK\\
J. Osborne, Dept. of Astronomy, UL, Leicester, UK\\
J. Paul, CEA, Saclay, France\\
E. Perinati, IAAT, T\"ubingen, Germany\\
P. Petitjean, IAP, Paris, France\\
F. Piron, LUPM, Montpellier, France\\
Y. Qiu, NAOC, Beijing, China\\
A. Rau, MPE, Garching, Germany\\
J. Rodriguez, CEA, Saclay, France\\
S. Schanne, CEA, Saclay, France\\
N. Tanvir, University of Leicester, UK\\
E. Vangioni, IAP, Paris, France\\
S. Vergani, GEPI, Meudon, France\\
F.-Y. Wang, NJU, Nanjing, China\\
J. Wang, NAOC, Beijing, China\\
X.-G. Wang, GXU-NAOC, Nanning, China\\
X.-Y. Wang, NJU, Nanjing, China\\
A. Watson, UNAM, Mexico, Mexico\\
N. Webb, IRAP, Toulouse, France\\
J. J. Wei, PMO, Nanjing, China\\
R. Willingale, Dept. of Astron., UL, Leicester, UK\\
C. Wu, NAOC, Beijing, China\\
X.-F. Wu, PMO, Nanjing, China\\
L.-P. Xin, NAOC, Beijing, China\\
D. Xu, NAOC, Beijing, China\\
S. Yu, SECM, Shanghai, China\\
W.-F. Yu, SHAO, Shanghai, China\\
Y.-W. Yu, CCNU, Wuhan, China\\
B. Zhang, UNLV, Las Vegas, USA\\
S.-N. Zhang, IHEP, Beijing, China\\
Y. Zhang, SECM, Shanghai, China\\
X.L. Zhou, NAOC, Beijing, China\\

\end{multicols}

\vspace{2.5mm}



\newpage
\tableofcontents 

\newpage
\begin{center}
\noindent {\Large\bf EXECUTIVE SUMMARY}
\end{center}

To take advantage of the astrophysical potential of Gamma-Ray Bursts (GRBs), Chinese and French astrophysicists have engaged the \textit{SVOM} mission (Space-based multi-band astronomical Variable Objects Monitor), aiming to:

\begin{itemize}
\item permit the detection of all known types of GRBs,
\item provide fast, reliable GRB positions,
\item measure from visible to MeV the spectral shape of the GRB prompt emission,
\item measure from visible to MeV the temporal properties of the GRB prompt emission,
\item quickly identify the afterglows of detected GRBs at X-ray and visible wavelengths, including those which are highly redshifted (z$>$5),
\item measure from visible to X-rays the spectral shape of the early and late GRB afterglow,
\item measure from visible to X-rays the temporal evolution of the early and late GRB afterglow.
\end{itemize}


Major advances in GRB studies resulting from the synergy between space and ground observations, the \textit{SVOM} mission implements space and ground instrumentation. 
The space segment includes:

\begin{itemize}
\item a wide field-of-view hard X-ray imager and spectrometer,
\item a wide field-of-view soft gamma-ray spectrometer,
\item a narrow field-of-view low-energy X-ray telescope,
\item a narrow field-of-view visible/near infrared (NIR) telescope.
The ground segment includes:
\item two follow-up telescopes (one of which featuring efficient NIR capabilities),
\item an array of wide field-of-view visible cameras.
\end{itemize}

The scientific objectives of the mission put a special emphasis on two categories of GRBs: very distant GRBs at z$>$5 which constitute exceptional cosmological probes, and faint/soft nearby GRBs which allow probing the nature of the progenitors and the physics at work in the explosion. These goals have a major impact on the design of the mission: the on-board hard X-ray imager is sensitive down to 4 keV and computes on line image and rate triggers, and the follow-up telescopes on the ground are sensitive in the NIR. 
\medbreak
At the beginning of the next decade, \textit{SVOM} will be the main provider of GRB positions and spectral parameters on very short time scale. The \textit{SVOM} instruments will operate simultaneously with a wide range of powerful astronomical devices. This rare instrumental conjunction, combined with the relevance of the scientific topics connected with GRB studies, warrants a remarkable scientific return for \textit{SVOM}. 
\medbreak
In addition, the \textit{SVOM} instrumentation, primarily designed for GRB studies, composes a unique multi-wavelength observatory with rapid slew capability that will find multiple applications for the whole astronomy community beyond the specific objectives linked to GRBs. For example, the \textit{SVOM} mission has been conceived to promptly scrutinize the celestial fields where sources have been detected by wide field-of-view astronomical devices such as the upgraded generation of gravitational wave detectors (advanced Virgo/LIGO) and high-energy neutrino detectors (KM3NeT, IceCube).
\medbreak
The following pages list the scientific themes that will benefit from observations made with \textit{SVOM}, whether they are specific GRB topics, or more generally all the issues that can take advantage of the multi-wavelength capabilities of \textit{SVOM}.


\section{Context}
\subsection{Time-domain astrophysics: the discovery space after \textit{Swift}}
\label{sec:timedomain}
Astronomy is truly undergoing a revolution in terms of our ability to
monitor the time-variability of the Universe in a continuous way using
new facilities coupled with fast computers. The opening up of the
temporal domain is transforming our knowledge of how the Universe
evolves, particularly for objects which are undergoing explosive
change, such as a supernova or a Gamma-ray Burst (GRB), 
e.g. \citet{kumar:2015}. These
explosive events can release enormous amounts of power both in
electromagnetic radiation and in non-electromagnetic forms such as
neutrinos and gravitational-waves and test our understanding of the
laws of physics under the most extreme conditions.
\medbreak
Observing facilities which are currently on-line enable the sky to be
monitored fairly continuously in real-time over large areas and across
the electromagnetic spectrum, capturing the temporal behaviour of the
Universe in a way previously unattainable. Examples facilities include
the LOFAR radio telescope \citep{vanHaarlem:2013}, the Pan-STARRs
optical facility (Chambers et al. in preparation) and the \textit{Swift} 
\citep{gehrels:2004}
and \textit{Fermi} 
high-energy
satellites. Non-electromagnetic facilities are also now observing,
particularly the Advanced LIGO-VIRGO gravitational-wave observatory
\citep{aasi:2015},
which recently found its first source \citep{Abbott:2016}, and the
IceCube neutrino experiment \citep{aartsen2015bis}. The data from all these facilities have
already opened up the temporal domain, but are just a foretaste of
what is to come.
\medbreak

Many of the previously developed theories have come under intense
strain by new observational results, such as the highly variable
emission seen at late times in GRBs, the discovery of extremely
luminous supernovae and the unexplained fast radio bursts \citep{lorimer2007}. Theoretical
models predict a variety of exotic explosions and stellar mergers,
together with their multiple signatures across the electromagnetic
spectrum. Theory also predicts that some will be accompanied by
gravitational wave, neutrino and high-energy particle emission. The
provision of \textit{SVOM} in the next decade will coincide with the
multi-messenger era and will provide a critical element of the era of
time-domain astronomy both by finding transients and by following-up
those from other facilities.
\medbreak

In the period when \textit{SVOM} will fly, the number of transients found will
increase by several orders of magnitude as even more powerful
facilities come on-line, in particular the Large Synoptic Survey
Telescope (LSST, \citealt{lsst:2009}) and the Square Kilometre Array
(SKA). The sheer grasp of the new facilities, which will produce
thousands of “alerts” per day from variable or transient sources, mean
we will require super-computers to process the data in real-time and
smart algorithms to broker which transients to focus on with follow-up
facilities. The importance of the temporal domain has been recognised
in recent reports by the Chinese Space and Technology Roadmap
\citep{china:2010}, the European Union ASTRONET group \citep{ASTRONET:FP7}
 and the USA
National Research Council Decadal Survey \citep{decadal:2010}.
\medbreak

While we can confidently predict the importance of \textit{SVOM} data based on
current observations, time domain experiments often make their most
startling discoveries in unexpected and serendipitous ways. This has
certainly been the experience from the \textit{Swift} mission and Pan-STARRSs,
for example. The power of \textit{SVOM} combined with the grasp of the future
facilities will without doubt stimulate a new revolution in astronomy.


\subsection{The astronomical panorama in 2020}
\label{sec:panorama}

\begin{figure}[hptb]  
   \begin{center}
	\includegraphics[trim=0 0 0 0,scale=0.7]{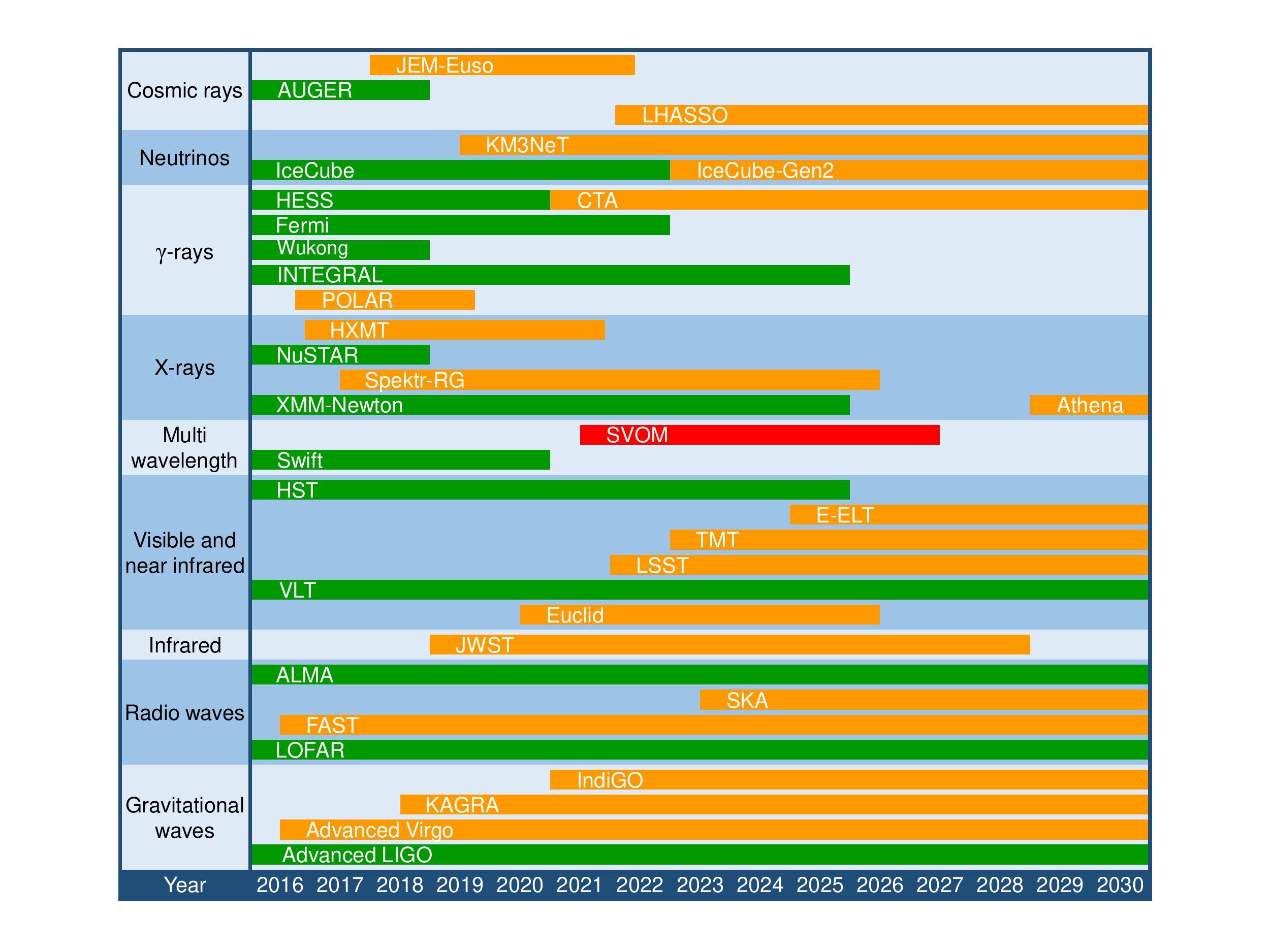}
	\caption{The observational panorama in the \textit{SVOM} era.}
	\label{fig:pano}
	\end{center}
\end{figure}

The astronomical panorama of the next decade will be shaped by new instruments developed to address various outstanding questions raised by present day astrophysics. 
This panorama encompasses large radio, infrared, visible and gamma-ray telescopes, advanced gravitational wave interferometers and neutrino detectors of the km$^3$ class (Fig. \ref{fig:pano}), as well as simulations with powerful computers. 
These instruments will revolutionize our understanding of astrophysics in fields as diverse as the first ages and the reionization of the universe, the nature of the dark universe (dark matter and dark energy), the demography and role of black holes, exoplanets and planetary formation, and fundamental physical processes. 
Young fields, like Time Domain Astronomy and Multi-Messenger Astrophysics are also expected to grow very fast, bringing new discoveries. 

In this panorama, \textit{SVOM} is expected to offer two unique capabilities, which are currently being provided by \textit{Swift}: monitoring the hard X-ray sky for high-energy transients and doing multi-wavelength follow-up of self-triggered high-energy transients and remarkable transients detected with other facilities (upon ToO request).
We have selected below some topics that will undoubtedly benefit from a close collaboration between \textit{SVOM} and other astronomical facilities.

\textbf{A universe of black holes:}
 \textit{SVOM} will provide crucial information on transient black hole (BH) activity.
The follow-up of a majority of \textit{SVOM} detected transients with large optical and radio telescopes on the ground, will permit measuring the history of stellar mass BH formation up to redshift z=10 and compare it with the star formation rate.
\textit{SVOM} will also increase the statistics on Tidal Disruption Events (TDEs), providing a better understanding of dormant massive BHs at the center of galaxies, while the optical to gamma-ray energy coverage of \textit{SVOM} will permit a better understanding of the physics at work in Active Galactic Nuclei (AGNs), especially blazars.

\textbf{The sources of gravitational waves:}
\textit{SVOM} will benefit from the operation of advanced gravitational wave (GW) detectors to explore the zoo of transient GW sources. 
These studies can elucidate the origin of short GRBs, identify which GRBs are coming from binary mergers and constrain their beaming angle. 
\textit{SVOM} will also help GW detectors to identify low significance signals from binary mergers.

\textbf{The epoch of reionization:}
GRBs provide unique glimpses into the epoch of reionization (z$\ge 6$), for instance they permit studying small high redshift galaxies, which cannot be observed by other means.
They also have the potential to constrain the contribution of massive stars to the reionization of the universe and tell us if the first stars (population III) produce GRBs.
These studies, which rely on early measures of the optical spectra of GRBs beyond z=6, emphasize the need to recognize high-z GRBs quickly, a task that will be eased by sensitive follow-up with \textit{SVOM} narrow-field instruments.

\textbf{Physics of relativistic jets:}
The physics of accretion/ejection is a fundamental problem in astrophysics. 
High-energy galactic transients, GRBs and AGNs offer complementary ways to capture the dynamic of this process. 
With built-in multi-wavelength capabilities and a close collaboration with radio and VHE telescopes on Earth, \textit{SVOM} will observe  transient activity from galactic and extragalactic relativistic jets, providing insight into their nature, origin, acceleration mechanism and radiation processes, with impacts on our understanding of the origin of VHE cosmic-rays.
\textit{SVOM} observations will benefit from complementary observations of GRB jets.
Polarimetry of the prompt emission (e.g. POLAR) will bring crucial insight into the magnetic field configuration of GRB jets, while km$^3$ neutrino detectors (ICECUBE, KM3Net) may provide unique clues on the nature and energy content of relativistic jets (hadronic vs leptonic or magnetically dominated).

\textbf{GRB progenitors:}
Our understanding of GRB progenitors will make great progress with \textit{SVOM}. 
The low energy threshold of \textit{SVOM} will permit the detection of nearby XRFs, clarifying the connection between "core-collapse GRBs" and SNIbc. 
The frequency of orphan afterglows detected with the LSST will permit measuring the true GRB rate and establishing meaningful comparison with the rate of SNIbc.
Regarding short GRBs, they will benefit from a double diagnostic: GW interferometers will be crucial for the identification of "merger GRBs" among \textit{SVOM} triggers, while optical observations with VT will permit the early detection of optical afterglows and searches for kilonova emission.

\textbf{Galactic monsters:}
The low energy threshold of ECLAIRs will permit the detection of thousands of transients from flare stars, white dwarves (novae), galactic neutron stars (X-ray bursters and magnetars) and active black holes (microquasars), and their detailed follow-up with MXT and VT. 
\textit{SVOM} observations, will permit powerful diagnostic of many important physical processes at work in stellar flares, nuclear burning on neutron stars and white dwarves, accretion/ejection or the origin of magnetar activity, especially if they are part of multi-wavelength campaigns involving ground based radio, NIR and VHE telescopes.

\textbf{Other transients:}
In the next decade, many astronomical instruments will monitor the sky for transients.
In the visible sky, LSST will detect each year dozens of orphan GRB afterglows and thousands of luminous supernovae and tidal disruption events, and much closer to us, thousands of gravitational lensing events in our galaxy. 
In radio, we anticipate the detection of thousands of fast radio bursts.
For the most interesting events, \textit{SVOM} may act as a follow-up machine combining X-ray and optical sensitivity, fast response (hours) and high availability.

In conclusion, it is clear that the highest benefit of \textit{SVOM} will be obtained in collaboration with the large astronomical facilities existing in the 2020 decade.
These collaborations will go in both directions: large facilities observing \textit{SVOM} high-energy transients, and \textit{SVOM} observing remarkable transients discovered by other facilities, with its narrow field telescopes.
The scientific outcome anticipated from such collaborations fully justifies efforts to establish strong collaborations well in advance.


\newpage
\section{\textit{SVOM} Mission}
\newcommand{\myRem}[1]{{ }}
\label{sec:SVOMmission}

\subsection{The \textit{SVOM} mission profile}

The \textit{SVOM} satellite will be launched by a Chinese launcher LM-2C from Xichang and will be inserted into a Low Earth Orbit with an inclination of 30$^o$, an altitude of 625 km and an orbital period of $\sim$96 min. With these parameters the satellite passes through the South Atlantic Anomaly several times per day, inducing an overall dead-time of 13 to 17\%.

\begin{figure}[hptb]
\begin{center}
\includegraphics[page=1,bb=0 170 720 540, width=1\linewidth]{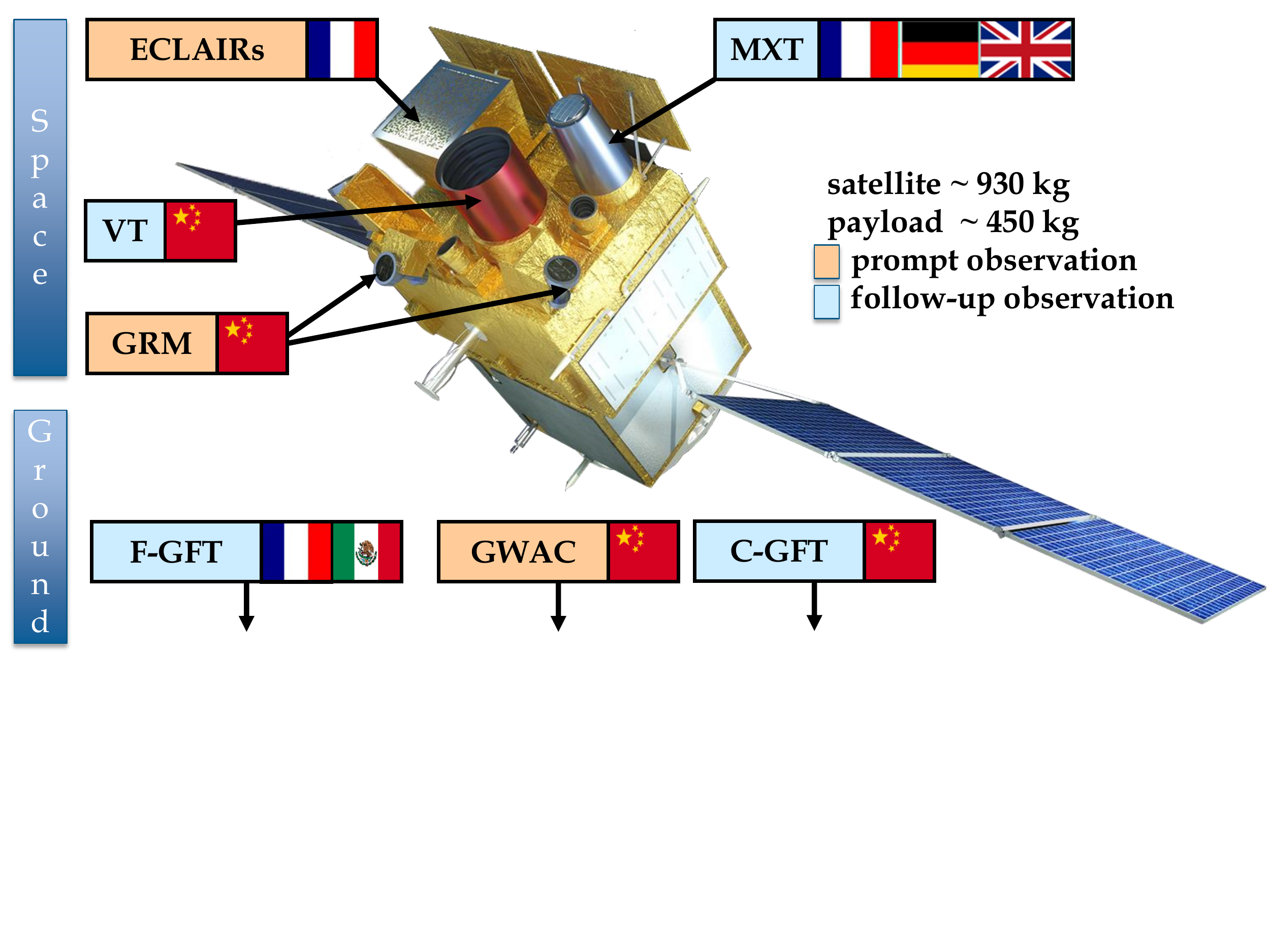}
\caption[\textit{SVOM} scientific instruments]{View of the \textit{SVOM} space-based and ground-based instruments.\label{fig:svom_instr}}
\end{center}
\end{figure}

\textit{SVOM} (Fig.~\ref{fig:svom_instr}) carries two wide field of view (FoV) high-energy instruments: a coded-mask gamma-ray imager (ECLAIRs), and a gamma-ray spectrometer (GRM), and two narrow field telescopes: a Microchannel X-ray Telescope (MXT) and a Visible-band Telescope (VT). The \textit{SVOM} ground segment includes additional instruments: a wide angle optical camera (GWAC) monitoring a part of the ECLAIRs FoV in real-time, and two 1-m-class robotic follow-up telescopes (the GFTs).

In order to facilitate redshift measurements of the GRBs detected by \textit{SVOM}, the satellite attitude will follow a predefined orientation, called B1 attitude law. Most of the year the optical axis of the \textit{SVOM} instruments will be pointed 45$^o$ offset from the anti-solar direction. This pointing includes avoidance periods in order to exclude the Sco-X1 source and the galactic plane from the ECLAIRs field of view. An additional constraint favors areas of the sky observable by large telescopes located in Chile, Hawa\"i and Canary Island. This strategy ensures that \textit{SVOM} GRBs will be detected towards the night hemisphere, quickly observable from ground by large telescopes, and optimizes the chances to detect the GRB counterparts and host galaxies. More details on the \textit{SVOM} pointing strategy can be found in \cite{cord}. 

As a consequence of the Low-Earth Orbit combined with a roughly anti-solar attitude law, the Earh occults each orbit the FoV of the \textit{SVOM} instruments. 
The mean duty cycle is about 65\% for ECLAIRs and 50\% for the narrow-FoV instruments MXT and VT.


\begin{figure}[ht] 
\begin{center}
\includegraphics[scale=0.4]{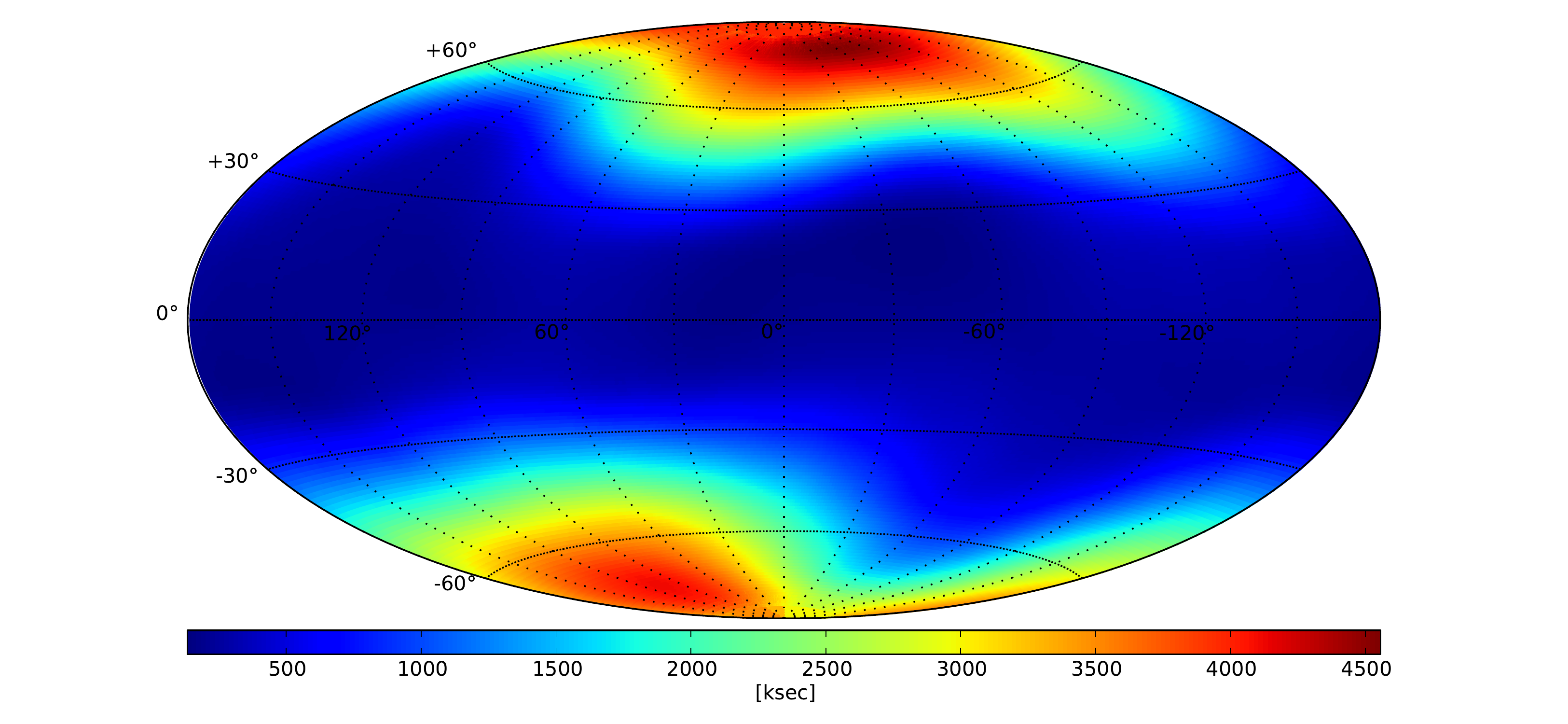} \hfill
\includegraphics[scale=0.36]{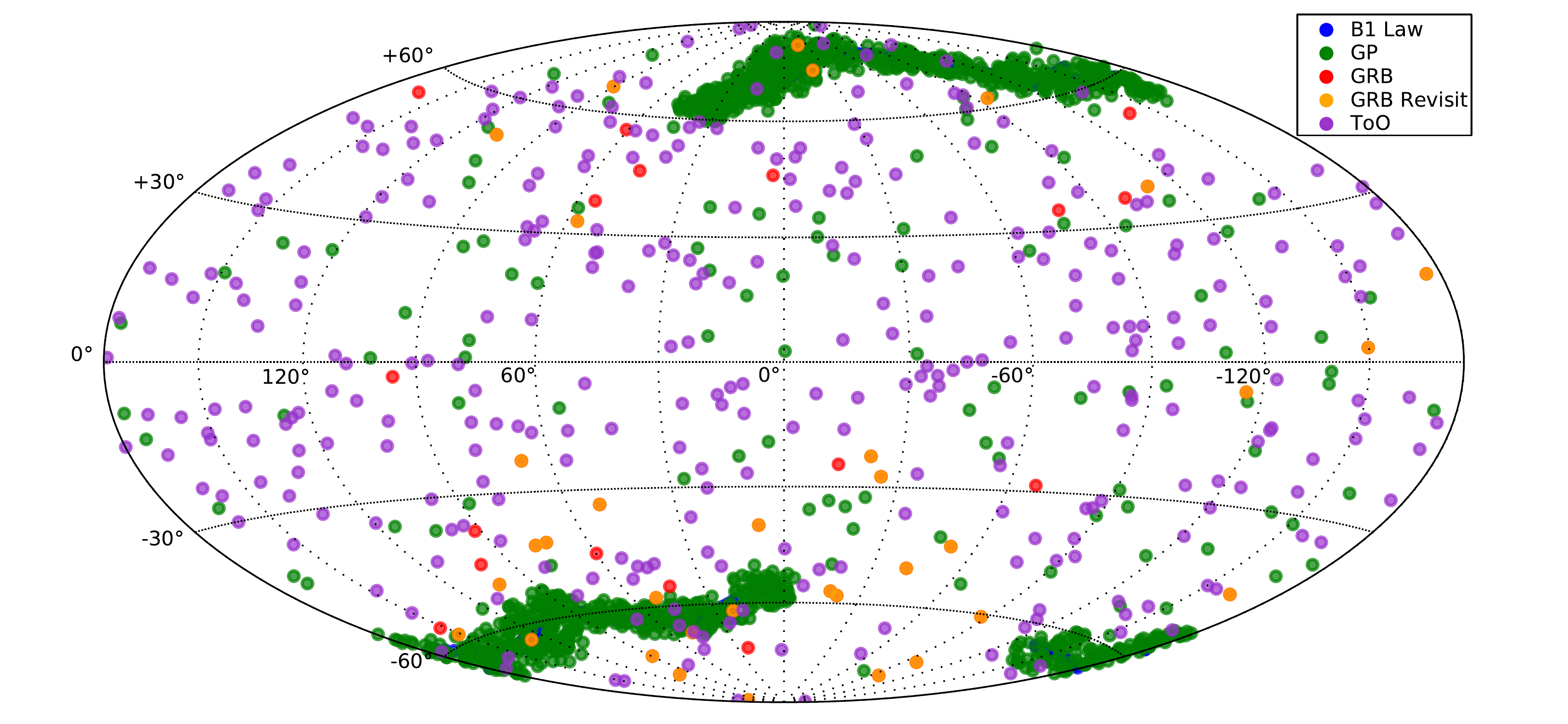}
\caption{Scenario of 1 year of observation following the B1 attitude law, simulating 65 GRBs, and one ToO per day. (Up) Sky exposure in ksec and in galactic coordinates for the ECLAIRs telescope (wide FoV). (Bottom) Map in galactic coordinates of the \textit{SVOM}  pointing direction for the same scenario, corresponding to the targets observed by the VT and MXT (narrow FoV).}
\label{fig:svom_b1law}
\end{center}
\end{figure}

Most often the \textit{SVOM} pointing follows the B1 attitude law (Fig.~\ref{fig:svom_b1law}), waiting for a GRB.
When ECLAIRs triggers on a GRB event, it communicates its position to the spacecraft, which slews autonomously within minutes to the source for rapid follow-up observations of the GRB afterglow emission with the MXT in X-rays and the VT in the visible band. The satellite stays pointed towards the source for 14 orbits ($\sim$1 day). The GRB position and its main characteristics determined by ECLAIRs are also quickly sent to the ground using the \textit{SVOM} VHF emitter. Refined positions of the X-ray counterpart detected by MXT are also down-linked via VHF. The VHF signal is received by one of the 40 to 50 ground stations, distributed on Earth under the satellite track. The system design ensures that 65\% of the alerts are received within 30 s at the Science Centers, which forward them via internet (using the GCN and VOEvent networks) to the \textit{SVOM} ground instruments (GWAC and GFTs) and to the scientific community. 


\begin{figure}[hptb]
\begin{center}
\includegraphics[trim=150 160 150 170, scale=1., clip=true]{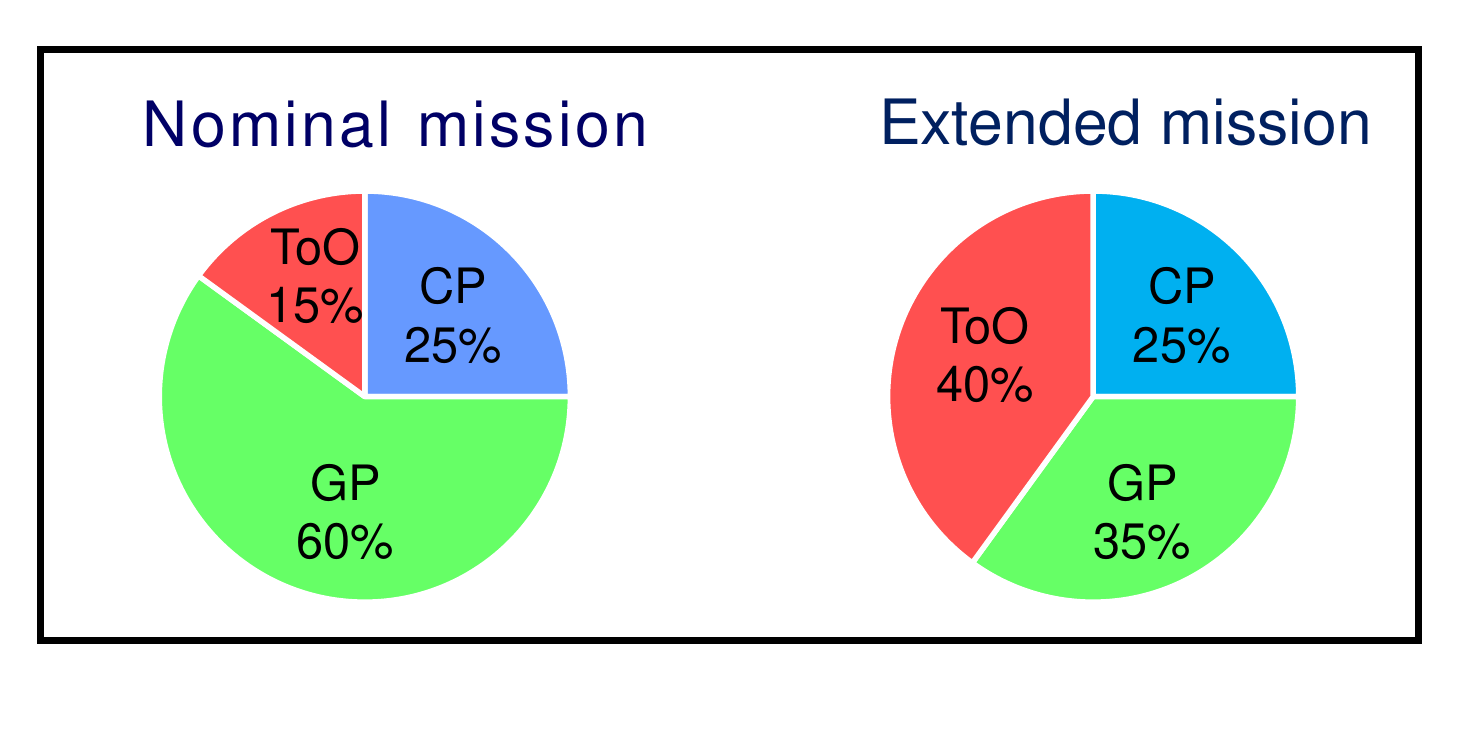}
\caption[\textit{SVOM} attitude law]{Allocated observation-time fraction for the three \textit{SVOM} scientific programs : Core Program (CP), Target of Opportunity Program (ToO), General Program (GP). The extended mission (3 years after the launch) involves a higher ToO fraction and more GP time outside the B1 law (from 10\% to 50\%).\label{fig:svom_timeallow}}
\end{center}
\end{figure}

The Core Program of the mission covers all \textit{SVOM} observations related to the detection and characterization of the prompt and afterglow emission of GRBs. With an expected GRB rate of around 60-70 per year, the Core Program accounts for about 25\% of the useful mission time.

The Target of Opportunity (ToO) program of \textit{SVOM} manages unplanned observations of transient and variable sources programmed from the ground.
All scientists will have the opportunity to apply for ToOs, evaluated by the PIs.
Accepted observations are performed immediately by the \textit{SVOM} ground instruments, and within some delay by the \textit{SVOM} space instruments, depending on the availability of the satellite up-link stations.
Observations are performed within 48 h for a standard ToO, and within 12 h for an exceptional ToO (e.g. galactic supernova or GW alert) using on request additional stations.
The typical observation duration is 1 orbit (45 min useful time) for a standard ToO, and 14 orbits ($\sim$10 h) for an exceptional ToO.
A significant portion of the useful time, 15\% at the beginning of the mission, is foreseen for ToO observations.

The General Program (GP) of \textit{SVOM} is devoted to pre-planned observations and complements the transient sky observations managed by the Core Program and ToOs.
The GP is built taking into account the system requirements of the GRB program, in particular the pointing strategy that optimizes the ground follow-up of \textit{SVOM} GRBs. 
Therefore the GP allows pointing at sources close (i.e. within 5$^o$ to 10$^o$) to the B1 attitude law.
In order to increase the scientific interest of the GP, it is foreseen to allow observations outside the B1 constraint during 10\% of the GP useful time.
The minimal duration of a GP observation is 1 orbit ($\sim$45 min).
The GP is open to all scientists responding to \textit{SVOM} calls for observation proposals, issued every year, and evaluated by a Time Allocation Committee based on their scientific merit.

In order to contribute to the opening-up of the time-domain astronomy scheduled for the next decade, an evolution of the \textit{SVOM} time allocation is planned (Fig.~\ref{fig:svom_timeallow}), increasing the ToO-time from 15\% to 40\%, allowing for 5 instead of 1 ToO per day.
In the same time the GP useful time permitted outside the B1 constraint increases from 10\% to 50\%.

The scientific products of the GRB Core Program and the ToOs are made public as soon as available.
GP data remain restricted during 1 year to the PI of each accepted proposal, after which they become public.

\subsection{The \textit{SVOM} instruments}

The \textit{SVOM} spacecraft consists of two wide-field instruments: ECLAIRs and the Gamma-Ray Monitor (GRM) for the observation of the prompt emission and two narrow field instruments: the Micro-channel X-ray Telescope (MXT) and the Visible Telescope (VT) for the observation of the afterglow emission.
\textit{SVOM} has also two sets of ground dedicated-instruments: a wide-field instrument GWAC for the observation of the optical prompt emission, and narrow-field instruments GFT for the follow-up observations in the visible and near-infrared domain.
This section describes individually each \textit{SVOM} instrument.



\subsubsection{The soft gamma-ray imager ECLAIRs onboard \textit{SVOM}}

The main goal of the hard X-ray coded-mask imager ECLAIRs
\footnote{institutes: APC Paris, CNES Toulouse, IAP Paris, IRAP Toulouse, IRFU/CEA Saclay, LUPM Montpellier (France)}, operating in the 4-150 keV energy range, is the detection and fast localization of hard X-ray transients onboard \textit{SVOM}.
Its low energy threshold of 4 keV will open to \textit{SVOM} the realm of extragalactic soft X-ray transients, such as X-Ray Flashes or SN shock breakouts, which are still poorly explored.

Coded-mask imaging is very efficient to survey large fractions of the hard X-ray sky, as demonstrated by instruments such as \textit{INTEGRAL}/ISGRI and \textit{Swift}/BAT \citep{Lebrun:03,Barthelmy:05}.
The ECLAIRs detection plane is made of 6400 pixels of CdTe ($4 \times 4$ mm$^2$, 1-mm thick) and its read-out electronics.
A $54 \times 54$ cm$^2$ coded mask with 40\% open fraction is located 46 cm above the detection plane to define a FoV of 2 sr (zero coded FoV) and a point spread function of 52 arcmin (FWHM).
A passive lateral Pb/Al/Cu-layer shield blocks the hard X-ray radiation originating from outside the FoV.

The on-board management and scientific trigger electronics (UGTS) configures the instrument and searches for new transient sources within reconstructed sky images of the ECLAIRs FoV, and alerts in near-real time the satellite about their localization.
Hard X-ray transients are identified as count-rate excesses on the detection plane (monitored on time scales from 10 ms to about 10 s on 4 energy bands and in 9 detector zones) followed by their sky imaging, or directly as new sources in sky images constructed cyclically (on time scales from 10 s to about 1300 s).
The UGTS stores each detected pixel-hit to mass memory, allowing detailed on-ground data analysis, among which offline searches of additional transients sources.

ECLAIRs has been optimized taking into account the limited resources (mass, power, volume) available on the \textit{SVOM} spacecraft, which allow a sensitive area of 1000 cm$^2$ and a point source localization error (PSLE) better than 12 arcmin (for 90\% of the sources at detection limit). 
Major challenges are the 4 keV energy-threshold and the on-board detection of transients on top of a strongly varying background (of about 3000 counts/s), modulated by the SAA passages and the Earth transit through the FoV every orbit. 
Detection-plane prototype measurements and simulations performed with the trigger software implemented suggest that those challenges are met by the instrument design \citep{Antier:15, Godet:14, Lacombe:14, Nasser:14, Schanne:14, Zhao:12}, and
predict a detection each year of 60 GRBs, several non-GRB extragalactic transients, dozens of AGNs and hundreds of galactic X-ray transients and persistent sources. 
Detailed descriptions of ECLAIRs are provided in a series of articles: \citet{Schanne:15, Godet:14, Lacombe:14, LeProvost:14, Schanne:14}. 

\myRem{ 

\begin{table}[htb] 
\label{tab:eclairs}
\begin{center}
\begin{tabular}{ll}
\hline 
\multicolumn{2}{c}{Detection plane}\\
\hline 
Energy range & 4-150 keV\\
Sensitive area & $1000\sim{\rm cm}^2$ \\
\hline
\multicolumn{2}{c}{Imaging}\\
\hline 
Field of view & 2.04 sr (zero response) \\
Mask open fraction & 0.4 \\
PSF$^a$ (FWHM) & 52 arcmin \\
PSLE$^b$ & 12 arcmin \\
\hline
\multicolumn{2}{c}{Trigger}\\
\hline 
Expected background$^c$ & $\sim 3000$ c/s\\
Trigger parameters & 20 time scales: $0.01 - 1300$ s \\
 & 4 energy bands \\
 & 9 detector zones \\
\hline 
\multicolumn{2}{l}{$^b$ The Point Spread Function measures the ability of an instrument to separate two nearby sources.} \\
\multicolumn{2}{l}{$^b$ Point Source Localization Error for a source at the limit of detection.} \\
\multicolumn{2}{l}{$^c$ Background for an open field of view, not occulted by the Earth \citep{Godet:08, Sizun:11}.} \\
\end{tabular}
\end{center}
\caption{Main characteristics of ECLAIRs}
\end{table}
} 

\myRem{ 
\begin{figure}[htb]
\centerline{\includegraphics[width=0.5\linewidth]{fig_eclairs.pdf}}
\caption{\small Schematic diagram showing the main components of the ECLAIRs instrument.}
\label{fig:eclairs}
\end{figure}

\begin{figure}[htb]
\includegraphics[width=0.50\textwidth]{fig_effarea.pdf}
\includegraphics[width=0.45\textwidth]{fig_psle.pdf}
\caption{\small Simulated performance of the ECLAIRs instrument.
\textbf{Left panel.} Effective area of the detection plane only (black) and of the complete instrument (red), from \citet{Sizun:11}.
\textbf{Right panel.} Point source localization accuracy as a function of the GRB signal-to-noise ratio. The orange, red and green curves  respectively show the radius of regions with confidence levels of 95\%, 90\% and 80\%.
}
\label{fig:eclairs_perfo}
\end{figure}
} 

\myRem{ 
\textbf{TBD Questions : \\
- Faut-il mentionner qui fait quoi, les laboratoires, le CNES ou bien est-ce deja fait dans un chapeau \textit{SVOM} ?\\
- Il faut donner un nombre de sursauts attendus qui soit coherent a travers tout le document, je n'ai rien mis ici pour l'instant.\\
}
} 

\subsubsection{The gamma-ray spectrometer GRM onboard \textit{SVOM}}

The Gamma-Ray Monitor\footnote{institute: IHEP Beijing (China)} 
(GRM) will contribute to GRB-related studies including a) GRB physics of progenitor, jet mechanism and components, energy dissipation mechanism and radiation mechanism, b) Multi-messenger studies through gravitational waves, neutrinos and high-energy cosmic rays, c) Cosmology and fundamental physics. 
In addition GRM will contribute to Terrestrial Gamma-ray Flashes (TGFs) studies by taking advantage of its low Earth orbit, wide FoV and energy range up to several MeV.

GRM consists of three detection modules (GRDs), each made of a NaI(Tl) scintillating crystal, a photomultiplier and its readout electronics. 
Each GRD has an area of 200 cm$^2$, thickness of 1.5 cm, FoV of $\pm$60 degrees with respect to its symmetry axis, 
dead time	$<$8 $\mu$s, temporal resolution $<$20 $\mu$s and energy resolution	of 16\% at 60 keV.
A plastic scintillator in front of the NaI(Tl) crystal is used to distinguish low energy electrons from gamma-rays. 
The 3 GRD-modules point at different directions to form a combined FoV of 2 sr, matching the ECLAIRs one, within which a rough localization ($\sim \pm$5$^o$) of transient sources can be achieved onboard. 
The energy range of the GRM is 15-5000 keV, extending towards high energies the range of ECLAIRs to measure $E_{peak}$ for a large fraction of \textit{SVOM} GRBs.
We expect that GRM will detect $>$90 GRBs/yr. 
GRM has a good sensitivity to short/hard GRBs, similar to GBM on \textit{Fermi}.
GRM generates by itself onboard triggers, whose time and crude localization is transferred to ECLAIRs for trigger enhancement, and to ground facilities (e.g. GWAC, GW experiments) for joint observations.

A calibration detector (GCD) containing a $^{241}$Am isotope is installed on the edge of each GRD for gain monitoring and energy calibration. 
In addition a charged particle monitor (GPM) will help to announce SAA entries and serve to protect the detection modules.

\myRem{ 
\begin{table}[htb]
\label{tab:grm}
\begin{center}
\begin{tabular}{ll}
\hline 
Detector	&	NaI(Tl) \\
Number of units	 & 3 \\
Energy range	& 15-5000 keV \\
Field of view	of each GRD & $\pm$60 degrees with respect to its symmetry axis \\
Field of view of all GRDs combined & $\pm$60 degrees with respect to +X axis \\
Sensitive area	& $>$200 cm$^2$  (each unit) \\
Dead time	& $<$8 $\mu$s \\
Temporal resolution	& $<$20 $\mu$s \\
Energy resolution	& 16\% at 60 keV \\
Burst observation rate &	$>$90 / year \\
GRB localization error	& $\sim$5 degrees \\
\hline
\end{tabular}
\end{center}
\caption{Main characteristics of GRM}
\end{table}
} 


\subsubsection{Combined ECLAIRs and GRM spectroscopy between 4 keV and 5 MeV}

The combined ECLAIRs and GRM dataset will allow a complete study of each GRB's prompt high-energy emission, from 4 keV to a few MeV, including characterization of its spectral shape.
Recent observations of prompt phases of GRBs in gamma-rays with \textit{Fermi}/GBM have revealed a complexity in their spectral shape, going beyond a simple phenomenological Band function (two smoothly connected power laws, \citep{1993ApJ...413..281B}), with the necessity to include an additional hard power law \citep{2013ApJS..209...34A,2010ApJ...725..225G,2010ApJ...716.1178A,2009ApJ...706L.138A}, whose extrapolation to low energy has also been observed in a few cases \citep{2013A&A...550A.102T}, or a thermal emission component \citep{2011ApJ...727L..33G,2013ApJ...770...32G}.
A GRB population simulation, based on the \textit{Fermi/GBM} spectral catalog \citep{2014ApJS..211...12G} and on specific GRBs with remarkable spectral characteristics, shows that the combination of ECLAIRs and GRM will constrain the spectral parameters (SED peak energy, spectral slopes, kT, etc.) with good precision, providing direct insights on the prompt emission mechanisms (e.g. \citealt{2003A&A...406..879G}).


\subsubsection{The ground-based visible telescope array GWAC}

The Ground-based Wide Angle Camera system\footnote{institute: NAOC Beijing (China)} 
(GWAC) of \textit{SVOM} is designed to observe in the visible domain the prompt phase of \textit{SVOM} GRBs.
During this phase, in which most of the energy is emitted in $\gamma$-rays, observations of emission at other wavelengths, particularly in the visible, are believed to provide clues to the understanding of the jet composition, as well as the energy dissipation, particle acceleration and radiation mechanisms.
Up to now, only in a few cases prompt emission has been observed from visible to $\gamma$-rays, e.g. in the naked-eye burst GRB080319B \citep{racusin2008} and GRB130427A \citep{vestrand2014} and a larger sample promises to revolutionize the understanding of GRB physics.
To do so, simultaneous observations at high cadence in visible and $\gamma$-rays are needed throughout the prompt and into the early afterglow phase.

\myRem{
The origin GRB prompt emission remains not fully understood after more than 40 years of observations. During the prompt phase, most of the outburst energy is emitted in the gamma rays, but the emission from other wavelength, particularly in optical, generated simultaneously with gamma-rays during a gamma-ray burst are believed to provide clues about the nature of the explosions that occur as massive stars collapse or merger of compact objects. In order to unveiling nature of the physics during the prompt phase of a gamma-ray burst, no time delayed temporal coverage relative to the GRB event onset and high cadence observations are needed throughout the prompt phase and well into the early afterglow phase. 

Up to date, only a few cases with well observations from optical to gamma-ray during prompt phase are well studies, like the naked-eye burst GRB 080319B (Racusin et al., 2008), the second brightest burst GRB 130427A (Vestrand et al., 2014). These optical prompt data provided clues to understand jet composition, energy dissipation mechanism, particle acceleration mechanism and radiation mechanisms, but lots of debates are declared since of the seldom cases. Detailed investigations with larger samples observed well from optical to gamma-rays in the future will no doubt help people to revolutionize their knowledge on GRB physics.
} 

GWAC is designed to observe more than 12\% of \textit{SVOM} GRBs in the visible band from 5 min before to 15 min after the trigger.
For a high observation efficiency, the total GWAC array will comprise 36 camera units, covering a total FoV of 5400 sq.~deg.
For a significant detection efficiency, the sensitivity (at 5$\sigma$ for 10 s exposures) will reach a limiting magnitude M$_{\rm V}$=15 under full Moon condition (M$_{\rm V}$=16 for new Moon).
Each camera unit will have an effective aperture size of 18 cm, a 4096$\times$4096 E2V back-illuminated CCD, operating in the 0.5 to 0.85 $\mu$m band, and a FoV of 150 sq.~deg, providing source localizations of 11 arcsec in 13 s exposure images.
GWAC will be saturated for objects brighter than M$_{\rm V}$=11.

A prototype version, the Mini-GWAC, is designed to observe bright GRBs, using 12 camera units covering about 5200 sq.~deg, in which each camera has an aperture of 7 cm and a FoV of $\sim$440 sq.~deg, reaching M$_{\rm V}$=11.5 under full Moon (M$_{\rm V}$=12.5 for new Moon), providing source localizations of 30 arcsec in 15 s exposure images.

Both GWAC and Mini-GWAC search for optical transients in real time around the  trigger coordinates from ECLAIRs, GRM, MXT, VT, GFTs, and other instruments.

Considering their super-large FoV and short cadence exposures, GWAC and Mini-GWAC are powerful instruments for searching short-time scale optical transients, such as counterparts of gravitational wave events, early phases of supernovae, counterparts of neutrino events, flare stars, near-Earth objects, etc.

GWAC and Mini-GWAC have also the capability to detect optical transients by themselves without external triggers.
Two 60 cm telescopes and several 30 cm telescopes will be setup beside to follow-up optical transient candidates detected by GWAC and Mini-GWAC, in order to check if they are real or false events, providing coordinates with $\sim$1 acrsec accuracy, and delivering multi-band light-curves for confirmed transients.

Both GWAC and Mini-GWAC have three observing modes. 
A first observing mode, designed for GRB prompt emission studies, points the cameras to the FoV of ECLAIRs and tracks the stars. 
In a second mode, designed to follow-up ToO requests, one or several cameras are pointed promptly to each accepted ToO target.
In a third mode, an even larger FoV is surveyed, by scanning several areas cyclically with each camera.

The GWAC, Mini-GWAC and follow-up telescopes will be divided into two sets, each of which having 18 GWAC cameras, 6 Mini-GWAC cameras, one 60 cm telescope, and several 30 cm telescopes. 
As a baseline, one set will be setup at Ali observatory, China, the other one at CTIO, Chile.

\myRem{

\begin{table}[htb]
\label{tab:gwac}
\begin{center}
\begin{tabular}{lll}
\hline 
 & GWAC	& Mini-GWAC \\
\hline
Number of cameras	& 36	& 12 \\
Number of CCDs	& 36	& 12 \\
Effective aperture	& 18 cm	& 7 cm \\
Field of view	& $\sim$5000 sq.~deg	& $\sim$5000 sq.~deg \\
Energy range	& 500-850 nm	& 500-850 nm \\
Limiting magnitude (5$\sigma$)	& M$_{\rm V}$=15 in 10 s (full moon) & M$_{\rm V}$=11.5 in 10 s (full moon) \\
																& M$_{\rm V}$=16 in 10 s (new moon)	& M$_{\rm V}$=12.5 in 10 s (new moon) \\
Temporal resolution	& $\sim$13 sec &	$\sim$15 sec \\
Absolute timing & 	$\sim$10 ms &	$\sim$10 ms \\
Source localization(5$\sigma$)	& 11 arcsec	& 30 arcsec \\
Burst observation rate	& $\sim$12\% of ECLAIRs triggers	& $\sim$12\% of ECLAIRs triggers \\
\hline
\end{tabular}
\end{center}
\caption{Main characteristics of the Ground Wide Angle Cameras}
\end{table}
} 


\subsubsection{The X-ray telescope MXT onboard \textit{SVOM}}

The Microchannel X-ray Telescope\footnote{
institutes: CNES Toulouse, LAM Marseille, IRAP Toulouse, IRFU/CEA Saclay, OAS Strasbourg (France), 
IAAT T\"ubingen, MPE Garching (Germany), 
University of Leicester (UK)
}
(MXT) is a very light ($\sim$35 kg), and compact ($\sim$1.2 m) focusing X-ray telescope operating in the 0.2-10 keV energy range.
With its sizable FoV for a pointed instrument (64$\times$64 arcmin$^2$) and its sensitivity below the mCrab level, MXT will identify X-ray transients in non-crowded fields, localize them to sub-arcmin precision, and provide detailed X-ray spectra.
The MXT subsystems are the optics based on square micro-channels, camera, carbon fiber structure, data processing unit, and radiator.
For more details on the MXT instrument see \citet{mxt-spie}.


The optics of MXT uses a ``Lobster Eye'' geometry, as first defined by \citet{angel}, and optimized for narrow-field use.
Grazing incident X-rays are reflected within the square pores (40 $\mu$m side) of the square micro-pore optics plates (MPOs, 40 mm side, 1-2 mm thickness).
The MPOs, produced by Photonis, are spherically slumped to a radius twice the focal length of MXT, and coated with Ir.
The optics focal length is 1~m.
Its front part is coated with 70 nm of Al for thermal insulation and to reduce the optical load on the detector.
The MPO Point Spread Function (PSF) is composed by a central spot and two cross arms: about 50\% of the incident X-rays are reflected twice and focused in the central PSF spot, X-rays reflected just once and focused in both PSF arms (2$\times$22\%), and the rest produces a diffuse patch.
In the ``Lobster Eye'' geometry the vignetting is very low, reaching 10--15\% at the edge of the FoV. 
Simulations indicate that the PSF reaches 4.5 arcmin FWHM at 1.5 keV.


The MXT camera uses a pnCCD developed by the MPE \citep{meidinger}, with an active area of 256$\times$256 pixels of 75 $\mu$m, and a reduced frame store area with 75$\times$51 $\mu$m pixels.
This CCD is fully depleted (450 $\mu$m depth) and has excellent low-energy response (45-48 eV FWHM at 277 eV), and energy resolution (123-131 eV FWHM at 5.9 keV), and is read-out rate every 100 ms.
The detector is actively cooled to -65$^\circ$C. 
A filter wheel allows to put a calibration source or additional optical/UV filters in front of the detector when needed. The GRB afterglow position is computed onboard in near real-time by the MXT Data Processing Unit.


\myRem{
\begin{table}[h]
\caption{MXT expected scientific performances.} 
\label{tab:scireq}
\begin{center}       
\begin{tabular}{|l|l|} 
\hline
Energy Range &	0.2-10 keV\\
\hline
Field of view &	64$\times$64 arc min\\
\hline
Point Spread Function &	4.5 arc min (FWHM @ 1.5 keV)\\
\hline
Sensitivity (5 $\sigma$) &	$\sim$10$^{-10}$ erg cm$^{-2}$ s$^{-1}$ in 10 s\\
 &  $\sim$2$\times$10$^{-12}$ erg cm$^{-2}$ s$^{-1}$ in 10 ks\\
\hline
Throughput &	1 mCrab  $\sim$0.20 ct/s  for N$_{H}$=4.5 $times$ 10$^{21}$ cm$^{-2}$, photon index = 2.08\\
\hline
Energy Resolution &	$\sim$75 eV (FWHM @ 1 keV)\\
\hline
Time Resolution	& 100 ms\\
\hline 
\end{tabular}
\end{center}
\end{table} 
}

The expected MXT effective area is 27 cm$^2$ at 1 keV for the central spot and 44 cm$^2$ including the PSF cross arms. 
Taking the expected background into account, this translates into a 5$\sigma$ sensitivity in the 0.3--6 keV energy band of of 8$\times$10$^{-11}$ erg cm$^{-2}$ s$^{-1}$ for a 10 s observation, and $\sim$10$^{-12}$ erg cm$^{-2}$ s$^{-1}$ in 10 ks.
The MXT throughput for 1 mCrab is $\sim$0.20 ct/s (for N$_{H}$=4.5$\times$10$^{21}$ cm$^{-2}$, assuming a photon index of 2.08).

The expected GRB afterglow localization performance, obtained by folding the entire \textit{Swift/XRT} afterglow dataset through the MXT response, shows that MXT is well adapted to study GRB afterglows: 50\% of the bursts will be localized better than 13 arcsec (statistical uncertainties only) within 5 min from trigger. 
This shows that despite a smaller effective area compared to \textit{Swift}/XRT, most afterglows will remain detectable up to about 10$^5$ s after trigger.

\myRem{
\begin{figure}[ht!]
\includegraphics[width=.5\textwidth]{sens.pdf}
\includegraphics[width=.5\textwidth]{loc.pdf}
\caption{Left: Estimated MXT sensitivity as a function of exposure time. The curves show the 0.3-6 keV absorbed flux detection limits for 3, 4 and 5 sigma significance above background as a function of integration time for a detection and background cell size fixed at 80\% of the enclosed PSF. The black curve is obtained for a photon index $\Gamma$=2.0, and an absorption column density $N_{H}$=3$\times$10$^{21}$ atoms cm$^{-2}$, while the red one for $\Gamma$=1.7 and $N_{H}$=3$\times$10$^{20}$ atoms cm$^{-2}$. Right: The cumulative distributions of the various 90\% confidence level error radii as a function of time since trigger for all bursts to which \textit{SVOM} has slewed.}
\label{fig:perf1}
\end{figure}
}



\subsubsection{The visible telescope VT onboard \textit{SVOM}}

The Visible Telescope\footnote{institutes: NAOC Beijing, XIOPM Xian (China)} 
(VT) is a dedicated optical follow-up telescope onboard \textit{SVOM}. 
Its main purpose is to detect and observe the optical afterglows of GRBs localized by ECLAIRs. 
It is a Ritchey-Chretien telescope with a 40 cm diameter and an f-ratio of 9. 
Its limiting magnitude is about M$_{\rm V}$=22.5 for a 300 s integration time.

VT is designed to maximize the detection efficiency of GRB's optical afterglows.
A dichroic beam splitter divides the light into two channels, in which the GRB afterglow is observed simultaneously, 
a blue channel with a wavelength from 0.4 to 0.65 $\mu$m and a red channel from 0.65 to 1 $\mu$m.
Each channel is equipped with a 2K$\times$2K CCD detector. 
The blue channel CCD is a normal thinned and back-illuminated one, while the red channel CCD is deep-depleted to obtain a high sensitivity at long wavelengths. 
The Quantum Efficiency (QE) of the red-channel CCD is over 50\% at 0.9 $\mu$m, which gives VT the capability of detecting GRBs with redshifts larger than 6.5.
The VT FoV is about 26$\times$26 arcmin$^2$, covering the ECLAIRs error box in most cases.

In order to promptly provide GRB alerts with sub-arcsec accuracy, VT performs data processing onboard.
After GRB localization by the coaligned MXT, lists of possible sources are extracted from successive VT sub-images, whose center and size is determined by the GRB position and error box provided by MXT. 
These lists are down-linked in near-real-time through the VHF network, 
to allow the ground software to produces finding charts 
and search the optical counterpart of the GRB, comparing the lists with existing catalogs.
If a counterpart is identified, an alert is distributed to the world-wide astronomical community, to trigger observations by large ground-based telescopes in order to measure the spectroscopic redshift of the GRB.

            
In the \textit{Swift} era, confirmed high-redshift GRBs are rare, in contrast to a theoretical predicted fraction of more than 5 to 7\%.
This is probably due to the fact that for most \textit{Swift} GRBs the early-time optical follow-up images are not deep enough for a quick identification, such that faint GRBs cannot be spectroscopically observed in time by the large ground-based telescopes. 
This situation will be significantly improved by \textit{SVOM}, thanks to the high sensitivity of VT, in particular at long wavelengths, and the generation of fast optical-counterpart alerts. 
Additionally, the anti-solar pointing strategy of \textit{SVOM} allows GRBs to be observed by large ground-based spectroscopic telescopes at early times.
Consequently more high-redshift GRBs are expected to be identified in the \textit{SVOM} era.

\subsubsection{The ground-based visible/infrared telescope F-GFT}


The French-Ground Follow-up Telescope\footnote{institutes: CPPM Marseille, IRAP Toulouse, IRFU/CEA Saclay, LAM Marseille,  OHP-OSU Pytheas (France), UNAM Mexico, UNAM Ensenada (Mexico)} 
(F-GFT) is a robotic 1-m class ground telescope dedicated to \textit{SVOM}. 
It is designed to provide the redshift determination of GRBs, which is fundamental for their understanding and their use in cosmology, and which is currently only possible from ground.
F-GFT measures the redshift photometrically with an accuracy of $\sim$10\%, using broadband optical/infrared photometry, and additionaly provides sub-arcsec localization of GRBs.
This information will be automatically delivered in less than 5 min (an order of magnitude quicker than current instruments), such that the most interesting GRBs (with high redshift, high extinction, etc) will be quickly identified and precisely localized, allowing rapid follow-up observations with the largest facilities (NOT, NTT, VLT, E-ELT, ALMA, etc) in order to obtain high-resolution afterglow spectra and very precise redshift determinations.

F-GFT is a collaboration between France and Mexico (UNAM and CONACyT). 
Mexico will host the F-GFT in the national observatory located in San Pedro M\'artir, Baja California, which offers very good observing conditions (median seeing of 0.8 arcsec and about 80\% of clear nights).


\textit{SVOM} sets strong constraints on the F-GFT technical specifications: very high availability for follow-up observations ($\sim$90\%), very good sensitivity (1.3 m mirror diameter), fast pointing speed (on target in less than 30 s), multiband photometry (from 0.4 to 1.7 $\mu$m, with at least two simultaneous bands), and FoV (13 arcmin in radius) covering the ECLAIRs trigger error box.
The sensitivity reaches magnitude 20.5 in r band and 19.0 in J band (AB system, 60 s exposure time, SNR=5), which allows to detect more than $\sim$95\% of the currently observed GRB dataset. 
Moreover $\sim$23\% of the \textit{SVOM} alerts will be immediately observable.
Seventeen hours after the ECLAIRs trigger, despite this long delay and decrease in the GRB brightness, the F-GFT has still sufficient sensitivity to allow the detection of $\sim$65\% of the GRBs in the infrared domain.

\myRem{ 
\begin{figure}[!b]
\centerline{\includegraphics[width=0.7\linewidth]{plot_mag_temps.png}}
 \caption{\small F-GFT sensitivity in the visible and near infrared bands as a function of exposure time.}
 \label{fig:F-GFT}
\end{figure}
} 


The unique combination of speed, FoV and sensitivity both in optical and infrared, permits the F-GFT to address a large number of scientific questions and provides a strong scientific return to the \textit{SVOM} mission, in particular: 
a) the fast F-GFT response-time, allowing quick follow-up observations, will permit the study of the GRB prompt emission mechanism and the transition between the prompt and afterglow emission, a largely unexplored domain up to now in the visible and near infrared, which F-GFT will cover with a time resolution adapted to the fast variability of the emission in this phase;
b) the unique sensitivity of F-GFT in the infrared will allow the search for highly obscured or redshifted GRBs (beyond redshift $\sim$6), invisible to the VT onboard \textit{SVOM}, important for the study of the young Universe and the epoch of reionization;
c) the sensitivity, availability and flexibility of the F-GFT will help searching for electromagnetic counterparts of high-energy neutrinos detected by ANTARES/KM3Net, or of gravitational wave events detected by advanced LIGO and Virgo.

\myRem{ 
\begin{itemize}
\item Its global response time allows it to start very quickly the observations. It will become possible to study the mechanisms associated to the prompt emission and also the transition between the prompt and the afterglow emissions. Up to now, there have been few observations deeper than magnitude 17 during the two first minutes after the start of the high-energy emission. The F-GFT is sized to explore this domain with a time resolution adapted to the fast variability of the emission in this phase. This area is largely unexplored and is precisely one of the main scientific objectives of the \textit{SVOM} mission.
\item Its sensitivity in the infrared domain is also unique. This capability is especially important for searching highly obscured or redshifted (beyond a redshift of $\sim$6) GRBs since they will remain invisible with the Visible Telescope onboard the satellite. The highly redshifted GRBs are for example especially important for the study of the young Universe and the epoch of reionization, and their fast identification obviously deserves special efforts. To achieve this requirement, the F-GFT will provide photometric redshifts in less than 5 minutes after the alert reception, which allows a reduction of this delay of an order of magnitude compared to current observations . It will be then possible to activate the largest facilities, as NTT, VLT, E-ELT, ALMA, etc., immediately from the F-GFT, which is of course critical for scientific exploitations of these events. 
\item Lastly, this facility will provide a service to astroparticle, as it could be used to search for the electromagnetic counterparts of high-energy neutrinos detected by ANTARES/KM3Net, or of gravitational wave candidates detected by advanced LIGO and Virgo. 
\end{itemize}
} 

\subsubsection{The ground-based visible telescope C-GFT}

The Chinese Ground Follow-up Telescope\footnote{institute: NAOC Beijing (China)} 
(C-GFT) is based on an existing 1-m telescope located at Xinglong observatory, China, 
which is an f/8 Ritchey-Chretien system providing fast repointing through an altazimuth mount.
It will be upgraded with a dichroic mirror system, 3 channels of SDSS g,r and i filters, a robotic control system, and a real-time data and communication system.

The 3 channels will be equipped with CMOS-CCD cameras with low read-out time, of size of 2K$\times$2K, corresponding to a FoV of 21$\times$21 arcmin$^2$, operating in the 0.4 to 0.95 $\mu$m band.
The C-GFT sensitivity will reach mag(r)=19 (at 5$\sigma$, AB mag) for 100 s exposure time during new Moon nights.

The C-GFT robotic control system will point the telescope automatically to the ECLAIRs trigger sky position received from the CSC over \textit{SVOM}net, and will also provide a user interface to change the pointing remotely.
The real-time data processing system will search GRB counterparts in the 3 channel images, after CCD data reduction (bias, dark, and flat field correction), cosmic ray removal, and astrometric calibration.
All detected GRB candidates will be provided with the parameters: sky localization (with accuracy of 0.5 arcsec), flux-calibrated magnitude, weight (or confidence) and finding chart.
If no GRB candidate is found, upper limits will be provided.
The data communication of C-GFT includes an instant data transfer network (based on the XMPP protocol) with VOEvent wrapped messages, transferring the detected GRB parameters.
A no-instant data transfer network, based on FTP servers, permits large volume data transfers of FITS images, without stringent temporal requirements.
The C-GFT telescope points towards the target, takes exposures, reads-out the data, processes it and reports the results 
all within respectively 5 min and 1 min after trigger reception.
More than 20\% of the ECLAIRs triggers will be observable by C-GFT.

In addition to the dedicated C-GFT at Xinglong observatory, two more 1-m telescopes will be set-up by the NAOC (National Astronomical Observatory of China) at Ali observatory, Tibet, western China, as a part of the LCOGT network. The NAOC will have access to $\sim$2500 h of observation time per year on LCOGT, which could be dedicated to follow-up \textit{SVOM} GRBs via LCOGT ToO calls with a typical response time of about 15 min.

\myRem{
The 1-meter telescope (Its dome is shown in Figure 1) is an f/8 Ritchey-Chretien system, which has characteristics of fast pointing for its altazimuth mount. To fulfil \textit{SVOM} observation requirements list in Table 1, we need to upgrade dichroic mirror system to make it has three channels of SDSS g,r, and i filter, robotic telescope control system, real-time data processing and data communication system. 

Figure 1 The dome of CGFT

\begin{table}[htb]
\label{tab:cgft}
\begin{center}
\begin{tabular}{ll|ll}
\hline 
Energy range	& 400-950 nm	& Report observational result	& $<$ 5 minutes \\
Field of View	& 21' x 21'	& Proposed location & Xinglong observatory \\
Aperture(diameter) & 1 m	& Reception of alert & \textit{SVOM}net \\
Channels & 3 channels of g,r,i	& Data communication & VoEvent + FTP \\
Detector & 3 CCD camera mounted	& Observational schedule	& Support remote update \\
Observation rate	& $>$20\% of ECLAIRs triggerss	& Localization accuracy	& 0.5 arcsec \\
Sensitivity	in 100 s & Mag(r) = 19 (new Moon, AB mag, 5$\sigma$) \\
\hline
\end{tabular}
\end{center}
\caption{The requirements on CGFT}
\end{table}

The future CGFT will have three channels mounted with 3 CCD cameras (preferable to have CMOS cameras to reduce read out time). Its preliminary schematic design is shown in Figure 2.Each detector has size of 2kX2k, which corresponding to field of view 21'x21' . The designed sensitivity of CGFT is mag(r)= 19 (at 5 sigma) for exposure time of 100 seconds at new Moon night.

Figure 2 The optical schematic design of the dichroic mirror system

CGFT will be a robotic telescope,the control system of which can server to point the telescope to required sky area according to the received trigger message from CSC by \textit{SVOM}net. The control system can provide interface for user to change observation strategy from remote.
The real-time data processing system will search GRB candidates from the three channels images after CCD basic reduction (bias, dark , and flat field correction), cosmic ray removing, and astrometric calibration. All the detected GRB candidates will  be provided with celestial coordinates position, flux calibrated magnitude, weight (or confidence) and finding chart. If no GRB candidate found, the upper limit will be also provided by the processing system. 
The data communication connections of CGFT includes instant data transfer and no-instant data transfer network. The instant data transfer network is based on XMPP protocol with VOEvent wrapped message. It transfers GRB triggers and detected results, such as position, magnitude, weight. While the no-instant transfer network is designed to transfer large volume data, which has no strict temporal requirement. It is based on FTP server. The fits images are transferred by this network. 
In the first 5 minutes after receiving the trigger, the CGFT need point telescope to the target, take exposure, read out data, data processing and report its results. All of these procedures have to be finished in the total 5 minutes. The provided scientific products are summarized in Table 2. 
In addition to 1-m CGFT at Xinglong observatory, National Astronomical Observatories China (NAOC) will set two 1-m telescopes at Ali observatory, Tibet, western China, as a part of LCOGT. NAOC will have observing time of about 2500 hour per year in the LCOGT 1-m telescopes. These observation time of LCOGT could be applied for to follow-up the \textit{SVOM} GRB afterglows as the role played in by 1-m CGFT at Xinglong observatory mentioned above. The typical response time of LCOGT for a ToO call is about 15 minutes.

\begin{table}[htb]
\label{tab:cgft}
\begin{center}
\begin{tabular}{l}
\hline 
Temporal requirement	Products	Unit 	Note \\
In 5 minutes after receiving trigger.	Detect GRB?	Yes	Detection time	Julian day	All the products are from  three channels \\
			brightness	Magnitude (flux calibrated)	\\
			Weight	Value in (0-1) \\
			Finding chart	Pngfile	\\
		No	Upper limit	Magnitude	\\
After one observation task	Light curve	Asciifile	Manual processing \\
	Refined GRB information	(like above, it is processed from combined image)	\\
	Temporal slope	--	\\
	Sed	--	\\
	Photo-Z (redshift indicator)	--	\\
\hline
\end{tabular}
\end{center}
\caption{The scientific products provided by CGFT}
\end{table}

}

\subsection{The \textit{SVOM} science ground segment}

The Chinese Science Center provides software tools for the General Program proposal writing and ToO generation.
They include up-to-date instrument performances delivered by the Instrument Centers, observation footprint and SNR estimators, and information on source accessibility, background maps and catalogs etc.
The Mission Center validates and coordinates the elaboration of observation requests made by the PIs for all observing programs of \textit{SVOM}.
The Control Center sends telecommands to be executed onboard the satellite.
The Chinese Science Center\footnote{institute: NAOC Beijing (China)} 
processes data of the Chinese instruments GRM and VT onboard \textit{SVOM} and the ground-based C-GFT and GWAC.
It also assists Instrument Scientists (ISs) of those instruments and Burst Advocates (BAs) on-duty on the Chinese side.

The French Science Center\footnote{institutes: APC Paris, CNES Toulouse, CPPM Marseille, GEPI Paris, IAP Paris, IRAP Toulouse, IRFU/CEA Saclay, LAL Orsay, LAM Marseille, LUPM Montpellier (France)}
processes data of the French instruments ECLAIRs and MXT onboard \textit{SVOM} and the ground-based F-GFT. 
Three Instruments Centers, one for each French instrument, are connected to the French Science Center and host experts in charge of monitoring and calibration of the French instruments, as well as assistance to the BAs.

The VHF network collects the triggers generated onboard \textit{SVOM} and sends them to the French Science Center, where the near real-time data analysis is performed locally or remotely by Chinese and French BAs.
The Chinese and French GFTs point towards the GRBs detected by \textit{SVOM} to produce results within 5 min.
The GWAC collects data from -5 to +15 min around the trigger time.

All data are examined and possibly re-processed by BAs and ISs. BAs can request revisits of interesting GRBs.
Data products are generated and re-processed as the pipelines improve throughout the mission lifetime, and will be made available by the Space Science Data Center.
Data of the French instruments will also be made available by the French Science Center.

\begin{figure}[htpb]
\begin{center}

\includegraphics[trim=60 80 60 80,scale=0.7,clip=true]{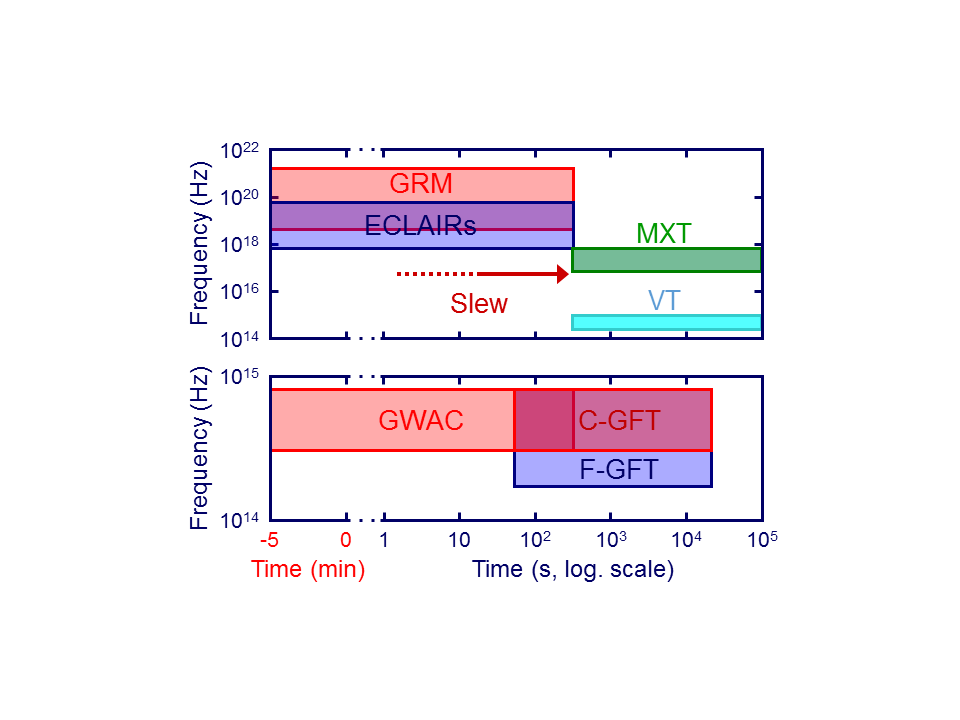}
 \caption{Spectral coverage of the GRB prompt emission and its afterglow with \textit{SVOM} instruments, as a function of time (the burst is detected at time t = 0). The upper panel refers to space-born instruments (GRM, ECLAIRs, MXT, VT), the lower panel to ground-based instruments (GWAC, C-GFT, F-GFT).}
 \label{fig:muti}
 \end{center}
\end{figure}


\myRem{ 

\pagebreak
\subsection{-----ALL THE REST IS TRASH-----}

\begin{itemize}
\item The only files to be modified are \verb+subsection-template.tex+ (i.e. your contribution) and \verb+template.bib+ (references). To check your contribution, compile:

 \verb+ latex main.tex+

\verb+ bibtex template.bib+

\verb+ latex main.tex+

\verb+ latex main.tex+

\item Text / math: use standard latex conventions, the package \texttt{amssymb} is loaded in the preamble of the main file. It allows to use many symbols, such as $\lesssim$ of $\gtrsim$. Full list of symbols available at

 \url{http://www.rpi.edu/dept/arc/training/latex/amssymblist.pdf}.  
\item Figures: the package  \texttt{graphicx} is loaded in the preamble of the main file. Use macro \verb+\reffig{}+ to insert in the text a reference to a figure, such as in the example below.

\verb+I include here a reference to \reffig{fig:example}.+

Leads to :

I include here a reference to \reffig{fig:example}.

\item Tables: use macro \verb+\reftab{}+ to insert in the text a reference to a table, such as in the example below.

\verb+I include here a reference to \reftab{tab:example}.+

Leads to :

I include here a reference to \reftab{tab:example}.

\item References: within the subsection, internal references can be generated using standard \LaTeX : \verb+\label{}+ and \verb+\ref{}+, or \verb+\reffig{}+, \verb+\reftab{}+.

To insert in the text a reference to another section or subsection in the White Paper, use macro \verb+\refwp{+\textit{keyword}\verb+}+ with the following keywords, like in the example below :

\begin{center}
\resizebox{\linewidth}{!}{\begin{minipage}{1.85\linewidth}
\begin{tabular}{|lc|lc|}
\hline
\textbf{Sub-section} & \textbf{Keyword} & \textbf{Sub-section} & \textbf{Keyword} \\
\hline
Introduction & \texttt{introduction} & GRBs to study the evolution of star formation & \texttt{grbstarformation} \\
Time-domain astrophysics: the discovery space after Swift & \texttt{timedomain} & GRBs to study galaxies & \texttt{grbgalaxies} \\
The astronomical panorama in 2020 & \texttt{panorama} & Absorption spectroscopy on the line of sight of GRBs & \texttt{grbspectroscopy} \\
\textit{SVOM} mission profile & \texttt{svommission} & GRBs as standard rulers & \texttt{grbrulers} \\
\textit{SVOM} instrument characteristics& \texttt{svominstruments} & Search for \textit{SVOM} counterparts on multi-wavelength triggers & \texttt{grbmultiwavelength} \\
The population of classical long GRBs: physical mechanisms & \texttt{grbphysics}  & Search for \textit{SVOM} counterparts on multi-messenger triggers & \texttt{grbmultimessengers}\\
The population of classical long GRBs: characterization of the population &  \texttt{grblong}& Active Galactic Nuclei & \texttt{agn} \\
The population of short GRBs & \texttt{grbshort} & Other extragalactic sources & \texttt{extragalactic}\\
The diversity of stellar explosions & \texttt{grbdiversity} & Galactic sources & \texttt{galactic}\\
GRBs as particle accelerators & \texttt{grbacceleration} & Miscellaneous topics & \texttt{other}\\
GRBs at high redshift & \texttt{grbhighredshift} & Studies of extrasolar planets and solar system bodies & \texttt{planets}\\
\hline
\end{tabular}
\end{minipage}}
\end{center}

\verb+I include here a reference to the subsection on \textit{SVOM} mission profile:+

\verb+see \refwp{svommission}.+

Leads to :

I include here a reference to the subsection on SVOM mission profile : see \refwp{svommission}.

\item References / bibliography. Include all the references in BibTex format in the separated file \verb+template.bib+. To insert citations, use \verb+\citet+, \verb+\citep+ and \verb+\citealt+ as in the examples below :

\verb+I want to include a reference to \citet{klebesadel:73}.+

Leads to:  I want to include a reference to \citet{klebesadel:73}. 

\verb+This has been known for a while \citep{klebesadel:73}.+ 

Leads to: This has been known for a while \citep{klebesadel:73}.
 
\verb+This has been known for a while (since 1967, \citealt{klebesadel:73}).+ 

Leads to: This has been known for a while (since 1967, \citealt{klebesadel:73}). 
 
\end{itemize}

\begin{table}[!b]
\begin{center}
\begin{tabular}{lc}
\textbf{SVOM Instruments} & \textbf{Energy range}\\
\hline 
ECLAIRs & 4-150 keV\\
GRM & 30-5000 keV \\
MXT & 0.2-10 keV \\
VT &  400 nm - 650 nm \\
  & 650 nm - 950 nm \\
\hline
GWAC & 500-800 nm\\
C-GFT & Visible\\
F-GFT & Visible and nIR\\
\end{tabular}
\end{center}
\caption{SVOM instruments and their energy range}
\label{tab:example}
\end{table}

\begin{figure}[!b]
\centerline{\includegraphics[width=0.5\linewidth]{fig_example.jpg}}

 \caption{\small Schematic showing the SVOM spacecraft with its multi-wavelength space payload. It consists of two wide-field instruments: ECLAIRs and the Gamma-Ray Monitor (GRM) for the observation of the prompt emission and two narrow field instruments: the Micro-channel X-ray Telescope (MXT) and the Visible Telescope (VT) for the observation of the afterglow emission.SVOM has two ground dedicated-instruments : a wide-field instrument GWAC for the observation of the optical prompt emission, and a narrow-field instrument GFT for the follow-up observations in the visible and near-infrared domain.}

 \label{fig:example}
\end{figure}


} 
\subsection{\textit{SVOM}: a highly versatile astronomy satellite}

%

With built-in multi-wavelength capabilities, flexible operations and ground follow-up opening a large discovery space, \textit{SVOM} will be a highly versatile Astronomy satellite. As depicted in Fig. \ref{fig:muti}, thanks to its unique instrumental combinations of wide-field (ECLAIRs + GRM + GWAC) and narrow-field (VT + MXT + C-GFT + F-GFT) instruments, \textit{SVOM}, already well suited to permit fruitful GRB studies, should also effectively contribute in the new fields of Multi-Messenger Astrophysics. 

\newpage
\section{\textit{SVOM} Advances on GRB Science (\textit{SVOM} core program)}


Gamma-ray bursts (GRBs) are short and intense flashes of photons of energy observed mainly from $\sim 100$ keV to $\sim 1$ MeV, which are non-thermal, have significant variability, and typically last less than $\sim 100$ s.  This GRB prompt emission is followed by long-lasting afterglow emission in X-ray, optical, radio. Observations of afterglows enable us to localize positions of GRBs accurately, to discover their host galaxies, and to measure their redshifts. Due to their cosmological distances, the isotropic-equivalent luminosities of GRBs can reach very high values (upper bound $10^{54}\, \rm erg ~s^{-1}$)  with a broad distribution of luminosities. Hence GRBs are the most powerful explosions in the universe, which are thought to be powered by the gravitational collapse of matter to form a black hole or a neutron star. 

GRBs can be separated into two different types by their durations with a rough separation line of $\sim 2$~s, since an obvious bimodal structure appears in the duration distribution \citep{Kouve93}. The number ratio of observed long GRBs with a duration $>2$ s to short GRBs for $<2$ s is roughly three to one. Generally, long GRBs tend to have a softer spectrum, while short GRBs are relatively harder. Long GRBs distribute in a very wide redshift range from 0.0085 to 9.4 and concentrate within $z\sim 1-2$, while short GRBs occur at closer distances with an average redshift of $\sim 0.5$ \citep{Berger:2014}.
There are several strong observational reasons for an association of long GRBs with the gravitational collapse of some massive stars \citep{Woo1993}, 
 while short GRBs are mostly likely associated with the mergers of binary compact stars \citep[see e.g.][]{perna:2002}. 

The above simple dichotomy for GRBs could still face some challenges. On one hand, the definition of GRB durations is actually instrument-dependent \citep[see e.g.][]{qin:2013}, which could involve some selection biases. On the other hand, it is indeed very difficult to assign some GRBs that have a short prompt spike and subsequent extended ($\sim10-100$ s) soft gamma-ray emission (e.g. GRBs 060505 and 060614) to the traditional short or long groups \citep{galyam:2006,ofek:2007}. Additionally, several ultra-long GRBs with a duration longer than thousands of seconds have also been detected \citep{campana:2011,virgili:2013}, which could just be the long-duration tail of normal long GRBs 
\citep{zhang:2014} or could indicate a distinct new kind of progenitors \citep{gendre:2013,levan:2014}. Therefore, multiple observational criteria may in principle be needed for a more physical classification of GRBs \citep[see e.g.][]{zhang:2006,lv:2010}.

\textit{SVOM} is well designed to study the physics of the GRB phenomenon in all its diversity, thanks to an excellent spectral and temporal coverage of the  prompt and afterglow emission combined with an optimized follow-up strategy aiming at the redshift determination for a large fraction of GRBs ($\sim 2/3$). The potential impact of \textit{SVOM} in the study of long GRBs is discussed in sections~\ref{sec:grbphysics}, \ref{sec:grblong} and \ref{sec:grbdiversity}, the special case of short GRBs is presented in section~\ref{sec:grbshort}. The synergy of \textit{SVOM} with other instruments in the study of GRBs as cosmic accelerators is treated in section~\ref{sec:grbacceleration}.

\subsection{Classical long GRBs: physical mechanisms}
\label{sec:grbphysics}

The physical understanding of GRBs is especially challenging, due to their extreme properties. This quest is however highly motivating as it offers the possibility to improve our knowledge of the final fate of massive stars, to understand astrophysical relativistic jets from GRBs and other sources such as microquasars or blazars, and to study an extreme case of cosmic accelerator. In addition, the use of GRBs as probes of the distant universe can be improved by a better understanding of the progenitors, and/or GRB emission processes. 
A "standard" scenario for GRBs is now well accepted \citep[e.g.][]{piran:1999}: an initial event such as the collapse of a massive star (long GRBs) or the coalescence of two compact objects (short GRBs, discussed below in section \ref{sec:grbshort}) leads to the formation of an accreting stellar mass compact source, most probably a black hole. This source releases a huge energy in the form of an ultra-relativistic jet. The prompt emission, which is non-thermal, is due to internal dissipation within the jet itself, for instance via internal shock waves \citep{rees:94}. The afterglow is associated to the deceleration of the jet by the ambient medium \citep{meszaros:93}, which leads to the propagation of a strong relativistic shock into the external medium (external shock) and of a reverse shock into the jet. Within this theoretical framework, many questions remain open.\\

\textbf{(1)~Nature of GRB progenitors and central engines.}
While the production of GRBs seems to be due to a more or less universal process -- the formation of short-lived relativistic ejecta emitted by accreting compact objects -- their progenitors may be quite diverse. 
Classical long GRBs are associated with the collapse of some massive stars (collapsar model).
However, the redshift distribution of \textit{Swift} GRBs show that the cosmic GRB rate does not follow the star formation rate, the efficiency of GRB production appearing to be higher at high redshift \citep{daigne2006}.  This shows that the production of a GRB depends on special conditions, which may be related to different properties of the progenitor: rotation, metallicity, binarity, etc. 
The progenitor stars should probably be very massive, typically heavier than $\sim 40 M_{\odot}$. 
Some may belong
to the first generation of stars in the Universe (see discussion in section \ref{sec:grbhighredshift}).
 A general picture is that the central iron core will collapse into a quickly spinning black hole. A high density transient accretion disk will form around the black hole, launching a pair of ultra-relativistic jets that move outward along the spin axis \citep{Woo1993, Ram2002, Zhang2003}. However, many details of the process are still highly uncertain. First, how are the jets launched, by the  Blandford-Znajek mechanism \citep{Bland1977} or another mechanism? Second, although a few numerical approaches have been devoted to this question \citep{Mac1999, Zhang2003}, it is still uncertain whether the jet can penetrate through the stellar envelope of the progenitor while maintaining an ultra-relativistic speed and keeping to be highly collimated. 
Morover, the nature of the central object resulting from the collapse is also debated: 
the iron core may first collapse into a hyper-massive neutron star, which then spins down and collapses into a black hole at a later stage \citep[supranova model,][]{Viet1999};  it is also possible that the remnant object is not a black hole, but a millisecond magnetar \citep{Klu1998, Whe2000, Dai2006, Bucc2009}, offering different mechanisms of  energy injection with possible observable signatures \citep{Dai2006, ZhangB2006, YuYB15}. 

The association of a GRB trigger with the detection of gravitational waves (GW) would provide a direct signature of the initial event leading to a GRB and the nature of the central object. Such an association is expected first for short GRBs (see section \ref{sec:grbshort}) as the GW emission from stellar collapses is expected to be weaker than in mergers \citep[e.g.][]{FryerLRR}. Indirect signatures on the nature of the progenitor and the central engine can also be obtained from electromagnetic observations.
Association or non-association of nearby GRBs with supernovae is an important clue (see section \ref{sec:grblong}). Early spectroscopy of the afterglow can probe the immediate environment of the GRB and contribute to the identification of the progenitors of GRBs \citep[e.g.][]{fox2008,castrotirado2010}.
 This necessitates accurate and rapid localizations. 
 More generally, the density profile of the circumburst medium derived from the afterglow modelling can bring also valuable information on progenitors.
 
 A broad spectral coverage of the prompt emission is necessary to investigate the existence of possible optical or X-ray precursors (a small fraction of long GRBs show a soft precursor emission before the main burst, \citealt{hu2014}), which also puts some constraints on the central engine (e.g. lifetime).
  Finally, an extension of the sensitivity of GRB detectors into the X-ray domain is needed to study and characterize the population of soft GRBs and low-luminosity GRBs, which holds the key to understand the continuum of events following the collapse of a massive star, from standard supernovae (no relativistic jet) to cosmic GRBs (highly energetic and relativistic jets); see section \ref{sec:grbdiversity}. \\
  
\textbf{(2)~Acceleration and composition of the relativistic ejecta, physical origin of the prompt emission.} 
GRBs give unique opportunities to study the physics of relativistic astrophysical jets. The nature of the physical mechanism that launches GRB ultra-relativistic ejecta is not elucidated, and the composition of the jet (electromagnetic or matter dominated) is currently the subject of an intense debate, as well as the internal dissipation mechanism responsible for the prompt emission (photospheric emission vs internal shocks vs magnetic reconnection, \citealt{kobayashi:97,daigne:98,meszaros:02,daigne:02,rees:05,beloborodov:10,spruit:01,zhang:11}). Many issues are also related to the microphysics in the emission regions, especially related to the acceleration of the electrons responsible for the non-thermal emission, and the identification of the associated radiative processes (synchrotron emission, inverse Compton scatterings, etc.). Progress in this field will require a better understanding of the prompt emission phase (energy reservoir and extraction mechanism, radiation processes) and especially to be able to observationally constrain the physical conditions in the jet when this emission is produced: radius, Lorentz factor, magnetization, magnetic field geometry, etc. 

Since the launch of \textit{Fermi}, knowledge of the high-energy part of the spectrum has increased. On the other hand, the prompt X-ray and optical emission remain poorly known. The prompt optical emission provides important clues about the emission region and the internal dissipation mechanism but has been detected only in a few cases: a larger sample, allowing a statistical description, is necessary to identify the underlying mechanisms. The prompt X-ray emission is important to better characterize the spectrum and identify the emission mechanism (related to the nature of the jet) and the dominant radiative processes. The identification of thermal components in the spectrum can be a signature of the photosphere of the relativistic jet and would therefore provide an important clue for its nature \citep{hascoet:13}. 

A better understanding of the physical origin of the prompt emission requires a sample of GRBs with well measured properties, with a good temporal resolution and the largest spectral range, from the optical to the high-energy gamma-rays, including X-rays and soft gamma-rays. Such a sample cannot be limited to the observation of the prompt emission, as the distance is mandatory to measure the radiated energy from the ejecta, and as the early X-ray afterglow contains important information about the end of the prompt phase \citep[e.g.][]{hascoet:12}.

Simultaneous detections of few GRBs with instruments providing a good description of the prompt emission and allowing the measurement of the distance, and instruments with polarization measurements capabilities, such as POLAR, would also provide a key GRB sample to study the nature of the relativistic outflow \citep[e.g.][]{granot:03}. In a similar way, the detection of high-energy neutrinos associated to a GRB would also provide important information on the composition of the jet and on the internal dissipation processes (see section \ref{sec:grbacceleration}). \\

\textbf{(3)~Interaction of the ejecta with the circumburst medium -- The physical origin of the afterglow.}
\textit{Swift} has brought much new and puzzling information on the afterglow phase, especially related to unexpected time variability \citep[see the recent review by][]{gehrels:13}. This led to the emergence of many variations of theoretical models accounting for these observations. Several emission regions are invoked: shocked external medium (forward shock), shocked ejecta (reverse shock), late internal shocks, etc. Some of these models have strong implications on the lifetime and energetics of the central engine as they imply late energy injection by the source or relativistic ejection at late times.

To distinguish between these models and better understand the physics of the interaction of the jet with its environment, and especially the different emission regions contributing to the afterglow and the nature of the circumburst medium, it is necessary to build a sample of GRBs where all information is available (a description of the afterglow at different wavelengths, but also the distance/redshift). A good description of the prompt emission in this sample is also necessary to put constraints on the lifetime and energetics of the central engine, and to test models linking some afterglow features with the prompt emission.\\

\textbf{\textit{SVOM} contributions.} To summarize, the understanding of the GRB physics requires observations in the largest spectral domain and in the largest temporal interval, from the possible precursor up to the transition between the prompt and afterglow emissions.
 Simultaneous observation of the prompt GRB event in the gamma-ray, X-ray and visible bands, combined with narrow field observations of the afterglow in X-rays, visible and near infrared immediately after the beginning of the event will enable a better understanding of mechanisms at work in such events. 
 New measurements (gravitational wave, gamma-ray polarization, neutrinos) may also become possible in the future, and their impact on the physical understanding of the GRB phenomenon will be maximized if these measurements are made for bursts whose "standard" properties, including the distance, are well measured.
 
\begin{figure}[htpb]
\centerline{\includegraphics[scale=0.4]{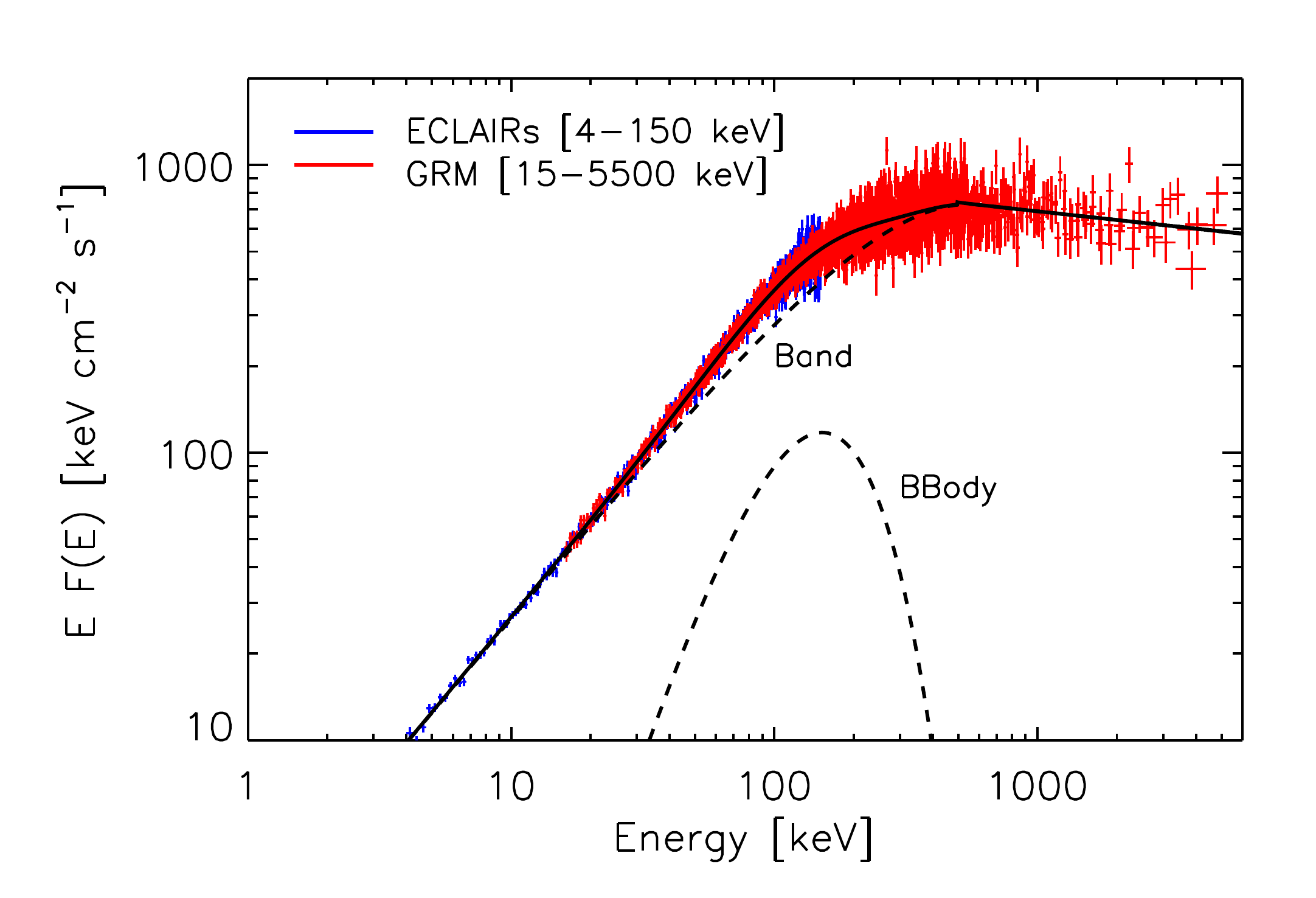}}
\caption{Simulation of the prompt spectrum of a long GRB in \textit{SVOM}/ECLAIRs+GRM.
The long GRB 100724B was detected by \textit{Fermi}/GBM and shows  a two components spectrum \citep{guiriec:2011}. 
It has been simulated in ECLAIRs and GRM assuming that the burst is on-axis in ECLAIRs and therefore 30$^\mathrm{o}$ offaxis in GRM. The resulting spectrum is plotted together with the best fit (solid line) by a combination of a non-thermal (BAND) and  quasi-thermal (BBody) components. The recovered parameters are equal within $1\sigma$ to those published for the same spectral model by \citet{guiriec:2011}.
This illustrates how well this complex shape is recovered by \textit{SVOM} thanks to the broad spectral coverage of the prompt emission allowed by the ECLAIRs+GRM.
}
\label{fig:grbphysics_spectrum}
\end{figure}

\begin{figure}[htpb]
\begin{center}
\includegraphics[scale=0.40]{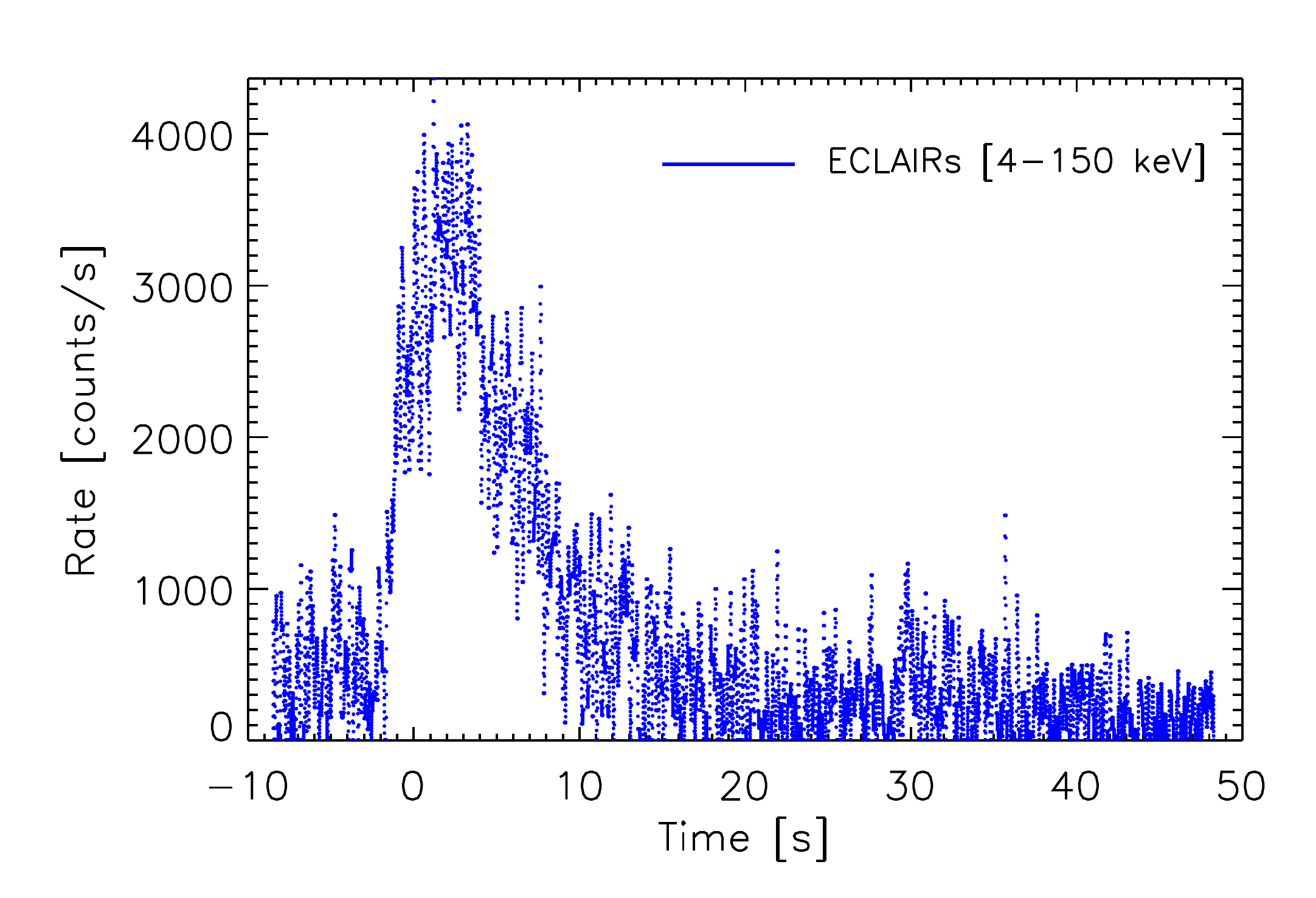}
\hfill
\includegraphics[scale=0.40]{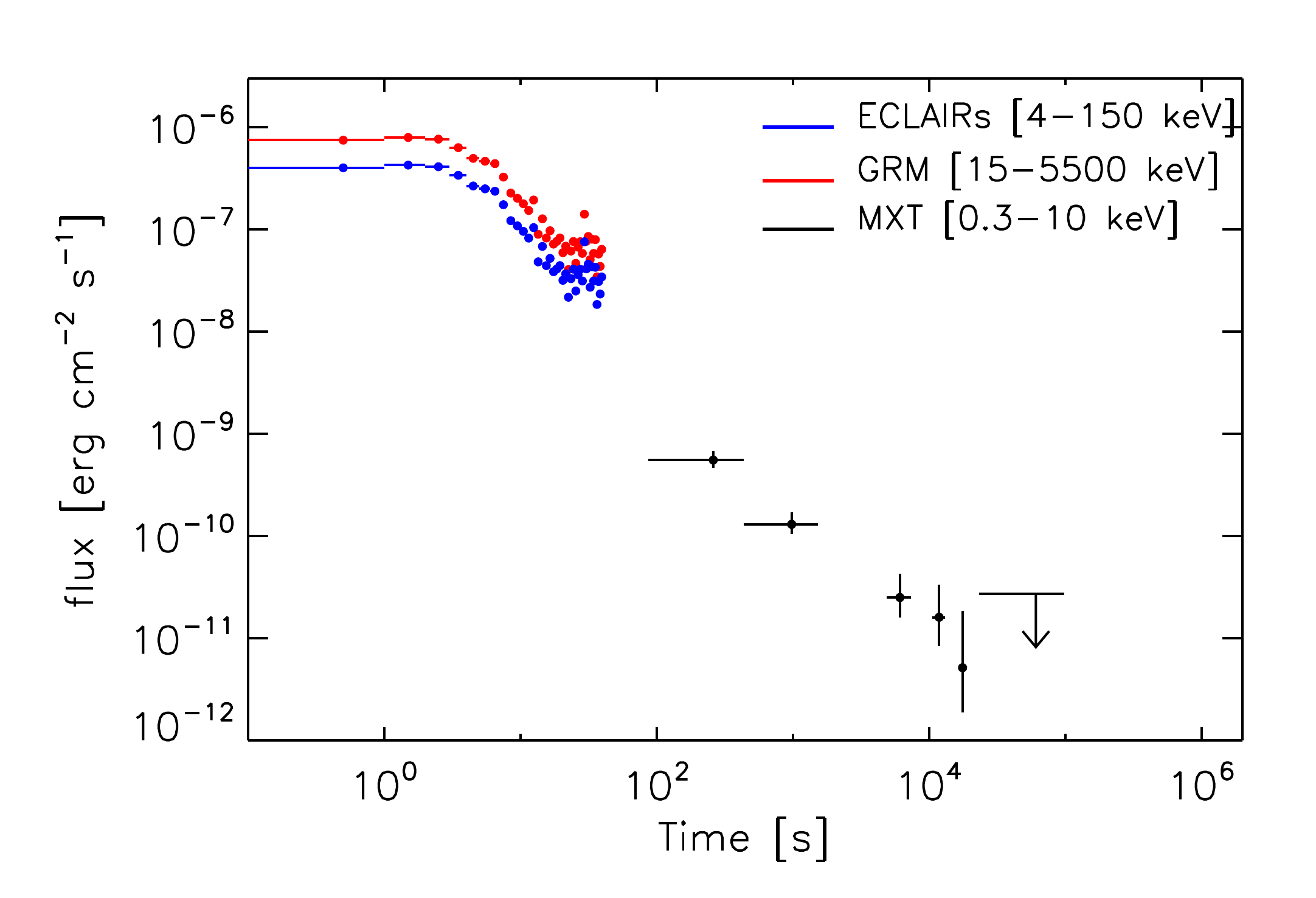} 

\caption{Simulation of the prompt and afterglow emission of a long GRB in \textit{SVOM} instruments.
GRB 091020 is a classical long GRB detected by \textit{Swift} and \textit{Fermi}/GBM. Using GBM data \citep{gruber:2014} for the prompt emission and \textit{Swift}/XRT data for the X-ray afterglow \citep{evans:2007,evans:2009}, this GRB has been simulated in ECLAIRs and GRM, 
assuming that the GRB is on-axis for ECLAIRs and therefore 30$^\mathrm{o}$ off-axis in GRM,  
 and in MXT where the afterglow is detected up to $\sim 10^4$ s. \textit{Left :} simulated prompt light curve in counts per seconds in ECLAIRs.  \textit{Right :} prompt and afterglow flux in log-log scale simulated in ECLAIRs, GRM and MXT.
}
\label{fig:grbphysics_prompt_afterglow}
\end{center}
\end{figure}

  The \textit{SVOM} mission is well adapted to these objectives. Compared to previous missions, it offers \textit{simultaneously} (i) the capacity to trigger on all types of GRBs (see section \ref{sec:grbdiversity}), (ii) an excellent efficiency of the follow-up and the redshift measurement; (iii) a good spectral coverage of the prompt emission by ECLAIRs+GRM allowing a detailed modeling, as illustrated in \reffig{fig:grbphysics_spectrum}. For a significant fraction of \textit{SVOM} GRBs, GWAC will provide in addition a measurement or an upper limit on the prompt optical emission; (iv) a good temporal and spectral coverage of the prompt and afterglow emission thanks to MXT, VT and the GFTs, as illustrated in \reffig{fig:grbphysics_prompt_afterglow}.


%

\subsection{Classical long GRBs: characterization of the population}
\label{sec:grblong}

%

The association of long GRBs (LGRBs) with core-collapse SNe is a well established fact (see \citealt{Hjorth2011,Cano2016}, for a review). Through photometry and spectroscopy it is possible to look for SN signatures in the afterglow light-curves and spectra. The spectral features indicate that these SNe are broad-line type Ibc. For LGRBs close enough not to have the associated SN flux too faint to emerge from the afterglow ($z\lesssim0.6$), all the searches for an associated SN have been successful, except for two LGRBs (GRB\,060605 and GRB\,060614; e.g.: \citealt{Fynbo2006}).  
Monitoring LGRB afterglows with the VT on board of \textit{SVOM} will make possible the systematic search for SN signatures in the afterglow light-curves, together with the identification of the rise time of the SN, allowing an efficient spectroscopic follow-up. 

A well sampled multi-wavelength coverage of the prompt and afterglow phases for a large fraction of GRBs, together with a high percentage of redshift determinations, are the conditions needed to make a step forward in the characterization of the LGRB population. \textit{SVOM} will have a great impact on this topic, thanks of the unique synergy of the space and ground segments (see section \ref{sec:SVOMmission}). Not only will \textit{SVOM} have a visible telescope onboard (the VT) capable to systematically follow the LGRBs detected by the higher energy instruments, but it will also benefit of the follow-up carried out by a dedicated network of telescopes included in the ground segment (GWAC, F-GFT and C-GFT).



Rapid identification of the LGRB counterparts in the optical and/or NIR frequencies in different filters will give a first redshift indication and ease the rapid spectroscopic follow-up by the larger telescopes. This is mandatory for the redshift determination of a significant fraction of the LGRB detected by \textit{SVOM}. Nowadays only $\sim30\%$ of LGRBs have an associated redshift. The goal of \textit{SVOM} is to double this fraction. LGRB samples complete in redshift are the fundamental ingredient to determine the rest-frame properties of LGRBs and their evolution with cosmic time, characterizing the LGRB population in terms of redshift distribution, luminosity function, progenitors and being able to use it to trace star-formation.

To achieve these goals it is mandatory to have a sample not biased against {\it dark} GRBs (see \citealt{Jakobsson2004,van-der-Horst2009}). As {\it dark} GRBs are optically dim, the NIR camera on the F-GFT will be particularly helpful to identify their afterglows and provide their precise positions, needed for the spectroscopic observations allowing the redshift determination. 


\textit{SVOM} will operate from 2021, an epoch were facilities such as \textit{SKA} (and its precursors)  will be available
(see section \ref{sec:panorama}).
Thanks to their improvement of sensitivity, \textit{SKA} will make possible a detailed study of the afterglow in the radio domain for a much larger number of LGRBs. \textit{SKA} should be able to detect all \textit{SVOM} LGRB afterglows and, in addition to an amazing great leap forward on the study of the afterglow emission phases, it will allow the determination of the density of the environment and jet opening angle of \textit{SVOM} LGRBs. This information will allow the characterization of the LGRB progenitors and a better determination of the true rate of LGRBs. 

The jet opining angle and the redshift determination are also a fundamental information to test the correlations found between some quantities related to the gamma-ray prompt emission of LGRBs. Indeed, \cite{Amati2006}, \cite{Yonetoku2004} and \cite{Ghirla2004}, found a correlation between the peak energy (E$_{peak}$) and the isotropic energy (E$_{iso}$), isotropic luminosity (L$_{iso}$) and isotropic energy corrected by the jet collimation (E$_{\gamma}$) of the gamma-ray prompt emission, respectively. These correlations can in principle bring to the use of LGRBs to determine cosmological parameters (see section \ref{sec:cosmo}), but their physical explanations are still unknown and there is still an open debate on their robustness.
A \textit{SVOM} sample of LGRBs complete in redshift (and including the jet opening angle determination for the E$_{peak}$ vs E$_{\gamma}$ correlation), will make us able to test these relations.

\subsection{The diversity of long GRBs: X-ray flashes, underluminous and ultralong GRBs} 
\label{sec:grbdiversity}


Long gamma-ray bursts exhibit a great diversity in their luminosities, spectral and 
temporal properties. The luminosity function of classical long GRBs
covers a wide range in isotropic luminosity from $<10^{50}~{\rm erg~s^{-1}}$
to $>10^{54}~{\rm erg~s^{-1}}$ \citep[e.g.][]{sun:2015}. But some nearby long GRBs
have luminosities as low as a few $10^{47}~{\rm erg~s^{-1}}$. These
underluminous GRBs must have a large event rate, and probably
form a distinct population \citep[e.g.][]{liang:2007,bromberg:2011}.
The peak energy of the spectrum goes from a few keV in X-ray flashes to several MeV in 
hard bright bursts and the spectrum extends to several tens of GeV in some events detected 
by \textit{Fermi}/LAT.
GRB light curves can be simple, showing one single pulse or extremely complex with many pulses and spikes, often 
overlapping. The class of ultra-long GRBs extends the maximum observed burst duration to more than one 
hour.  
 
\textit{SVOM} will take advantage from its wide spectral coverage of the prompt emission
(from X/$\gamma$ with ECLAIRs and GRM to the optical with GWAC) to explore and better understand 
this diversity, which is a source of complexity but can also provide clues to decipher the various mechanisms
at work during the prompt emission of GRBs. A few examples of expected \textit{SVOM} contributions are given below. 

%

\begin{figure}[ht]
\centerline{\includegraphics[scale=0.45]{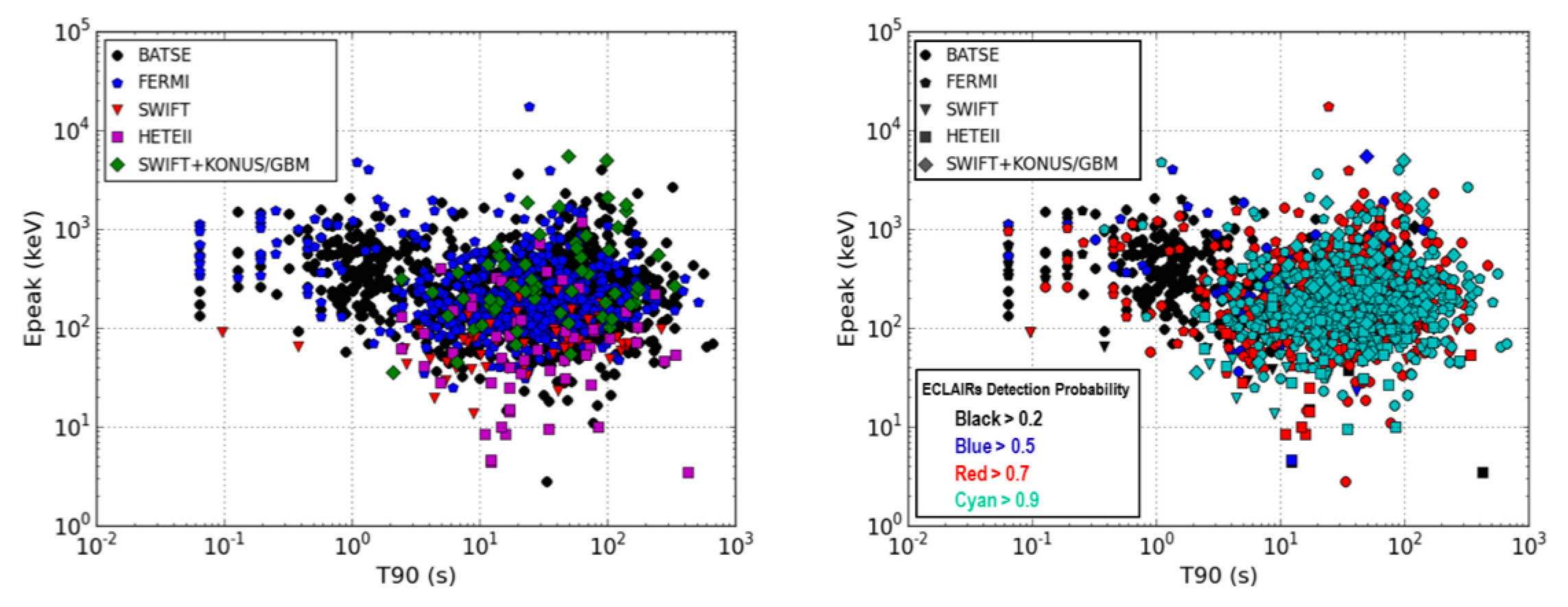}}
\caption{Exploration of the GRB diversity by \textit{SVOM}. This figure illustrates the capacity of ECLAIRs to trigger on the various known classes of GRBs. \textit{Left :} the sample of known GRBs with a prompt spectral measurement is plotted in the hardness (peak energy $E_\mathrm{peak}$) -- duration ($T_{90}$) plane. Each symbol correspond to a different GRB catalog : \textit{CGRO}/BATSE \citep{goldstein:2013}, 
\textit{HETE2} \citep{sakamoto:2005},
\textit{Swift} GRBs with a spectrum measured by either \textit{Konus} or \textit{Fermi}/GBM (sample taken from \citealt{heussaff:2013})
 and 
\textit{Fermi}/GBM \citep{gruber:2014}. 
The classical long GRB population is well covered by all instruments but the number of GRBs with both an accurate prompt spectrum and an afterglow detection -- corresponding approximatively to the \textit{Swift}+\textit{Konus}/GRB sample -- remains relatively small. 
Short hard GRBs (small duration, high peak energy) are mainly detected by BATSE and GBM thanks to a broader spectral range of the trigger instrument. The artificial vertical line at $T_{90}=64\, \mathrm{ms}$ is due to the time-resolution of the sample and corresponds to the shortest GRBs. X-ray rich GRBs and X-ray flashes (long duration, small peak energy) belong mainly to the \textit{HETE2} sample. 
\textit{Right :} each GRB in the total sample of the left panel has been simulated in ECLAIRs. The resulting detection probability (averaged over the whole field of view, assuming an isotropic distribution in the sky) is color-coded. This illustrates the capacity of \textit{SVOM} to trigger on all classes of GRBs. The detection probability is especially good for classical long GRBs, X-ray rich GRBs and X-ray Flashes. The efficiency for short GRBs may be improved using ECLAIRs+GRM trigger which is more sensitive to hard bursts.}
\label{fig:grbphysics_exploration}
\end{figure}


\textbf{Spectral diversity: X-ray rich GRBs and X-ray Flashes.}
Based on the peak energy distribution ($E_p$) or hardness ratio of GRBs, GRB-related
high-energy transients are sometimes classified into several sub-types \citep[e.g.][]{sakamoto:2005}, including typical GRBs ($E_p>50$ keV), X-ray rich GRBs
(XRR, 30 $\le$ $E_p$ $\le$ 50 keV), and X-ray flashes (XRFs, $E_p < 30$ keV). 
The relatively narrow spectral coverage of {\it Swift} and the {\it Fermi} GBM trigger starting at 15 keV and 8 keV respectively
have limited the ability of 
these two missions to efficiently observe soft events. 
Conversely ECLAIRs, with a  
trigger starting at 4 keV will increase the samples of X-Ray rich GRBs and X-Ray 
Flashes (see \reffig{fig:grbphysics_exploration}), providing redshifts and a detailed spectral information on these soft events. 

\textbf{Spectral diversity:  spectral differences between various classes of GRBs.}
Apart from the class of XRR and XRFs, differences in spectral properties are also expected in relation with the luminosity, the temporal properties 
(highly variable vs smooth light curves, ultra-long events) or the source distance. The extended spectral coverage of \textit{SVOM} and the image mode available with
ECLAIRs will allow studying in details these spectral differences 
in e.g. the ratio between thermal and non thermal components or the level of additional power-laws contributing at low or/and high energy. In addition 
the ability to efficiently obtain the burst redshift from the same plateform will be extremely valuable for the physical interpretation of the results.  

\textbf{Temporal diversity: ultra-long GRBs.} 
A new class of events, characterized by a very long duration, has recently been
identified \citep[e.g.][]{levan:2014a}. They have been detected via the image mode of {\it Swift}. 
Similarly the image trigger available on \textit{SVOM} will detect ultra-long GRBs. \textit{SVOM} will offer
a very complete monitoring of these objects with the three onboard instruments and the GWAC on the ground. Together 
with its good localization capability \textit{SVOM} 
will contribute   
shedding light on their physical origin.

\textbf{Temporal diversity: various progenitors.}
Apart from the known separation between short and long GRBs expected to 
come respectively from compact star mergers and collapsars, other subtypes of events
may have specific progenitors. For example, while typical long-duration GRBs may 
arise from successful jets from the core collapse of Wolf-Rayet stars, low-luminosity
GRBs may be related to unsuccessful choked jets \citep[e.g.][]{bromberg:2011}. Ultra-long
GRBs have been suggested to probably have a blue supergiant progenitor \citep{levan:2014a}.
The very complete spectral coverage 
available with \textit{SVOM} will help identifying thermal and non thermal 
components (possibly peaking at widely separated energies) of different sub-types
of GRBs, providing clues that will help to discriminate among various possible scenarios.  

\textbf{\textit{SVOM} contributions.}
By offering a broad spectral coverage and localization on a single plateform \textit{SVOM} combines
several {\it Swift} and {\it Fermi} capabilities and offers new ones, such as the continuous monitoring
of the ECLAIRs FoV in the optical. 
\textit{SVOM} is well designed to explore the GRB diversity and will fly when GW observatories 
and new facilities (\textit{JWST}, \textit{LSST}) will be fully operational.

\subsection{Short GRBs}
\label{sec:grbshort}

Short GRBs were discovered by \citet{mazets:1981}, and are generally defined as those having $T_{90} < 2$ s \citep{Kouve93}. They appear to form a population distinct from the longer GRBs as they form a secondary peak in the burst duration distribution measured by high energy detectors; they also have generally harder prompt emission, have an energy output $\sim$ 1\% of long GRBs, and are seen to occur at closer distances than long GRBs (short $\left\langle z\right\rangle\sim 0.5$, long $\left\langle z\right\rangle\sim 2.0$). For a review see \citet{Berger:2014}. In distinction to the long bursts, which occur in regions of high specific star formation rate, short GRBs occur in a broader range of environments, including elliptical galaxies; the short GRB rate is ten times higher per unit galaxy mass in late-type than in early-type galaxies. This variety of host galaxy types indicates a range of burst delay times since the initial star formation. 

The distinction between long and short GRBs is not without problems however. It has been argued that the band-pass of the detector should be considered in setting the threshold $T_{90}$ value \citep{bromberg:2013}, as bursts are known to last longer at lower energies. In addition, when the cosmological time dilation of the bursts is accounted for the durations of long and short bursts no longer appear distinct, and it seems as if short GRBs are rather like the first 0.3 rest-frame seconds of the long GRBs \citep{ghirlanda:2015}. GRB060614 and GRB060505 were nearby bursts with $T_{90}> 2$ s but without the supernovae characteristic of long GRBs, and are well-known examples of bursts that are difficult to assign to the short or long burst groups \citep{galyam:2006,ofek:2007,zhang:2007}.

Physical classification schemes have been proposed based on the widely held view that short GRBs have a formation mechanism independent of that of the long GRBs \citep{zhang:2007,zhang:2009,lv:2010,bloom:2008}: while long GRBs are considered to be the result of the collapse of a massive star, short GRBs are thought to be due to the collision of a neutron star and a black hole or a pair of neutron stars in a binary system due to orbital decay caused by the emission of gravitational radiation. Consistent with this picture, long bursts are seen to be accompanied by a type Ibc supernova if they are close enough, whereas no supernova occurs with short bursts \citep[e.g.][]{kann:2011}, and short GRBs can occur at a significant distance from their host galaxy, as expected if their progenitors are the very long-lived NS-NS or NS-BH systems which have received a significant kick at the time of their creation. The merger of white dwarf binaries has also been proposed to be responsible for un-kicked short GRBs in both early and late-type galaxies \citep[e.g.][]{levan:2006}. For both long and short GRBs a highly relativistic jet along the rotation axis is pointed in our direction; it is the emission from this that we see as the gamma-ray burst, and this common factor is likely responsible for the difficulties in classifying burst types. 

\begin{figure}[ht]
\centerline{\includegraphics[scale=0.5]{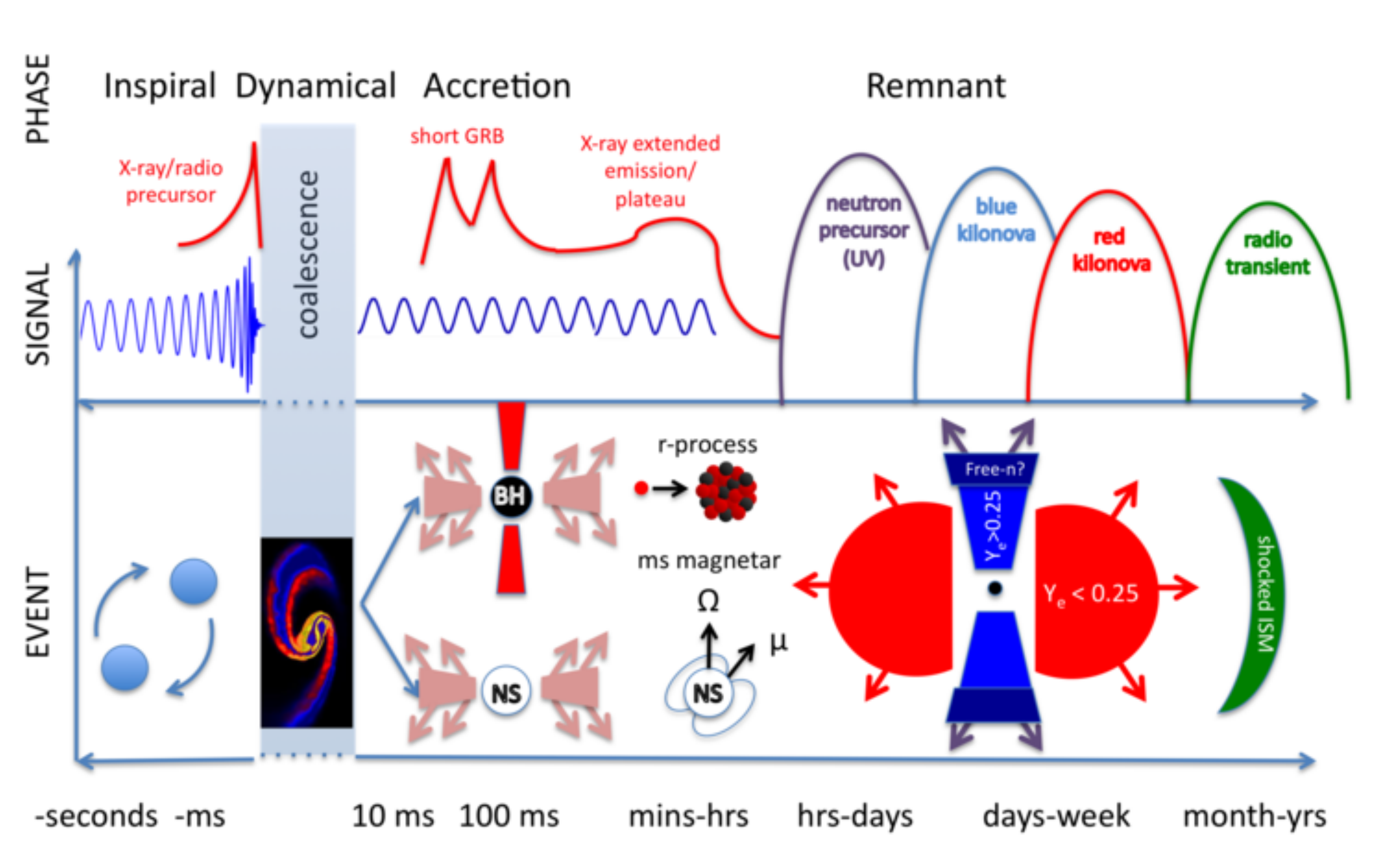}}
\caption{Expected electromagnetic counterparts to gravitational waves in a NS/NS or NS/BH merger.
This figure, taken from the review by \citet{fernandez:2015} illustrates the expected time sequence: gravitational wave emission during the inspiral, merger and post-merger phase, associated short GRB and its afterglow, kilonova and radio afterglow. The uncertainties related to the final state of the merger are also indicated: see \citet{fernandez:2015} for a full description of the figure.  }
\label{fig:grbphysics_short_gw}
\end{figure}

The compact star merger model of short GRBs leads directly to two predictions: the simultaneous production of strong gravitational wave emission from the final stages of orbital decay and merger, and the substantial production of r-process elements in the neutron-rich merger ejecta. The recent detection of gravitational wave emission from a massive BH+BH binary by the advanced LIGO interferometers \citep{abbott:2016a} gives great hope that the NS+NS predictions of  \citep{abbott:2016b}, although uncertain, will eventually be realised at close to $\sim$40 detections per year. Decay of radioactive r-process elements in the rapid ejection of up to a few percent of a solar mass of neutron star material is expected to give rise to an unbeamed "€˜kilonova"€™ peaking in the near IR a few days after the merger \citep{li:1998,barnes:2013,kawaguchi:2016}, a prediction apparently confirmed for the short GRB 130603B by \citet{tanvir:2013}. X-ray heating may be a more efficient kilonova powering mechanism \citep{kisaka:2016}. Candidate kilonova emission has also been found in GRB 060614 \citep{yang:2015} and GRB 050709 \citep{jin:2016}. The final stellar outcome is expected to be a black hole \citep{fryer:2015}, although a rapidly spinning magnetar may be formed \citep[e.g.][]{rowlinson:2013}, perhaps as an intermediate product before final collapse, in which case the detectable electromagnetic signatures from the system can become much richer and brighter \citep{Gao15}. \citep{fernandez:2015} review the electromagnetic emission from colliding neutron star systems (see \reffig{fig:grbphysics_short_gw}).



\textit{Swift} has discovered just over 8 short GRBs per year, leading to substantial progress, for example by discovering the first short GRB afterglow \citep{gehrels:2005}. It has shown that $\sim$45$\%$ of short GRBs fade beyond observational limits in X-rays within $\sim$17 minutes. Median r-band AB magnitudes are $\sim$23 at 7 hours, an order of magnitude lower flux than from long GRB afterglows (in spite of their much closer distance). Because of their relative faintness short GRBs are hard to study.

\begin{figure}[ht]
\centerline{\includegraphics[scale=0.5]{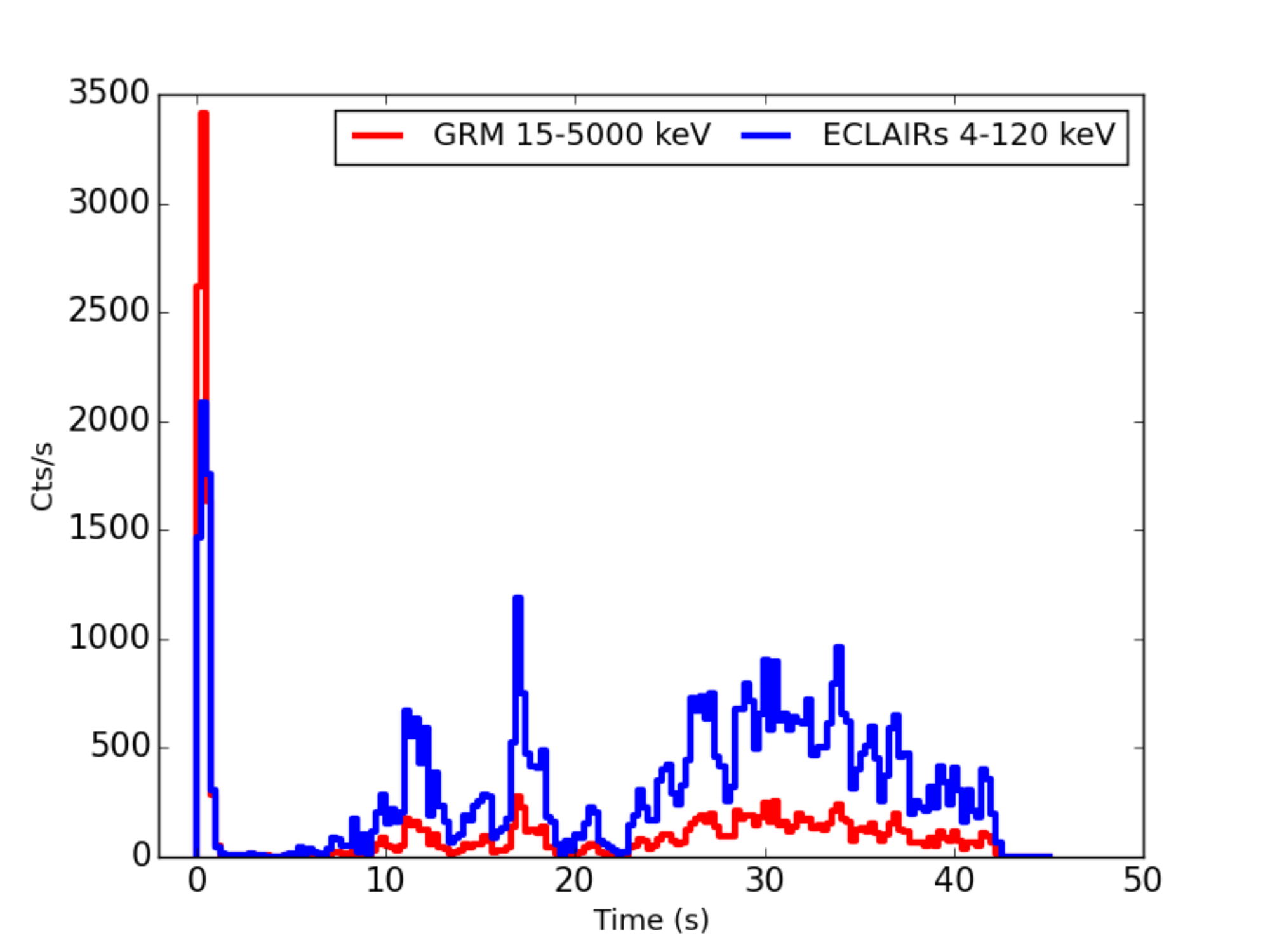}}
\caption{Simulation of the light curve of a short GRB with a soft extended emission in \textit{SVOM}/ECLAIRs+GRM.
GRB 990712A shows an initial short and hard spike, leading to the classification as a short hard GRB, followed by a soft extended emission 
\citep{norris:2006}. The simulated light curve is plotted in counts per second in ECLAIRs and GRM, based on the time-dependent spectral analysis published by \citet{kaneko:2006} and assuming that the GRB is on-axis for ECLAIRs and therefore 30° off-axis in GRM. 
During the initial hard spike,  GRM is more sensitive than ECLAIRs:
it is detected at $62\sigma$ in GRM and $41\sigma$ in ECLAIRs. 
Therefore, in the case of weaker short GRBs, the detection by GRM may be used to increase confidence in a marginal ECLAIRs detection and to lower the slew threshold. During the soft extended emission, ECLAIRs becomes more sensitive: detection at $77\sigma$ in ECLAIRs and $15\sigma$ in GRM. Therefore the combination of the two instruments will allow to increase the sensitivity of \textit{SVOM} to short GRBs and the capacity to slew and catch their early afterglow.     }
\label{fig:grbphysics_short}
\end{figure}

\textit{SVOM} has much to add, with its complementary wide-field and wide band-pass suite of sensitive instruments. The \textit{SVOM} pointing scheme is optimised for ground-based follow-up, so it will enable the measurement of a much higher fraction of redshifts than has been possible so far, helping to overcome the selection effects evident in the current sample. Knowledge of the burst distance is key to establishing the energetics and physical mechanisms responsible for the emission. In addition, the large FoV of the \textit{SVOM} GWAC is expected to provide the very first measurements of the prompt optical emission, while the GFTs may cover the so-called extended emission phase, so helping to constrain outflow and emission models. The VT on \textit{SVOM} is more sensitive than the \textit{Swift} UVOT, and so it will locate a much higher fraction of short GRB afterglows than has been possible so far. The combination of GRM and ECLAIRs is well suited to investigation of the prompt initial hard spike and the following softer extended emission which is observed in a significant fraction of short GRBs (see \reffig{fig:grbphysics_short}). This may be due to magnetar spin-down, or to accretion gating by a magnetic propeller and a signature of a significant mass ratio in the progenitor binary leading to a large ejecta mass and so a larger fall-back accretion disk \citep[e.g.][]{fan:2006,gompertz:2014}. The relationship between the optical and X-ray emission in the minutes after the burst certainly require clarification. \citet{rowlinson:2013} show that the magnetar signature seen in X-rays is not necessarily seen in the optical. 

The 50 keV - 5 MeV sensitivity of the 3 GRM modules provides a radical new capability compared to \textit{Swift}, the combination of GRM and ECLAIRs is a powerful tool that will allow strong constraints on the low energy prompt spectrum. This will enable newly sensitive searches for optically thick thermal components in \textit{SVOM} bursts, and also will strongly constrain the low energy part of the Band function, which has been shown to be a defining characteristic of short GRBs (\citealt{ghirlanda:2015}, see however \citealt{guiriec:2010}).



Kilonovae may be the most commonly found electromagnetic counterparts to gravitationally detected neutron star mergers.
The frequency of these associations will allow the evaluation of contribution of mergers in the nucleosynthetis of the heaviest elements in the Universe \citep[e.g.][]{vangioni:2016}.
The red and blue bands of the VT will be useful in defining the early decay that will help in the identification of new kilonovae. Short GRBs are also expected from appropriately oriented mergers; the wide FoV of the MXT will be especially useful in searching the large error regions of the advanced LIGO and VIRGO gravitational wave interferometers (X-ray search strategies are proposed by \citet{evans:2016} and \citet{gehrels:2016}). The combination of gravitational wave signal and X-ray light curve from the MXT will be key driver of new science by identifying the EM counterpart, linking the NS equation of state to the X-ray light curve and establishing whether a mergernova or a magnetar phase exists in the collapse 
\citep{lasky:2014,dallosso:2015,li:2016}. 
\citet{bloom:2009,metzger:2012} and \citet{li16,chu16} provide overviews of the multiple and important benefits of joint GW-EM short GRB measurements; these include increased detection sensitivity, more complete diagnosis of the binary parameters, establishment of the astrophysical evolutionary context, constraints on the equation of state of the neutron stars, cosmological constraints through the measurement of both redshift and luminosity distance, as well as a strong constraint on the speed of gravitational waves.

\subsection{GRBs as particle accelerators}
\label{sec:grbacceleration}


Electrons and hadrons are accelerated on a very short timescale at the shock fronts in the jets 
to ultra-relativistic speeds. In the standard framework, the gamma-ray emission is generaly explained by the synchrotron emission of the relativistic electrons. The origin of the 
high-energy gamma-ray emission ($\gtrsim 100 $ MeV) measured by \textit{Fermi}/LAT is still debated: 
both leptonic and hadronic models can be invoked to explain the spectral shape of the 
extra-components. The combination of multi-wavelength and multi-messenger studies are 
required to answer to this question.

The observation by the \textit{Fermi}/LAT of several GRB photons with energies reaching 10-100 GeV 
in the source frame~\citep{Ackermann2013}, as late as $\sim1$ day after the trigger in the 
case of GRB130427A, is encouraging for GRB detections at very high energies with the 
current ground-based imaging atmospheric Cherenkov telescopes and with synoptic detectors 
such as \textit{HAWC}~\citep{Meszaros2015} and \textit{LHAASO}~\citep{DiSciascio2016}. None of the past and 
current experiments ever succeeded in capturing a high-energy signal from a GRB
with their present operational threshold energies of $\sim100-200 $ GeV and the attenuation by the 
extragalactic background light (EBL) in this band at the typical distances of GRBs could be 
hindering their detection. The Cherenkov Telescope Array (\textit{CTA})~\citep{Inoue2013} is the next 
generation ground-based facility in the northern and southern hemispheres, which when combined 
will cover the entire sky over a broad energy range from tens of GeV up to $\sim100 $ TeV, with a 
sensitivity considerably improved with respect to existing instruments. Moreover, the large 
field of view will maximize the chance to discover a GRB compared to present experiments. 

The 
\textit{Fermi}/LAT has detected an attenuation in the high-energy power-law component of some GRBs, 
which has been attributed to internal gamma-ray opacity to pair production. Therefore, a simple 
extrapolation of the spectra to the VHE range is uncertain and makes the detections rate 
prediction difficult ($\sim$1 VHE GRB per year depending on the considered experiment \citep{Abeysekara2012,Inoue2013,Gilmore2013,Taboada2013}. Joint time-resolved 
spectral analyses based on \textit{SVOM}, \textit{HAWC}, \textit{LHAASO} and \textit{CTA} data will help to pinpoint the nature and 
the origin of the acceleration and emission processes at high energies in GRB jets. 

Such analyses 
are also crucial to study the most energetic photons from a burst. Precious information on the 
gamma-ray opacity has been  provided by the detection of high-energy photons, allowing to set 
stringent limits on the jet bulk Lorentz factor. Similarly, variability studies and spectral 
shapes in the GRB prompt emission phase from MeV to TeV energies will help to distinguish between 
leptonic and hadronic models, and to answer the long-standing question of the origin of the cosmic 
rays observed on Earth with energies up to $10^{20}$ eV. The detection of \textit{SVOM} GRBs at energies 
greater than $\sim10 $ GeV will also provide crucial tests of the amount and of the origin of the EBL 
at high redshifts ($z > 2$), beyond the reach of blazar active galactic nuclei. It will bring 
additional constraints on the intergalactic magnetic fields through their impact on the 
propagation of the photon/pair cascades created by the interaction of VHE photons with the EBL. 
By studying the delay between the low and high energy emissions, \textit{SVOM} GRBs will also be used as 
probes of the Lorentz invariance violation, which can manifest as a dependence of the speed of 
light in vacuum on its energy. \textit{SVOM} will provide GRB alerts with very similar characteristics 
as \textit{Swift} alerts at a rate of 60-70 GRBs/year and excellent localization well within the \textit{CTA} 
field-of view. More generally, \textit{SVOM} will provide the low-energy context which is fundamental 
for any broad-band multi-component spectral analysis, in particular for the understanding of 
GRB properties in the gamma-ray extreme energy range.

GRBs have been proposed as one of the potential sources for ultra-high energy cosmic rays 
(UHECRs) with energy up to $10^{20}\,{\rm eV}$ \citep{Vietri1995,Waxman1995}. The 
baryon interactions with the fireball photons produce a burst of neutrinos with energies 
above 100~TeV in the standard internal shock model \citep{Waxman1997,Hummer2010}. 
In the external shock region, GRBs may also produce neutrinos with even higher energies 
\citep{Waxman2000,Li2002}. A lower energy neutrino component (few TeV) may be 
also expected through hadronic interactions with the stellar envelope 
material~\citep{Meszaros2001,Murase2006}. The detection of an high-energy neutrino 
(HEN) signal in coincidence with GRBs would be a direct proof of the existence of an hadronic 
component in the jets. The ANTARES and the IceCube detectors are the current most sensitive 
neutrino telescopes in operations in the Northern and Southern hemispheres, respectively. During 
the \textit{SVOM} operations, the KM3Net detector in the Mediterranean Sea will achieve an instantaneous 
sensitivity larger than the current IceCube telescope~\citep{Km3net2016}. After 2025, 
the IceCube Collaboration also plans to extend its array to $10\,\mathrm{ km^{3}}$ \citep{ICGen22014}. A 
first HEN signal has been detected by IceCube and found to be consistent with an isotropic 
flux~\citep{Aartsen2014} with energies above few tens of TeV. The sources of these neutrinos 
are currently unknown. Their potential origins as GRBs have been discussed by several 
authors~\citep[see e.g.][]{Liu2013,Murase2013}

Searches for a neutrino signal from individual bright GRBs and/or through the stacking of a 
large number of GRBs have been unsuccesful so far \citep{Abbasi2010,Antares2013}. 
This non-detection suggests that the standard GRB population is not the major contributor to 
the diffuse HEN flux \citep{IceCube2016}. Low-luminosity GRBs and chocked GRB jets, largely 
missed by current gamma-ray satellites (see section \ref{sec:grbdiversity}) may contribute significantly to the 
HEN diffuse ~\citep{Wang2007}. Less powerful jets or denser external material are more favorable for VHE 
production \citep{Ando2005,Gupta2007,Murase2008}. \textit{SVOM} may be more 
efficient in the detection of such GRB populations thanks to the ECLAIRs low threshold 
energy of 4 keV (see \reffig{fig:grbphysics_exploration}), therefore providing a new sample  to search for a correlated neutrino 
emission. Predictions for HEN emissions from GRBs are still uncertain. Improving the modeling 
requires gamma-ray spectra, multi-band afterglow light curves and redshift measurements. 
Thanks to the performance of its instruments, to their large multi-wavelength coverage and 
to the excellent space-ground synergy, \textit{SVOM} will provide a sample of well characterized GRBs, 
which is primordial for the search of their potential HEN counterpart.

\subsection{\emph{\textit{SVOM}} and the high-redshift universe}
\label{sec:grbhighredshift} 
Thanks to their extreme high luminosities
and the spectroscopy of the optical afterglows, GRBs can be detected and studied up to high redshifts.
Thus, high-$z$ GRBs have been considered as a new powerful tool to explore the early Universe.
\emph{\textit{SVOM}} will be ideal in that its capabilities are optimized to detect GRBs at high-redshifts. 
Indeed, because GRBs are associated with the death of massive stars, they are expected to
be detected up to extremely high redshift, $z>10$, where bright quasars may not have time to build up
their central black hole. They are thus unique probes of the ISM of the first galaxies and of 
the reionization epoch. In this regard, there are potentially important gains from rapid response of 
follow-up telescopes, ideally getting on target in some tens of minutes.


 In order to exploit the full potential of GRBs as a probe of the distant Universe, a 
larger sample of high-$z$ GRBs is needed compared to what is available now. 
The detection of high-$z$ GRBs is one of the essential goals
driving the design of future GRB missions, including the forthcoming \emph{\textit{SVOM}} satellite. It has been suggested that
the best strategy for detecting a large number of high-$z$ GRBs is to design a facility operating in the soft
X-ray band with a high sensitivity \citep{2015MNRAS.448.2514G,2015JHEAp...7...35S}.
ECLAIRs has a wide FoV of
2 sr ($89^{\circ}\times89^{\circ}$). The sensitivity of ECLAIRs is expected to be $7.2\times10^{-10}$ erg $\rm cm^{-2}$ $\rm s^{-1}$
($\sim30$ mCrab, 5 $\sigma$ detection level in 1000 s, \citealt{2011CRPhy..12..298P,2014SPIE.9144E..24G,2015arXiv151203323C}).
It will be sensitive from 4 to 150 keV. Thanks to the low energy threshold of 4 keV,
ECLAIRs will be sensitive to soft GRBs like X-ray Flashes and highly redshifted GRBs.

Our detailed simulations of the number of GRB detections expected with \emph{SVOM}/ECLAIRs are described as follows
(see \citealt{2015MNRAS.448.2514G}, for more details):

(1). GRBs are distributed in redshift (up to $z=20$) following a modified comoving
star formation rate: 
%
\begin{equation}
\psi(z)\propto (1+z)^{\delta} \psi_{\star}(z)
\label{z1}
\end{equation}
where $\psi_{\star}(z)$ denotes the cosmic star formation rate
\citep{2006ApJ...651..142H,2008MNRAS.388.1487L} in units of ${\rm M}_{\odot}$ ${\rm yr}^{-1}$  ${\rm Mpc}^{-3}$:
\begin{equation}
\psi_{\star}(z)=\frac{0.0157+0.118z}{1+(z/3.23)^{4.66}}.
\label{SFR}
\end{equation}
%
\citet{2012ApJ...749...68S} found that strong evolution in the GRB rate density $(\delta=1.7\pm0.5 (1\sigma))$
is required in order to account for the observed differential number counts of BATSE and
the observed $z$ distribution of the $Swift$ complete sample. Here we adopt their value of $\delta$.

(2). The GRB luminosity function (LF) has also been well constrained by \citet{2012ApJ...749...68S}
through the BAT6 sample, which is a broken power law function,
\begin{equation}
  \phi(L_{\rm iso}) \propto \left\lbrace \begin{array}{ll}\left(L_{\rm iso}/L_{\rm cut}\right)^{x}, ~~~~~~~L_{\rm iso} <L_{\rm cut} \\
                                          \left(L_{\rm iso}/L_{\rm cut}\right)^{y}, ~~~~~~~L_{\rm iso} >L_{\rm cut}\;, \\
\end{array} \right.
\label{lf}
\end{equation}
where $x=-1.50$, $y=-2.32$, $L_{\rm cut}=3.8\times 10^{52}$ erg $\rm s^{-1}$, and $L_{\rm iso}$ is the peak isotropic luminosity.

(3). To calculate the peak flux of bursts in a given energy range, we need to assign a spectrum to each mock GRB.
Assuming that all bursts are well fit with the Band function \citep{1993ApJ...413..281B} and the photon spectral indices
in the energy bands lower and higher than the peak energy $E_{\rm peak}$ are $-1$ and $-2.3$, respectively
(e.g., \citealt{2006ApJS..166..298K,2011A&A...530A..21N,2012ApJS..199...19G}).

(4). The spectral peak energy $E_{\rm peak}$ for each burst is obtained through the Yonetoku relation \citep{2004ApJ...609..935Y}.
We adopt the $E'_{\rm peak}-L_{\rm iso}$ relation obtained with the complete BAT6 sample \citep{2012MNRAS.421.1256N}:
\begin{equation}
\log{(E'_{\rm peak})} = -25.33 + 0.53 \log{(L_{\rm iso})}\;,
\label{Yonetoku}
\end{equation}
where $E'_{\rm peak}=E_{\rm peak}(1+z)$ is the rest frame peak energy.

(5). With the assumptions above, the peak flux $P$ of each simulated burst in a
given energy range $\Delta E=[E_{1},E_{2}]$ can be expressed as:
\begin{equation}
P_{\Delta E}=\frac{L_{\rm iso}}{4\pi d_{L}(z)^{2}}
\cdot \frac{\int_{E_{1}}^{E_{2}} N(E)dE}{\int_{1{\rm keV}/(1+z)}^{10^4{\rm keV}/(1+z)} E\, N(E)dE}\;,
\label{flux}
\end{equation}
where $N(E)$ is the photon spectrum and $d_{\rm L}(z)$ is the luminosity distance of the GRB.

We normalize the simulated GRB population to the real population of bright $Swift$/BAT bursts,
as \citet{2015MNRAS.448.2514G} did.
The BAT6 sample of \citet{2012ApJ...749...68S} is constructed by considering only bursts with
favorable observing conditions, which are bright in the 15--150 keV $Swift$/BAT band, i.e.,
with peak photon flux $P\ge 2.6$ ph cm$^{-2}$ s$^{-1}$.
This sample is composed of 58 GRBs and reaches a completeness level of 95\% in redshift.
Among the current $Swift$/BAT sample (773 events detected up to 2014 July with duration exceeding
$T_{90}\ge2$ s)\footnote{http://swift.gsfc.nasa.gov/docs/swift/archive/grb\_table/},
204 bursts with peak flux $P \ge 2.6$ ph cm$^{-2}$ s$^{-1}$ (i.e. the same flux
threshold used to define the BAT6 sample) have been extracted.
Considering the $Swift$ FoV of 1.4 sr and the observational period of $\sim$9.5 years,
a detection rate by $Swift$/BAT is estimated to be $\sim$15 events yr$^{-1}$ sr$^{-1}$
with peak flux $P\ge$2.6 ph cm$^{-2}$ s$^{-1}$ \citep{2015MNRAS.448.2514G}.
The simulated population of GRBs is normalized to this rate.

\begin{figure}[h]
\centerline{\includegraphics[width=0.6\linewidth]{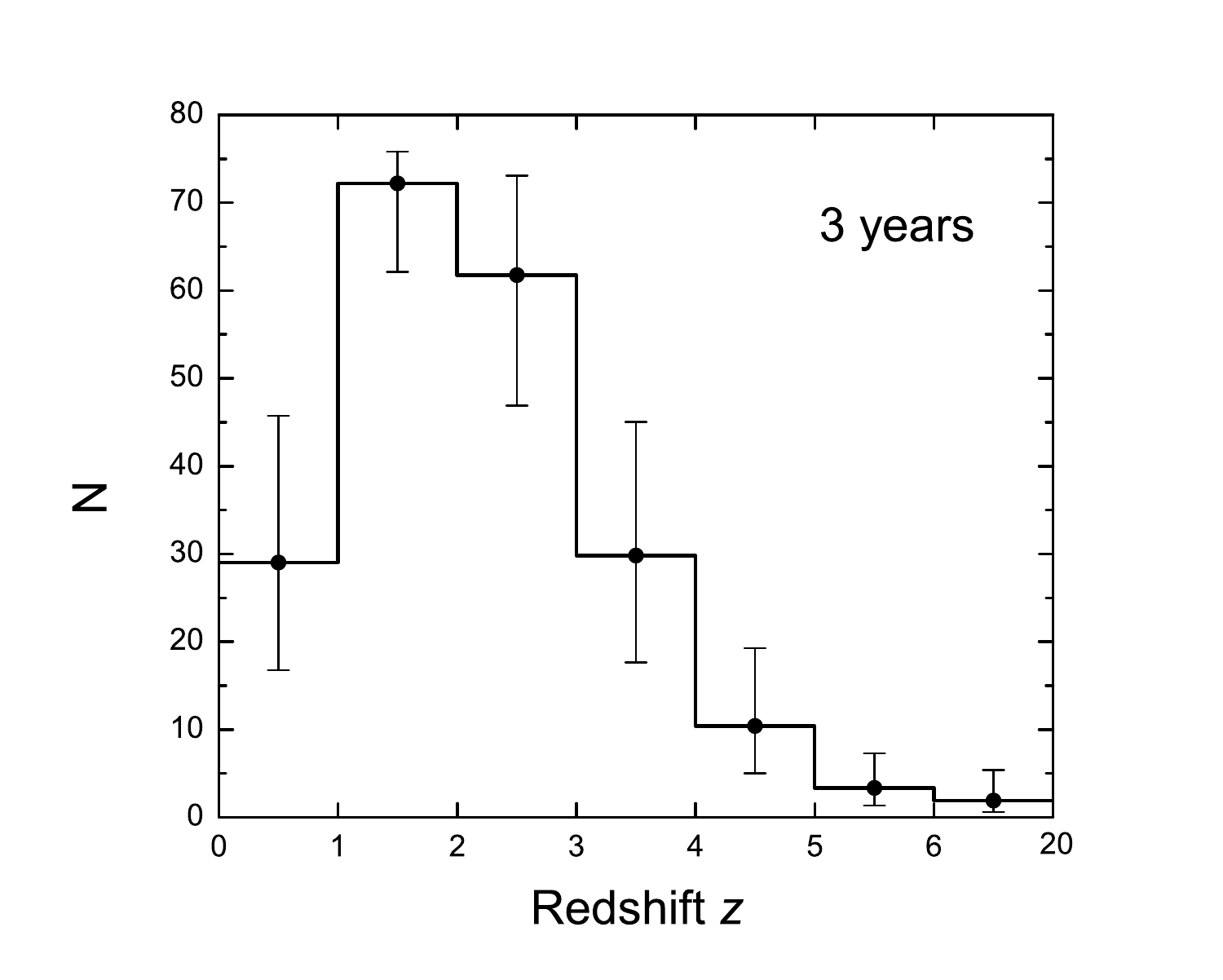}}
 \caption{\small Simulated redshift distribution of GRBs to be detected by ECLAIRs during the 3 year mission lifetime.
 Error bars show the $2\sigma$ uncertainties of the assumptions on the GRB population (i.e., the $2\sigma$ uncertainties of
 the GRB rate density evolution, $\delta=1.7\pm2\sigma$) in the redshift bin.}
 \label{fig:highz}
\end{figure}

Using the fraction of observed sky per year (i.e., the ECLAIRs FoV of 2 sr) and the expected ECLAIRs
sensitivity of $\sim7.2\times10^{-8}$ erg $\rm cm^{-2}$ $\rm s^{-1}$ for an exposure of 10 s
in the 4--150 keV energy range, our simulations show that ECLAIRs will detect 70--77 GRBs/yr
depending on the exact assumptions on the GRB population (see the simulation procedures). Over the 3 yr
lifetime of the mission, ECLAIRs is expected to be able to detect $\sim$ 200--230 GRBs.
\reffig{fig:highz} shows the differential distribution of the GRBs expected to be detected by ECLAIRs
as a function of redshift. One can see from this plot that nearly 4\%--5\% of ECLAIRs GRBs are expected
to be high redshift GRBs ($z>5$), which corresponds to a detection rate of about 3--4 GRBs/yr at $z>5$,
including $\sim$ 2--3 GRBs/yr at $5<z<6$ and $\sim$ 1 GRBs/yr at $6<z<7$.
These results are in good agreement with the predictions from \citet{2014SPIE.9144E..24G}
and \citet{2015arXiv151203323C}.

In the {\it \textit{SVOM}} era we expect to rapidly identify candidate
high-$z$ bursts (VT non-detections), and benefit from the greatly enhanced imaging and spectroscopic capabilities of 
{\it JWST} and subsequently the 30\,m class ground-based telescopes.
This will mean not only much better redshift completeness at high-$z$ than has been possible for the {\it Swift} sample, 
based on afterglow spectroscopy (and host spectroscopy if the afterglow was localised, but no spectrum secured),
but also both extremely high signal-to-noise afterglow spectra in many cases, 
together with much deeper imaging to search for hosts in emission than is possible with current technology.
Thus, in the three year core mission, we expect to roughly double the current known sample of $z\gtrsim6$ GRBs, but, crucially, for most of these
also to obtain much more precise measures of host luminosity, column density and metallicities.

\subsubsection{GRBs to probe reionization}
\begin{figure}[!b]
\centerline{\includegraphics[width=0.5\linewidth,angle=270]{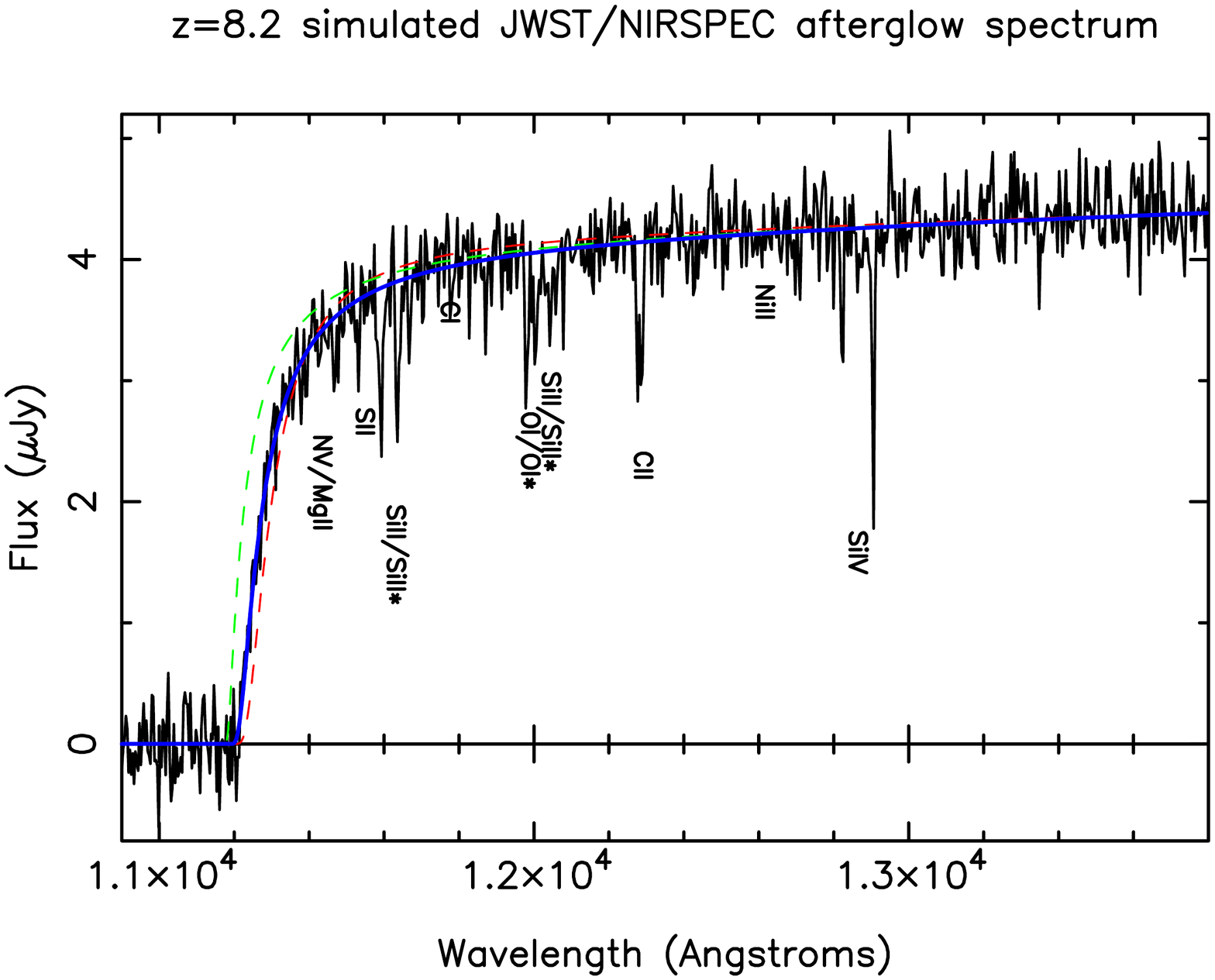}}
 \caption{\small  Simulated {\it JWST}/NIRSPEC spectrum of a GRB afterglow based on GRB\,090423, as it would have been
observed 4 days post-burst in a 5\,hr exposure.  This clearly shows that high $S/N$ spectroscopy will be possible
even for faint afterglows using {\it JWST}. The simulation assumed a 100\% neutral IGM and a host HI column density of
log$(N_{\rm H}/{\rm cm^{-2}})=21$.  The continuum cannot be fit by either a pure neutral IGM model (green line), or a pure host model
with an ionized IGM (red line).  Instead the two components can be decomposed, as illustrated with the blue fit.
}
 \label{fig:DLAIGM}
\end{figure}

After the Big-Bang, expansion made the universe temperature decrease rapidly. After about 380,000 years, the universe became 
transparent, radiation and matter decoupled and hydrogen recombined. Under the action of gravitation,
the first objects formed slowly and, either stars or quasars or both, started to reionized the universe.
This phase change is a crucial period and GRBs offer a unique opportunity to probe the ionization state of the gas
at that time.

Therefore, a very important observation to be made in the spectrum of high redshift ($z>6$) GRB afterglows is 
to search for the red damping wing of the absorption trough produced by neutral hydrogen
in the IGM. As it is known that reionization ends around $z\sim6$ \citep{Becker2007},
this signature would be definite evidence for this crucial phase of the universe history. 
GRBs have the advantage that no large scale proximity effect is expected contrary to QSOs that ionize
the IGM to a distance of several Mpc \citep{Guimaraes2007}. Their disadvantage is that
usually neutral gas from the host galaxy located 
in front of the GRB already produces a damping wing which is difficult to 
disentangle from the IGM effect \citep{Patel2010}. This is possible however if a high SNR 
spectrum of the afterglow is available.
At a redshift of $z=8$, the contribution
of the IGM is predominant as is shown in \reffig{fig:DLAIGM}. At the time \emph{SVOM} will be launched,
\textit{JWST} will be the best facility to perform such observations waiting for the advent of E-ELT.


It has long been suspected that the primary driver of reionization was extreme-UV radiation emitted from early generations
of massive stars \citep{2001ARA&A..39...19L}.  
This explanation requires both that sufficient star formation was occurring at $z\gtrsim6$ and also that
a sufficient proportion of the UV radiation that they produced escaped from their host galaxies into the surrounding medium.
Establishing whether these combined requirements are met in reality is fraught with difficulties.  

Determinations of the global star formation density at high-$z$ generally relies on
estimating the star formation rates in individual detected galaxies, and then extrapolating the luminosity function (LF) to account
for galaxies below the detection threshold.
Various studies have suggested that the galaxy luminosity function steepens at high redshift, to the point that
the overall star formation is likely dominated by the faint galaxies, below the depth of {\it HST} deep field observations \citep{2015ApJ...803...34B}.  
This adds considerable uncertainty to the extrapolation
since the form of the LF must be postulated 
\citep[usually taken to be a Schechter function;][]{1976ApJ...203..297S},
the faint end slope estimated from observations at the bright end, and
some cut-off chosen below which it is assumed that the lower rates of star formation could not be sustained.

The average escape fraction, $f_{\rm esc}$, of ionizing radiation presents an even tougher problem. It is not feasible to measure directly at
$z>6$ because the high opacity of the intergalactic medium itself will absorb almost all the radiation that might
escape any given galaxy.  
Most direct searches for Lyman continuum emission at lower redshift  have suggested that the escape fraction is only a few percent, or even less,
although, again, there are large uncertainties and potential systematics \citep{2016A&A...587A.133G}.
It is generally found that $f_{\rm esc}$ must increase to values of at least $\sim10$\% at $z\gtrsim6$, for stars to have been the agents of reionization \citep[e.g.,][]{2015ApJ...802L..19R}.  

Observations of high redshift gamma-ray bursts and their hosts provide powerful alternative routes to addressing both of these problems, and
hence offer exciting opportunities for {\it \textit{SVOM}}.

\subsubsection{The galaxy luminosity function in the era of reionization}

By conducting deep searches for the hosts of GRBs at high-$z$ we can directly estimate the ratio of star-formation occurring in detectable and undetectable galaxies to some limiting magnitude.
Thus, without fixing the luminosity function itself, we obtain the correction factor by which detected star formation must be multiplied
to infer the total star formation. Note, this is an essentially independent route to the global star formation rate compared to that
discussed in section~\ref{CSFR}, and does not  suffer from uncertainties in the evolution of GRB properties with redshift.
The sole assumptions are that GRB-rate is proportional to star-formation rate and that redshift incompleteness is
independent of host star-formation rate.  At high redshifts, where the large majority of star formation is at sub-solar metallicities, and most galaxies are
relatively free of dust, these assumptions are likely to be valid \citep[e.g.,][]{2016ApJ...817....8P}.

Since the position and redshift of each host is known from the GRB afterglows, 
the follow-up observations can be much shorter than equivalent multi-filter deep field searches for Lyman-break galaxies.
Early application of this technique to small samples high-$z$ hosts, none of which were detected in deep {\it HST} or VLT imaging, 
confirmed that the majority of star formation above $z >5$ occurred in galaxies below the effective detection limit of {\it HST} 
\citep{2012ApJ...754...46T,2012A&A...542A.103B,2012ApJ...749L..38T}.
Subsequently, new {\it HST} observations have resulted in the first detections of $z\sim6$ hosts  
\citep[][see Fig.~\ref{fig:NRT_NH}]{2015arXiv151207808M},
although the total sample remains modest (9 {\em Swift} GRBs at $z\gtrsim6$ at the time of writing).

\subsubsection{The escape fraction of ionizing radiation}

Ground-based spectra of afterglows at $z\gtrsim2$ usually show prominent troughs due to Lyman-$\alpha$ absorption
(in most cases resulting in a damped line)
by neutral hydrogen in the host galaxy, the median column density being
${\rm log}(N_{\rm H}/{\rm cm}^{-2})\sim21.5$.  
Since column densities of ${\rm log}(N_{\rm H}/{\rm cm}^{-2})\gtrsim18$ are essentially opaque to ionizing radiation, we can already say that the large 
majority of sight-lines to GRBs would have $f_{\rm esc}=0$.  
Of course, it is possible that in any given galaxy, another sight line
would have lower column density, but by assuming that GRB progenitors are representative, in terms of their locations, of typical
very massive stars (which are responsible for the bulk of UV production), 
we can infer statistically the average $f_{\rm esc}$ from a sample of GRBs \citep{2007ApJ...667L.125C}.
This has been performed for GRBs at  $z = 2$--4, indicating an upper limit of $f_{\rm esc}< 7.5$\% \citep{2009ApJS..185..526F}.

\begin{figure}[!b]
\centerline{\includegraphics[width=0.7\linewidth]{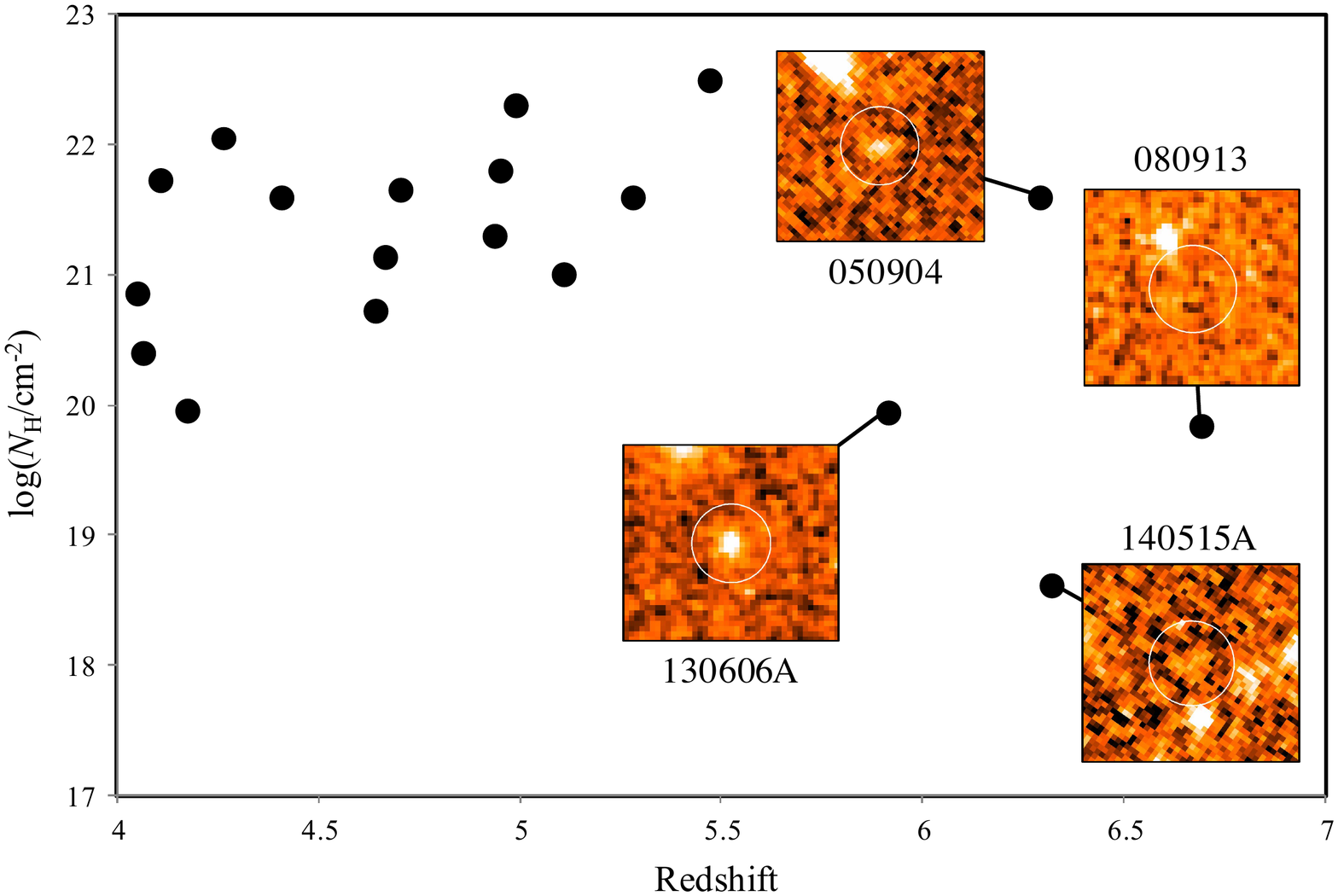}}
\caption{\small Plot of HI column density for the sample of {\it Swift} GRBs
\citep[taken from various sources, primarily][]{2013MNRAS.428.3590T,2014arXiv1405.7400C}, showing marginal evidence
for a decline at $z>5.5$, potentially indicating an increase in the escape fraction of ionizing radiation.
Deep {\it HST} imaging of the four highest redshift GRBs in this figure are shown as inset panels. Two hosts (of GRBs\,130606A and 050904)
are clearly detected, one (GRB\,140515A) is marginally detected, and the last (GRB\,080913) is undetected. 
}
\label{fig:NRT_NH}
\end{figure}

At higher redshifts, the flux blueward of the Lyman-$\alpha$ line becomes increasingly attenuated by  absorption due to 
the Lyman-$\alpha$ forest. 
This begins to present a technical challenge, as the contribution of the neutral fraction of the IGM must be disentangled
from the absorption due to the host, but this is possible given sufficiently high quality spectra \citep[e.g.,][]{2015A&A...580A.139H}.
So far there are only hints
of reducing HI column density in hosts at $z\gtrsim5.5$ \citep[][see Fig.~\ref{fig:NRT_NH}]{2014arXiv1405.7400C} 
compared to lower redshifts, 
but the sample remains too small for firm conclusions, and still favours a low value of $f_{\rm esc}$.

\subsubsection{Catching PopIII stars}
Gamma-ray bursts also offer the exciting opportunity to search for the first stars (hereafter, Pop III-primordial metal-free) formed in the Universe.
Numerical simulations show that Pop III stars form in primordial minihaloes at $z \sim 20$, with virial mass $M_{\rm vir}$ $\sim$ 10$^6$ M$_\odot$ 
and temperatures $T_{\rm vir}$ $\le$ 10$^4$  K.
These primordial stars are considered to have played a crucial role in the early cosmic evolution by emitting the first light and producing the 
first heavy elements.

The theoretical efforts developed to explain the formation and fate of PopIII stars remain largely untested because
there is no direct observation of these stars up to now. 
Different techniques may soon shed light on this important issue. 
For example the analysis of the morphology of 21cm spots could help discriminate the contribution of the different populations 
at different redshifts and carry wealth of information on Pop III stars. 
The \textit{JWST} will also allow  to directly observe the formation of the first galaxies, thus opening an amazing window 
for understanding the nature of the first stars and the evolution of populations at high redshift.

Another technique for observing these primordial stars may be to use GRBs. Indeed Pop III stars may also produce collapsar 
gamma-ray bursts whose total isotropic energy could be $\sim$2 orders of magnitude larger than average 
\citep{2006ApJ...642..382B,2003ApJ...591..288H,2011MNRAS.413..543S}. Even if the Pop III star has a supergiant hydrogen envelope, the GRB 
jet can break out of it because of the long-lasting accretion of the envelope itself. In this context the minimum energy expected for a GRB 
triggered by a Pop III star is near the maximum energy recorded for any GRB. Therefore, any GRB at $z\geq6$ and 
$E_{\rm iso}$ $\geq 8\times10^{54}$~erg would potentially have a Pop III progenitor.

Unfortunately it is not so easy to associate with certainty a distant GRB to a Pop III star. A promising method is to exploit the 
remarkable brightness of the GRB afterglow triggered by a Pop-III star. In the K band, the JWST will be able to detect GRBs, and to 
conduct spectroscopy on their afterglows out to $z\sim16$, even after 1 day. In the M band, the redshift horizon is extended further still, 
to $z\sim35$.
%


\subsection{The cosmic Star Formation Rate evolution}
\label{CSFR}

\begin{figure}
\centering
\includegraphics[width=0.7\textwidth]{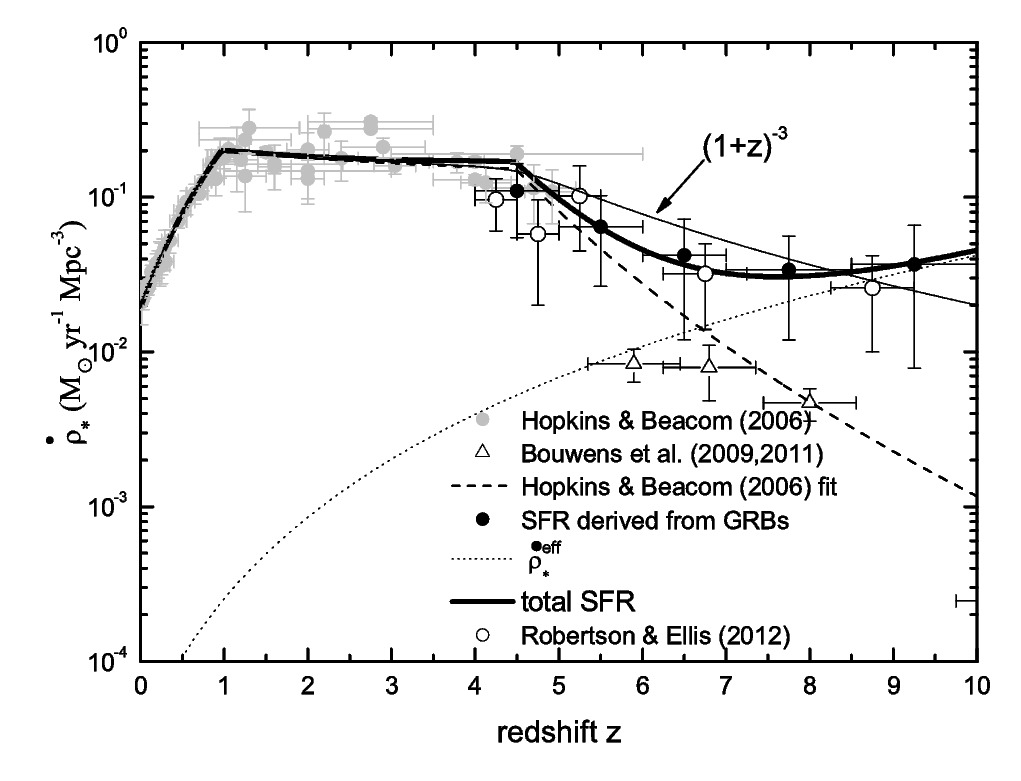} 
\caption{\label{figcosmicsfr} The cosmic star formation history.
The grey points are taken from Hopkins \& Beacom (2006), the dashed
line shows their fit. The triangles are from
\citet{Bouwens2009,Bouwens2011}.
The open circles are taken from
Robertson \& Ellis (2012). The filled circles are the SFR derived
from GRBs in Wang (2013). Adapted from \cite{wang2013}.} \label{SFR}
\end{figure}

Long-duration GRBs triggered by the collapse of massive stars,
provide a complementary method for measuring the global star formation rate density
\citep{Totani1997,Wijers1998,Bromm2002,vangioni15b,wang15,Petitjean2016}. The
short life-time of such stars makes them natural tracers of the recent
star formation. As a result, the rate
of long GRBs ($\dot{N}$) is linked to the star formation rate (SFR):
\begin{equation}
\label{EqGRBSFR} \dot{N}(z) = \epsilon \times \dot{\rho}_*(z)
\label{equaefficiency}
\end{equation}
$\epsilon$ is the efficiency with which GRBs occur (and are
detected) by unit of star formation rate $\dot{\rho}_*(z)$. 
Since long GRBs can be detected up to very high redshift (at $z\gtrsim 6$), if $\epsilon$ was perfectly 
known, we could use equation \ref{EqGRBSFR} to measure the cosmic SFR in the very early
history of the universe.
However, several works have shown that the rate of GRBs does not strictly follow the
SFR history \citep{Le2007,Salvaterra2007,Kistler2008,Yuksel2008,Wang2009a,robertson12,wang2013}.
For example, the SFR density inferred from high-redshift ($z>6$) GRBs has consistently been significantly 
higher than obtained through galaxy studies \citep{Bouwens2009,Bouwens2011}, although the finding of 
a very steep faint end of the galaxy luminosity function has begun to ease this tension 
\citep{Trenti2013}.  
%
It is however still possible to use GRBs to measure the
evolution of the cosmic SFR by assuming an evolution of the efficiency with redshift:
\begin{equation}
\epsilon=\epsilon_0 (1+z)^\delta.
\label{equaefficiencyevolution}
\end{equation}
By combining this form with a cosmic SFR history,
GRB detector efficiency, and luminosity function, it is possible to
fit simultaneously for the various parameters \citep[see e.g. the review of ][and references
therein]{wang15}. The redshift dependence ($\delta$) can be
constrained at $z<4$ where other tracers of the cosmic SFR are
available. The GRB rate at $z>4$ then provides a measurement of
the cosmic SFR in the first galaxies (see Fig. \ref{figcosmicsfr} for an example).
Some variants of this method were used e.g. in
\citet{daigne06,lapi08,kistler09,robertson12,salvaterra12,wang2013,Trenti2015}
finding $\delta$ in the range 0.5 to 1.5. Some studies suggest
however a perfect correspondence of the GRB rate and SFR, i.e.
$\delta \sim 0$ \citep[e.g.]{elliott12}. 

Possible reasons for the over-production of long GRBs at high redshift are 
discussed in details in \citet{wang15}. This issue is interesting to investigate
as this may yield astrophysical requirements for GRB production.
The metallicity evolution of the universe may well play an important role
since theory and observation both support that long GRBs prefer low-metallicity environments.
However, a few GRB hosts have high metallicity, so that the role of metallicity is still debated
\citep{Wolf2007,Kocevski2009,Graham2009,Svensson2010}. 
Other possibilities are the evolution of the stellar initial mass function
(IMF), whose effect on the GRB rate are studied in  \cite{Wang2011b} ;
the evolution of the break in the luminosity 
function of long GRBs \citep{Virgili2011} ; or even cosmic strings 
(linear topological defects formed in very early universe) serving as 
GRB central engines \citep{Cheng2010}.

The relation between GRB rate and SFR is also discussed at redshift lower than 1.
Several selection effects affect the observed redshift
distribution of GRBs \citep{Coward2007}, and so their intrinsic rate.
The most important one is the flux sensitivity of detectors. 
To correct for this effect, the  Lynden-Bell's $c^{-}$ method \citep{Lynden-Bell1971}
has been applied to the \emph{Swift} long GRBs \citep{Yonetoku2004,Kocevski2006,Wu2012,Yu2015}. 
The obtained trend suggests that the formation
rate of GRBs does not directly trace SFR at low redshift $z<1.0$ \citep[see however][]{pescalli16}.

Finally some insight on the relation between SFR and GRBs can come from the study of GRB host
galaxies. A strong dependence of the GRB rate
on host-galaxy properties out to $z\sim 1$ is found by
\cite{Perley2013}. \cite{vergani15} found
that the mass distribution of long GRB host galaxies is different than
the expected one for star-forming galaxies at $z<1.0$. In the same redshift range, 
\citet{boissier13} found that $\epsilon$ depends on the SFR or stellar mass of the host galaxy (and is possibly related to a metallicity effect). 
Such a trend is also found in other works
\citep{vergani15,kruhler15}, but questions remain due to the
possibility of biases, the unknown nature of dark bursts, sample
sizes \citep[e.g.][]{hunt14}. Other studies argue that GRBs trace
star formation without any bias \citep[e.g.][]{michalowski12}. At higher
redshift ($3<z<5$), \citet{greiner15} argues for a good
correspondence between GRB rate and SFR, i.e. a constant $\epsilon$
\citep[see also][]{2016ApJ...817....8P}. On the other hand, \citet{schulze15}
still argue for a metallicity varying $\epsilon$ in the redshift
range ($0<z<4.5$).

\emph{SVOM} will allow us to improve over the legacy
of \emph{Swift} observations 
on several aspects:

\noindent $\bullet$ The error-bars at redshift larger than 5 will simply diminish owing
to the number of GRBs that will be detected during the nominal mission
(see header of this section).

\noindent $\bullet$ 
Based on the predictions for the \emph{SVOM}
mission, \cite{Wei2016} performed Monte Carlo simulations of a sample of
450 GRBs (5 yr observations). Using simulated GRBs with $z<4$ and
$L_{\rm iso}>1.8\times10^{51}$ erg $\rm s^{-1}$, they show it is possible to 
better constrain $\delta$ in equation \ref{equaefficiencyevolution}
and other parameters 
than with the current \emph{Swift} sample.

\noindent $\bullet$ 
While the study of host galaxies could bring important insight, they have yielded 
up to now to conflicting results (see above). This is likely due to the small statistics 
and biases affecting the current samples.
 
Owing to its quick follow-up in the optical and in the near infrared, \emph{SVOM} will allow
us to construct large, complete, and unbiased samples of long GRBs. The comparison of the 
their host galaxies with the usual population of galaxies will
allow us to constrain how $\epsilon$ varies with physical 
properties of the host galaxy, and more generally to 
learn about the long GRB progenitors (e.g. to confirm or not a metallicity effect). This
will help us to better understand the connection between star formation and long GRBs, so 
that the cosmic SFR can be studied from the long GRB rate on a sound basis.

\subsection{Studies of GRB host galaxies}
%
\begin{figure}[htpb]
\centerline{\includegraphics[width=1.\linewidth]{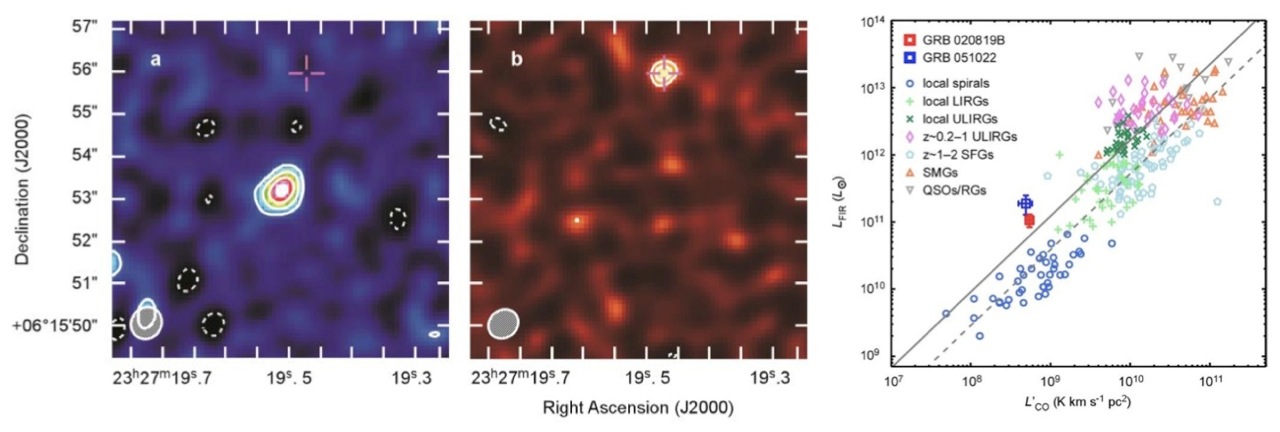}}
 \caption{\small The cold dust ({\sl left panel}) and CO molecular gas ({\sl central panel}) emission maps of the GRB\,020819B host galaxy at z=0.4. The star-forming 
region where the  GRB was discovered is indicated by the cross symbol and shows very low gas-to-dust ratio compared to the central region of the galaxy. 
{\sl Right panel}: CO versus total Infrared luminosities for 2 GRB hosts, compared to  a variety of galaxy samples at low and high redshift. 
The Schmidt-Kennicutt law for star-forming disks is represented by the dashed line, while the solid line depicts the relation followed by starburst 
sources. For their CO luminosities, the 2 GRB hosts discussed here display more intense star-forming activity than disk galaxies, pointing to more efficient 
processes at converting gas into stars. Adapted from \citet{Hatsukade14}.}
  \label{fig:COGRB}
\end{figure}
%

\subsubsection{GRBs as probes of  star-forming processes and physical conditions  in  very young stellar environments}

Studying  the host-galaxies of GRBs 
supplements in a very original way what is being learned from the deep cosmological surveys of galaxies selected by their continuum emission.
First, the selection of cosmic sites with LGRBs is done
independently of the properties of their underlying host galaxies, hence yielding an identification  of star-forming sources that is 
not affected by galaxy continuum detection limits. Besides, their redshift can be measured from absorption lines detected in the spectrum
of the afterglow, even when the host is too faint to be followed-up with 8m-class telescope spectroscopy. LGRBs thus allow probing the very 
faint end of the galaxy distribution with accurate spectroscopic redshifts, which can not be performed with other techniques. 
Finally, because of the very short lifetime of their progenitor, the selection of star-forming environments with LGRBs can be sensitive to 
much shorter time-scales (typically $\sim$10\,Myrs) than what is typically probed with continuum-selected galaxy surveys (e.g., $\gtrsim$100\,Myrs 
at Ultraviolet and far-Infrared wavelengths). In fact, high resolution imaging obtained with the {\it HST\,} has shown that 
the localization of LGRB afterglows within their host galaxies is more closely connected to the highest surface brightness regions than what can 
be observed from the spatial distribution of more ordinary core-collapsed Supernovae \citep{Fruchter06}. Although it
could be simply due to increased star formation density leading to a higher probability for producing  long GRBs \citep{Kelly14,Blanchard16}, 
it may also be explained by an aging effect if LGRBs have progenitors with shorter lifetimes than SNe and explode in young and luminous regions 
where the bulk of massive stars has not yet vanished \citep{Raskin08}. Irrespective of these possible interpretations, LGRBs offer an original 
view on high-redshift sources, enabling the census of star-forming environments with markedly different properties than those characterizing 
the bulk of galaxies selected with other techniques.

In this regard, the most exciting aspect making the use of GRBs unique is the possibility to combine constraints on the neutral gas
obtained from absorption spectroscopy on the line of sight of the afterglow (see next section) with constraints on the ionized gas obtained with 
subsequent spectroscopic follow-up of the host galaxy in emission.  This has proven to be technically difficult so far, not only because  LGRB 
host galaxies tend to be generally  faint but also because long-slit spectroscopy of distant sources do not  resolve emission line properties 
on the scale of individual star-forming regions, unlike the information encoded in absorption on the line of sight of the GRB. In the near future, 
\emph{SVOM} will yet benefit from a unique synergy with facilities like \textit{JWST} and possibly the Extremely Large Telescopes, which will  
allow us to constrain at sub-kpc scales the ionized gas properties (density, temperature, ...) of GRB-selected star-forming environments 
thanks to spatially-resolved IFU spectroscopy. Such analysis has already been carried for few  GRB host galaxies restricted to the low 
redshift Universe \citep[e.g., GRB\,980425:][]{Christensen08}. Beyond $\sim$2020, it will be possible to   extend these studies with larger 
GRB samples up to z\,$\sim$\,2, probing sources at the peak of galaxy formation history. Besides, 
with the future generation of large millimeter and radio interferometers available in the \emph{SVOM} era  (e.g., ALMA, SKA), it will also be possible to
measure the atomic and molecular gas masses as well as the dust luminosity in the close vicinity of GRBs.
Combining all these follow-up campaigns will provide a unique opportunity to characterize both the ionized and neutral gas phases, the 
local gas conditions, the metal enrichment and the nature of star formation processes within individual star-forming regions.

\begin{figure}[htpb]
\centerline{\includegraphics[width=0.7\linewidth]{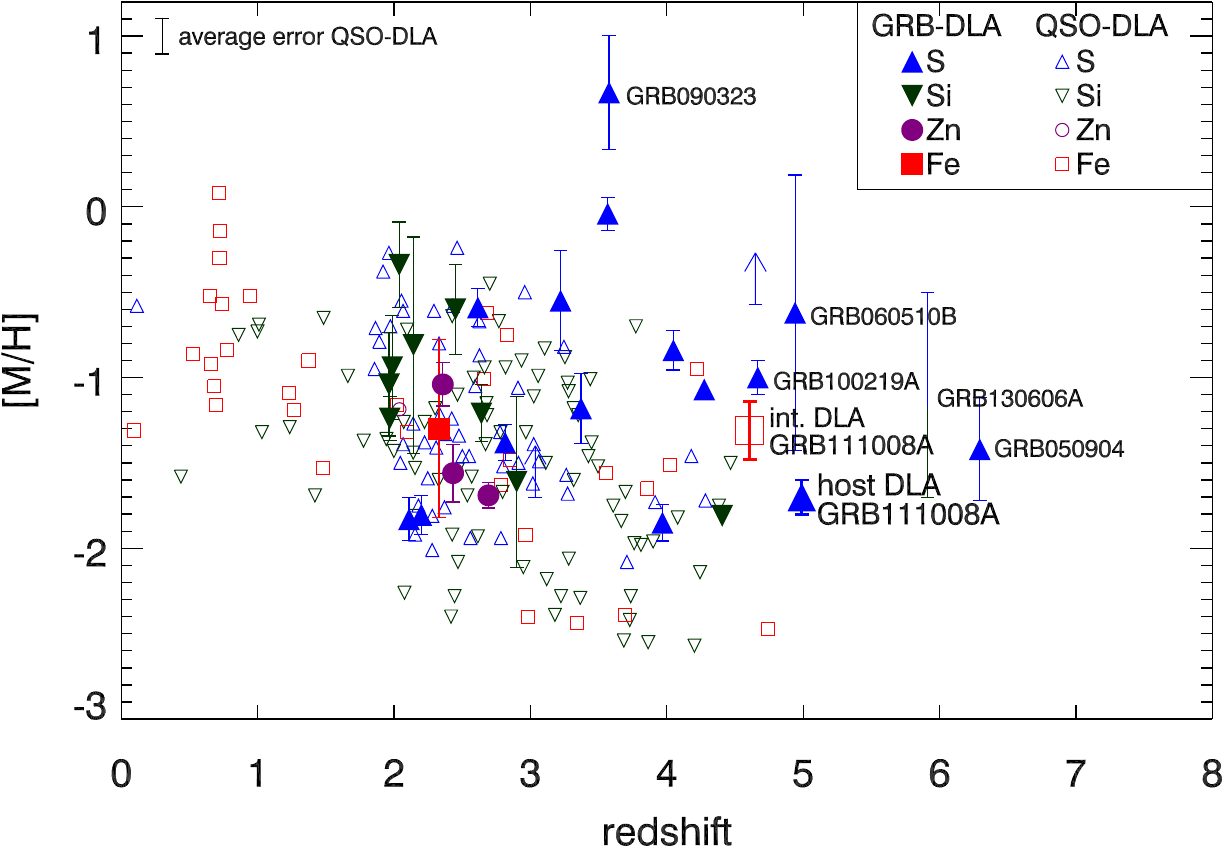}}
 \caption{\small  Metallicities measured in DLAs versus redshift.
Red squares are for intervening DLAs observed along the line of sight to quasars 
(see \citealt{Rafelski2012}) and blue squares are for GRB-DLAs (Figure from \citealt{Sparre2014}).
}
 \label{fig:metals}
\end{figure}

In the case of LGRB sites experiencing very young star formation, this will allow us to quantify the efficiency of  converting the gas into stars, 
especially in the early phases of star formation just when the gas is undergoing its first gravitational collapse. Hydrodynamical simulations of 
gas-rich disk galaxies at intermediate redshifts indeed suggest that this very early phase could be characterized by very high star formation 
efficiency during the first $\sim$\,15\,Myrs, before evolving to a more quiescent mode of star formation following the standard Schmidt 
law \citep[e.g.,][]{Bournaud14}. However,  observational constraints supporting this scenario are still missing, not only because of the difficulty 
of identifying such environments of early star formation in the distant Universe, but also for the limited performance of current facilities at 
long wavelengths. Similarly, models suggest that the relative velocity dispersions of the ionized and molecular gas phase components can put 
tight constraints on the strength of stellar feedback, which still remains today a key but unconstrained ingredient in our understanding of 
galaxy formation. Combining the properties derived in absorption using the \emph{SVOM} GRB afterglows with the host galaxy follow-ups carried with JWST, 
SKA and ALMA will thus produce spatially-resolved constraints on stellar feedback and star formation efficiency that will not be rivaled by any other methods.

Finally, recent observations of CO emission carried out at (sub)millimeter wavelengths  have suggested that galaxies hosting LGRBs, or at least the 
star-forming regions where they are produced, could be deficient in molecular gas \citep{Hatsukade14,Stanway15}. This was directly observed in 
the host of GRB\,020819B at z\,$\sim$\,0.4, where the GRB occurred in a star-forming region with low gas-to-dust mass ratio (see \reffig{fig:COGRB}). 
Similarly, the amount of molecular gas inferred from the very few H$_2$ absorption lines detected so far in the spectra of GRB afterglows appears to be 
weak with respect to the neutral gas content \citep{Ledoux2009}. This apparent lack of gas in the molecular phase within LGRB hosts is far from being understood 
at this stage. It could be due to an extremely high efficiency in turning  molecular gas into stars in the close environment ($\sim$\,100\,pc) of GRBs, 
but why these conditions are specifically found among GRB hosts is not clear. Interestingly, 21cm observations of nearby LGRB hosts have revealed high 
levels of neutral HI gas, typical of ordinary star-forming galaxies with comparable stellar mass and star formation rate \citep{Michalowski15}. Given recent 
evidence for metallicities below solar in GRB host environments,  this could indicate either a much shorter timescale for gas cooling and star formation than 
the timescale needed for the conversion from HI to H$_2$, or that star formation in these sources is directly fueled by large reservoirs of atomic gas 
from the intergalactic medium \citep{Krumholz12}.
This interpretation is however based on a limited number of GRB hosts at low redshifts, and it is also not clear if this small sub-sample is 
fully representative of the whole LGRB host population, especially for sources at higher redshifts. \emph{SVOM} will provide a more homogeneous 
sample of long GRBs with accurate localizations and spectroscopic identifications, which will be perfectly suited for follow-up with ALMA and SKA 
(at least up to $z$\,$\sim$\,1) to characterize their gas and dust content at the scale of individual star forming regions. In parallel, the more 
rapid and more systematic identification of GRB afterglows with \emph{SVOM} will also lead to more uniform constraints on the metal and gas content 
seen in absorption, which will result in a better understanding of the physical processes triggering star formation in the distant Universe.

\subsubsection{Absorptions in afterglow spectra}

Afterglows of GRBs can be very bright and therefore are ideal targets to study the ISM 
of high-redshift galaxies. 
Since long GRBs are produced by the deaths of massive stars, the spectrum of the afterglow
yields unique information on the host galaxy and possibly the star-forming region
where the afterglow explodes. The most prominent absorption line seen in most high-redshift GRB
afterglow spectra is a strong Lyman-$\alpha$ H~{\sc i} absorption associated with a wealth
of metal absorption lines from which it is possible to derive accurate metallicities (see \reffig{fig:metals}). 
H~{\sc i} column densities are much larger than in most Damped Lyman-$\alpha$ (DLAs) 
systems observed on the line of sight to quasars often reaching log~$N$(cm$^{-2}$)~$>$~22.
%
It can be seen that GRB DLAs probe higher redshifts. The fact that metallicities seem
to be higher toward GRBs indicates as well that the two populations do not probe the
same gas. 
Since GRBs are located close to the region of star formation, one can expect
to probe higher column densities and thus higher metallicities. Recent
searches for high-column density QSO-DLAs show that at higher column densities, metallicity
are on average higher \citep{Noterdaeme2015}. GRBs are thus ideal beacons with which it is possible to 
study the physical conditions, formation of molecules and properties of dust
in the inter-stellar medium for different environments (star formation activity, 
low-metallicities etc...). All this information gathered along the line of sight can 
be combined with the properties of the host-galaxy gathered once the afterglow has faded. 

In addition, the powerful UV flash following the GRB ionizes and excites the gas in
the circumburst environment and the interstellar medium surrounding the GRB up to
200~pc from the GRB \citep{Vreeswijk2013}.
Absorption from excited levels of species like Fe~{\sc ii}, Ni~{\sc ii}, Cr~{\sc ii} 
are detected in afterglow
spectra taken just after the burst. Since the GRB afterglow fades rapidly, recombination 
prevails and the populations of these levels change (most of the time decrease but sometimes 
can increase for a while). This induces 
these absorptions in the afterglow spectrum to vary and at the end to disappear.
Detection of these time-dependent processes, with timescales ranging from seconds 
to days in the observer frame can lead very interesting information on the 
burst itself and the ISM of the host \citep{Vreeswijk2007,Cia2012}. 

Finally, one could search for absorption signatures from the relics of gas expelled by 
the GRB progenitor. However, the task seems difficult and 
no clear and robust signature has been found yet probably because of the high ionization 
state of this gas \citep{fox2008}.

%

\subsection{Cosmology and fundamental physics}
\subsubsection{Could GRBs be used as standard rulers ?}
\label{sec:cosmo}
\begin{figure}[htpb]
\centerline{\includegraphics[width=1.0\linewidth]{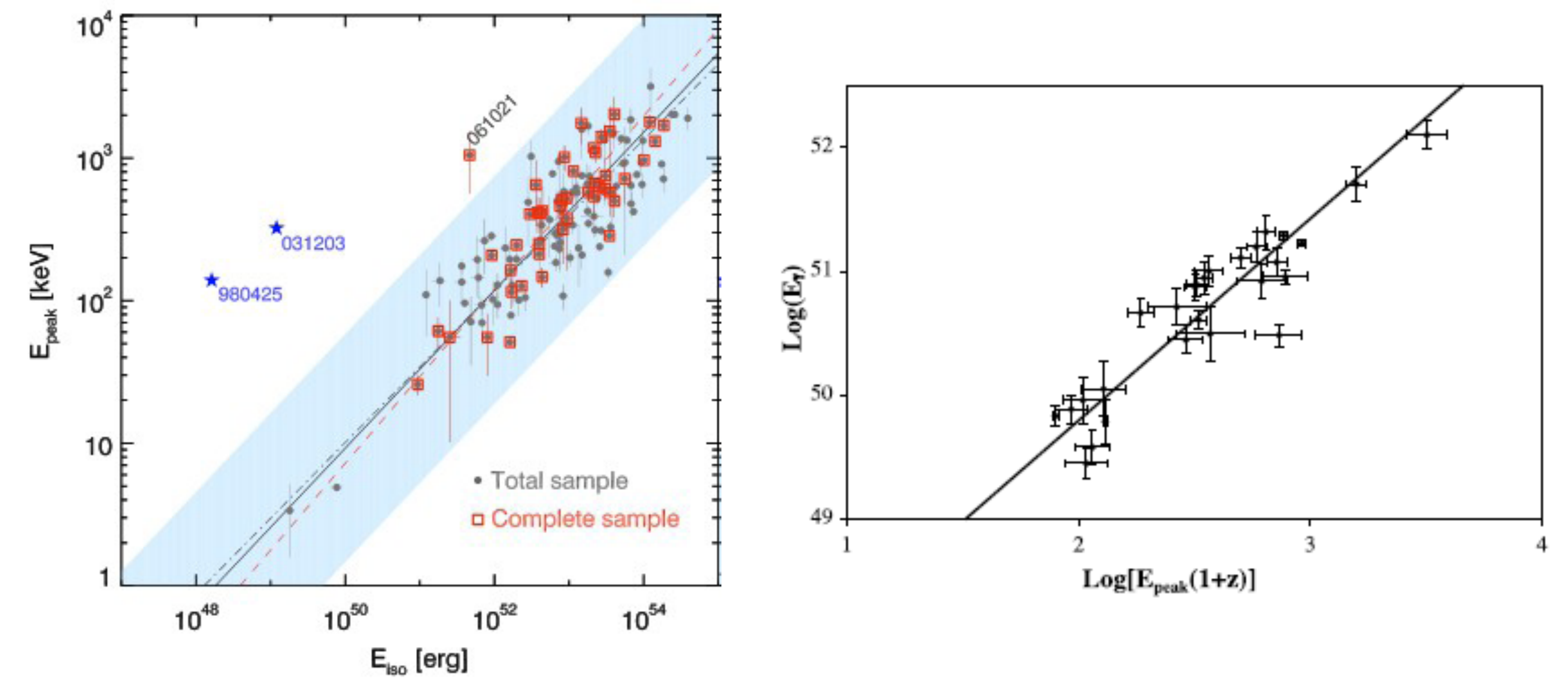}}
\caption{\small {\sl Left-hand panel}: The Amati relation
\citep{2012MNRAS.421.1256N}
showing the spread in the bulk correlation and obvious
outlier including short bursts. {\sl Right-hand panel}: The so-called
Ghirlanda relation for 27 GRBs \citep{2007ApJ...660...16S}.
$E_{\gamma}$ is the burst energy calculated using the jet-angle,
estimated from the so-called jet-break time in the optical afterglow, and
$E_{\rm iso}$.}
\label{fig:amati_ghirlanda}
\end{figure}

The isotropic energy outflow from GRBs,
estimated using the redshift and the integrated gamma-ray fluence,
is enormous, up to $E_{\rm iso}\sim10^{54}$ ergs, and even if the outflow is
collimated in jets the total energy involved is still huge,
$E_{\gamma}\sim10^{51}$ ergs. 
The possibility that  GRBs tap
a standard energy reservoir to provide this prodigious output
has been pursued by many authors following
the initial suggestion from \citet{2001ApJ...562L..55F}. If the total energy
available were roughly constant or predictable by some means
and we could reliably estimate the collimation,
then GRBs could be used as a cosmological probe to very high redshifts,
\citep{2003ApJ...594..674B,2004ApJ...616..331G}.
Early analysis of {\em BATSE} data revealed
a correlation between the characterisic photon energy  $E_{\rm p}$,
 the peak of observed $E.F(E)$ spectrum, and the fluence
\citep{1995ApJ...454..597M,2000ApJ...534..227L}. When redshifts
became available for long bursts
the isotropic energy, $E_{\rm iso}$, could be estimated from
the fluence and the peak photon energy could be transformed into the source
frame, $E_{\rm pz}$ and the so-called Amati relation, a correlation
between $E_{\rm iso}$ and $E_{\rm pz}$ in the sense that more energetic
bursts have a higher $E_{\rm pz}$, was discovered using data from
{\em BeppoSAX} \citep{2002A&A...390...81A}.
This correlation has subsequently been
confirmed and extended although there remain many significant
outliers, including all short bursts, \reffig{fig:amati_ghirlanda}.
The physical origin of the correlation may be associated
with the emission mechanisms operating in the fireball but the theoretical
details are far from settled
(see the discussion by \citet{2006MNRAS.372..233A} and references therein).
More recently a tighter correlation between $E_{\rm iso}$, $E_{\rm pz}$
and the jet break time, $t_{\rm break}$, measured in the optical afterglow
has been reported by \citet{2004ApJ...616..331G},
see \reffig{fig:amati_ghirlanda}.
This is explained in terms of a modification to the Amati relation in which
$E_{\rm iso}$ is corrected to a true collimated energy, $E_{\gamma}$,
using an estimate of the collimation angle derived from $t_{\rm break}$.
The details of the
collimation correction depend on the density and density profile
of the circumburst medium, \citet{2006A&A...450..471N} and references therein.
Multivariable regression analysis was performed by \citet{2005ApJ...633..611L}
to derive a model-independent relationship,
$E_{\rm iso}\propto E_{\rm pz}^{1.94}t_{\rm zbreak}^{-1.24}$, indicating that the
rest-frame break time of the optical afterglow, $t_{\rm zbreak}$
was indeed correlated with the prompt emission parameters.

Other studies have concentrated on the properties of the
isotropic peak (maximum) luminosity of the GRB, $L_{\rm iso}$
ergs s$^{-1}$, measured over some short time scale $\approx1$ s,
rather than the time integrated isotropic energy, $E_{\rm iso}$.
\citet{2003ApJ...583L..71S} noted a possible correlation between $L_{\rm iso}$ and
$E_{\rm pz}$ and later \citet{2004ApJ...609..935Y}
published such a correlation for 16 GRBs with firm redshifts. A correlation
between $L_{\rm iso}$ and the spectral lag was first identified by
\citet{2000ApJ...534..248N}
and explained in terms of the evolution of $E_{\rm peak}$ with time.
The shocked material responsible for
the gamma-ray emission is expected to cool at a rate proportional
to the gamma-ray luminosity and it has been suggested that
$E_{\rm peak}$ traces the cooling \citep{2004ApJ...602..306S}.
A similar correlation between $L_{\rm iso}$ and the variability of the GRB ($V$) was
described by \citet{2000ApJ...539..712R},
and \citet{2001ApJ...552...57R} and a related correlation between $E_{\rm pz}$ and
$V$ was described by \citet{2002ApJ...576..101L}.
The origin of the $L_{\rm iso}-V$
relation is likely to be related to the physics of the relativistic shocks
and the bulk Lorentz factor of the outflow. It could be that
high $\Gamma_{\rm outflow}$ results in high $L_{\rm iso}$ and $V$ while
lower luminosity and variability are expected if $\Gamma_{\rm outflow}$ is low
\citep{2002ApJ...578..812M}.
A rather bizzare correlation involving $L_{\rm iso}$, $E_{\rm pz}$ and duration
was found by \citet{2006MNRAS.370..185F}.
They employed the ``high signal'' time,
$T_{45}$, as formulated by \citet{2001ApJ...552...57R} in their study of
variability, and showed that $L_{\rm iso}\propto E_{\rm pz}^{1.62}T_{45}^{-0.49}$
for 19 GRBs with a spread much narrower than that of the Amati relation.
There is currently no explanation for such a correlation although, if it is not 
simply an artifact of the small sample size,
it may be connected with the spectral lag and variability
correlations and the Amati relation.

The correlation between $E_{\rm iso}$ and $E_{\rm pz}$ supplemented by additional
empirical information can be
used in pseudo redshift indicators, as for example \citet{2003A&A...407L...1A},
but the intrinsic spread in the correlation
and uncertainty about the underlying physical interpretation
introduce large errors, typically
of a factor $\sim2$. It may be possible to reduce the errors by
simultaneous application of several independent luminosity/energy
correlations, and an extension of the Hubble Diagram to high redshifts
using GRBs has been attempted \citep{2007ApJ...660...16S}.
However, it is not clear that the correlations briefly described above
are truly independent and there may be some underlying principle or
mechanism which connects them all together. Recently, and more controversially,
\citet{2007ApJ...671..656B} have raised serious doubts about the validity
of these correlations suggesting that it is likely that they
are introduced by observational/instrumental bias and have nothing to
do with the physical properties of the GRBs and hence they conclude
that GRBs are probably useless as cosmological probes. At very least their arguments suggest 
that selection effects must be carefully accounted for if these relations are applied in practice.

In order to estimate the cosmological parameters, these
correlations must be calibrated in a cosmological model independent way
to avoid the obvious circular argument problem. This could be done
in two ways; using a solid physical interpretation of these relations to
determine the slope independently from cosmology or
calibration of the relations using many low redshift GRBs
\citep{2006NJPh....8..123G}. Furthermore it is likely that the physical
mechanism which gives rise to the correlations yields parameters which are
dependent on the redshift. The luminosity, total energy,
characteristic photon energy and temporal lags etc. may well depend
on metallicity, the circumburst environment
and the structure of the progenitor stars all of which
evolve with time. Analysis of {\em BATSE} and more recently {\em Swift} data
indicates that the luminosity function of GRBs evolves with time
\citep{2012ApJ...749...68S, 2015ApJ...806...44P,2015MNRAS.447.1911P}.

Independent of whether we can use GRBs as a probe of high redshift 
cosmology or not, a better understanding of the luminosity evolution
of GRBs is crucial if we are to use them in the study of the cosmic star
formation rate, the period of reionisation and other properties of the
high-z Universe.  
What is required to make progress in our understanding of
the correlations and the physics which lies behind them is
simultaneous multiwavelength observations of the prompt and afterglow
phases of GRBs. The combination of the GRM, ECLAIRs, MXT, VT and GWAC on
{\em \textit{SVOM}}
can provide excellent spectral and temporal coverage significantly
superior to previously available instrumentation. They will
provide a large sample of GRBs for which we know the redshift,
the prompt spectrum from gamma rays to soft X-rays
and in some cases extending to the optical, the spectral-temporal
structure of the prompt and soft X-ray and optical coverage
extending from the end of the prompt into the later stages of the
X-ray and optical afterglow. These data will greatly improve our understanding
of the presently known correlations and the physical mechanisms behind them.

\subsubsection{Variation of fundamental constants}

Metal  lines  of  absorption  systems  due  to  intervening  galaxies
along the line of sight towards distant sources provide insight into the atomic structure 
at the cosmic time and location of the intervening  object.  All  atomic  transitions  
depend  on  the  fine-structure constant, offering a way to probe possible variations
in its value in space and time (\citealt{Rahmani2013} and references therein). These absorption lines are readily observed
in the spectra of GRB afterglows and could be used to probe these possible variations.
In case of the detection of H$_2$, the variations of the electron to proton mass ratio
can be probed as well. Both measurements could be performed at the same time.

GRB afterglows, if well selected, can be unique targets in this field after the advent
of extremely large telescopes (TMT and ELT) since
GRB afterglows can be very bright even at high redshift and yield spectra of the highest SNR.
The key here is that the probability that the line-of-sight intersects a diffuse molecular 
cloud is much higher in the case of GRBs compared to QSOs. Even though detection of H$_2$ 
in GRB afterglows has been rare up to now, it may be due to a selection effect because only bright afterglows
can be followed up at high spectral resolution when the presence of H$_2$ implies the presence of dust
and thus attenuation of the afterglow \citep{Petitjean2006}. The situation may change with the advent of larger telescopes,
and with the VT on \textit{SVOM} giving quicker identification of potentially obscured bursts.

Note that these studies will be possible only if a fast response mode on a high resolution spectrograph 
is available (e.g. \citealt{Vreeswijk2007}) but also if wavelength calibration is controlled
to a very high level of precision. 




\newpage

\section{\textit{SVOM} Advances on Rapid Follow-Up Observations (\textit{SVOM} ToO program)}

\subsection{Introduction}
\textit{SVOM} will become a premier time-domain machine in the early part of
the next decade, an era when time domain astronomy will truly come of
age in terms of multi-wavelength, wide-field sky coverage
plus multi-messenger information. The
advent of SKA (radio), LSST (optical) and CTA and HAWC (very high energy) on
the ground, for example, will provide a very large increase in the
number of rapidly available triggers for a wide variety of source
types. Other electromagnetic facilities, such as the \textit{SVOM} GWAC, will also find many
transients. 

The launch of \textit{SVOM} will also coincide with an era where there will be a significant improvement in the capability of multi-messenger observatories.
The gravitational wave observatories will have improved sensitivity and provide localisations of size 5-10 square degrees \citep{Aasi:2013wya} compared to the current several hundred square degrees. \textit{SVOM} can tile such regions quickly and efficiently.
The first phase of the new KM3NeT neutrino facility will complement IceCube and also provide much improved localisations which the \textit{SVOM} narrow-field instruments can observe in a single pointing (for track events).
 
The much larger volume of triggers will provide a challenging scientific
opportunity for \textit{SVOM} which will have the on-board capability to obtain
multi-wavelength follow-up observations.


\begin{figure}[!h]

\centerline{\includegraphics[trim=52 250 52 250,scale=1.,clip=true]{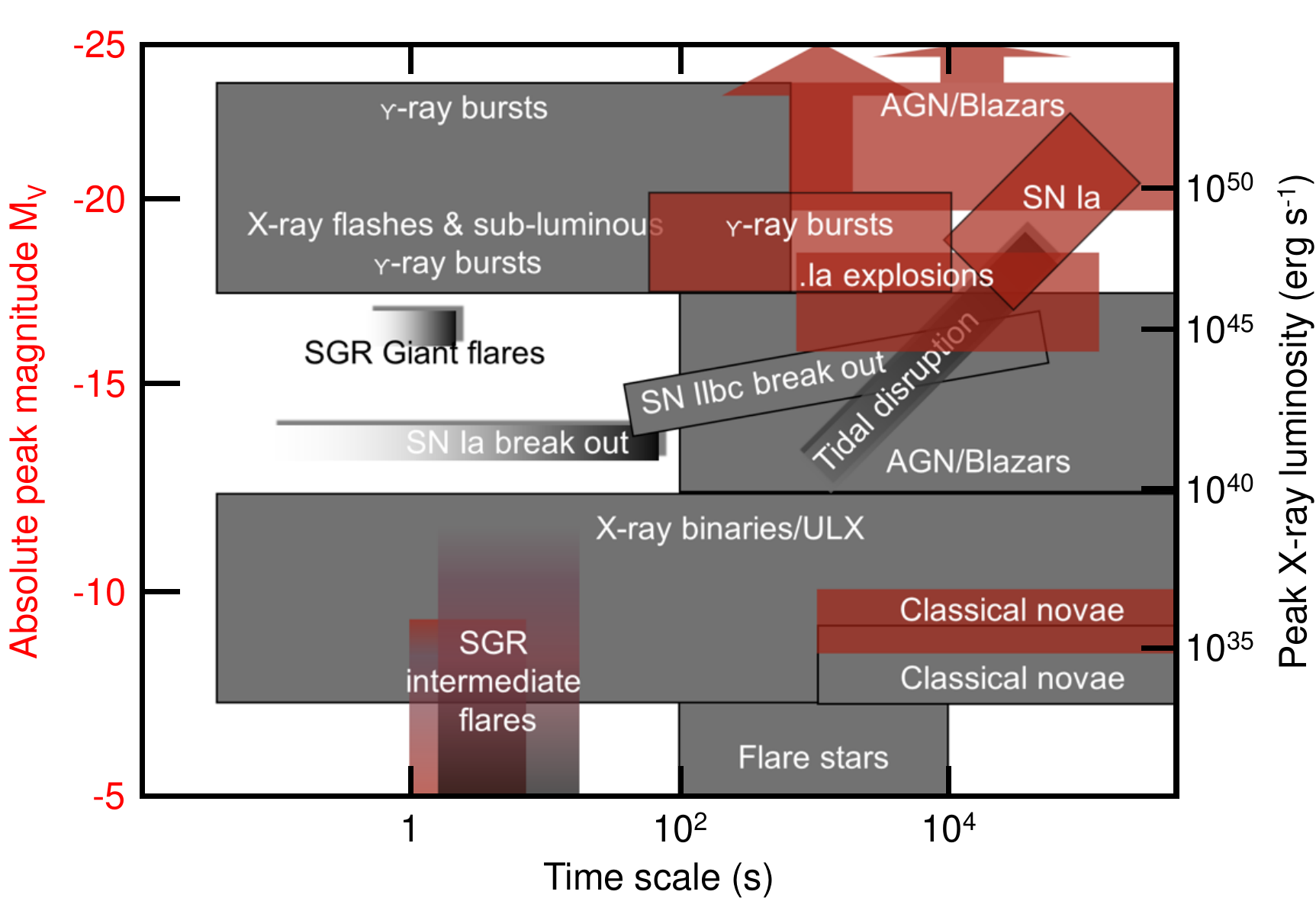}}
 \caption{\small The characteristic absolute optical magnitude (red) and X-ray luminosity (grey) versus characteristic timescale for various classes of object which could be studied by \textit{SVOM}. The optical/X-ray scales are on the left/right axes. Sources/phenomena that have been relatively poorly studied are plotted without border and with fading shading. Some object classes have a large range in characteristic timescales due to different physical processes. Adapted from \citet{Jonker:2013yyp}.}
 \label{fig:timescale}
\end{figure}

\subsection{Search for \textit{SVOM} counterparts to multi-wavelength triggers}
\label{sec:grbmultiwavelength}\medbreak

\textit{SVOM} has multiple roles to play in the new era. It can: (1) follow-up triggers
from other facilities, including multi-messenger facilities, and any
candidate counterparts found by other electromagnetic facilities; (2)
trigger multi-wavelength follow-up of \textit{SVOM} triggers, including faint
sources found in ground analysis which did not result in an on-board
trigger and sources found with the ground-based \textit{SVOM} facilities; (3) monitor sources likely to undergo a transient; and
(4) survey classes of transients to provide population
information. These types of observation will in some cases require
MoUs for mutual benefit but will also require detailed planning to
avoid overwhelming the observational limits of \textit{SVOM} given the sheer
number of triggers that will emerge from the wide-field transient
machines. The latter issue is perhaps the greatest challenge facing
astronomy -- how to decide what to observe?


\medbreak

In terms of likely (known) classes of source there are several of
particular importance well suited to the sensitivity of \textit{SVOM}. Some of the explosive transient event types available for \textit{SVOM} are shown in \reffig{fig:timescale}. Example classes of object include:


\begin{itemize}

\item Gamma-Ray Bursts: \textit{SVOM} can localise and provide prompt follow-up for
GRBs not automatically triggered on board, including those from other
facilities and those found in \textit{SVOM} ground analysis which may be of
particular interest (low luminosity, high redshift etc.).




\item Supernova: LSST and other optical facilities will find very large
numbers of supernova, including rare classes demanding prompt
follow-up. \textit{SVOM} can observe in the optical and X-ray to look for
information on the environment and progenitor, such as observing
luminous supernova that may be powered by a central engine possibly in
analogy with GRBs.

\item Tidal Disruption Events: \textit{SVOM} can probe the physical properties of
TDEs which have emission which is either relativistic jet or accretion
flow dominated. \textit{Swift} has found a few jet-dominated cases raising the
probability of discovery across the electromagnetic spectrum, but the
rate of discovery should rapidly increase with the new survey
facilities.

\item Active Galactic Nuclei: \textit{SVOM} can target AGN undergoing a flaring
event, including Seyfert galaxies and quasars where the
combination of optical and X-ray data can constrain synchrotron versus
synchrotron-self-Compton emission. 

\item Blazars: \textit{SVOM} can provide crucial information on the spectral energy distribution of blazars. Future facilities will increase the
number of high-redshift blazar candidates and provide many potential
events with large variability amplitudes possibly due to state
changes.

\item Galactic transients: Many types of transients in the Milky Way are
potential \textit{SVOM} targets, such as flaring magnetars (rare but provide
crucial insight into emission mechanisms), X-ray binaries and ULX sources (constrain
jet and disk emission) and accreting white dwarfs (probe novae). \textit{SVOM}
itself will find some of these in outburst but can also follow-up
triggers from other facilities.

\item Unknown: The most exciting discoveries will come from transients which
are poorly understood, such as fast radio bursts, and those yet to be
found.  For example, CTA will probe the very high energy sky in much
greater depth and over larger areas than currently achievable and
likewise SKA in the radio and LSST in the optical.

\end{itemize}
\subsection{Search for \textit{SVOM} counterparts to multi-messenger triggers}
\label{sec:grbmultimessengers}

Multi-messengers astronomy has been discussed for a long time for its ability to shed light on the physical process giving birth to GRBs in the case of gravitational waves (section \ref{sec:grbshort}) or on the acceleration mechanisms in the jets for neutrinos and gamma-rays (section \ref{sec:grbacceleration}). In the very first moments following the explosion, the photons do not escape the dense medium, we will have to rely on some new messengers to get some information. The detectors for those messengers are becoming mature. Gravitational waves have finally been discovered and astrophysical neutrinos have been observed. The multi-messenger astronomy era has really begun. 

\textit{SVOM} with its ground and space instruments will offer a large and complementary follow-up capability through ToOs (Target of Opportunity). GWAC with its 5000 sq. deg. coverage can start the observation since the alert reception. The GFTs with their small field of view will confirm GWAC candidates and will be able to do follow-up for well localized events.
To activate the satellite instruments, we will rely on a specific ToO program to send the observation program using S-band stations. This program guarantees less than 12 hours between the alert and the start of space observations (less can be expected for most cases) and can be activated around 20 times per year.
From space, MXT and its 1~sq.~deg. field of view will have the possibility to cover larger sky portion using a specific tiling procedure.

\subsubsection{Gravitational waves} 

On September 14, 2015 at 09:50:45 UTC the two detectors of the Laser Interferometer Gravitational-Wave Observatory LIGO simultaneously observed a transient gravitational-wave signal \citep{2016PhRvL.116f1102A}. The signal matches the waveform predicted by general relativity for the merger of a black hole binary. This first detection, followed by a second one in december 2015, opens new prospects and the beginning of an exciting new era of astronomy.

While an electromagnetic emission is not expected from binary black hole mergers, it is not completely excluded yet. However, one of the most promising sources for joint electromagnetic-gravitational wave observations are coalescing binaries including at least a neutron star. Those sources are considered to be good progenitor candidates for GRB of short duration (section \ref{sec:grbshort}).

In 2020 there will be five active gravitational-wave detectors on Earth including Virgo in Italy. This world-wide network is expected to observe few to hundreds gravitational-wave events per year associated to neutron-star binary mergers during the future science runs.

The long baseline of the intercontinental network will yield a much more accurate localisation of the source (below tens of square degrees) than the one obtained by LIGO alone \citep{Aasi:2013wya}. As shown in \reffig{fig:Multi-Messengers}, the GWAC will be able to cover rapidly the sky error region from the ground. From space, MXT will be able to cover a significant amount of the error box with the help of multiple tilings.

\subsubsection{Neutrinos} 

IceCube has demonstrated the existence of neutrinos of astrophysical origin using the outer layer of the IceCube detector as a veto and searching for events starting inside the inner volume (HESE sample) (\citet{IceCube2013}, \citet{IceCube2014}). Two event topologies are detected with such detectors:  track and cascade event resulting for the muon neutrino charged current interaction and the electron/tau neutrino interaction and muon neutrino neutral current interaction, respectively. Due to the different topologies of the events, the angular resolution is roughly 10-15 deg. for cascades and 0.5 deg.  for muons. Since 2008, a follow-up of multiplet events, two times per month, is working with optical and X-ray telescope \citep{IceCube2013}. The HESE events are now also sent to follow-up facilities through the GCN network (\citet{IceCube2007}, \citet{IceCube2015}). The typical rate of the HESE events is around 5 track events  and 10 cascade events per year.  The IceCube Collaboration is planning an expansion of the current detector, IceCube-Gen2, including the aim of instrumenting volume of clear glacial ice of the order of 10 km$^{3}$ at the South Pole~\citep{ICGen22014}

\begin{figure}[htpb] 

\centerline{\includegraphics[trim=0 250 0 250, scale=1., clip=true]{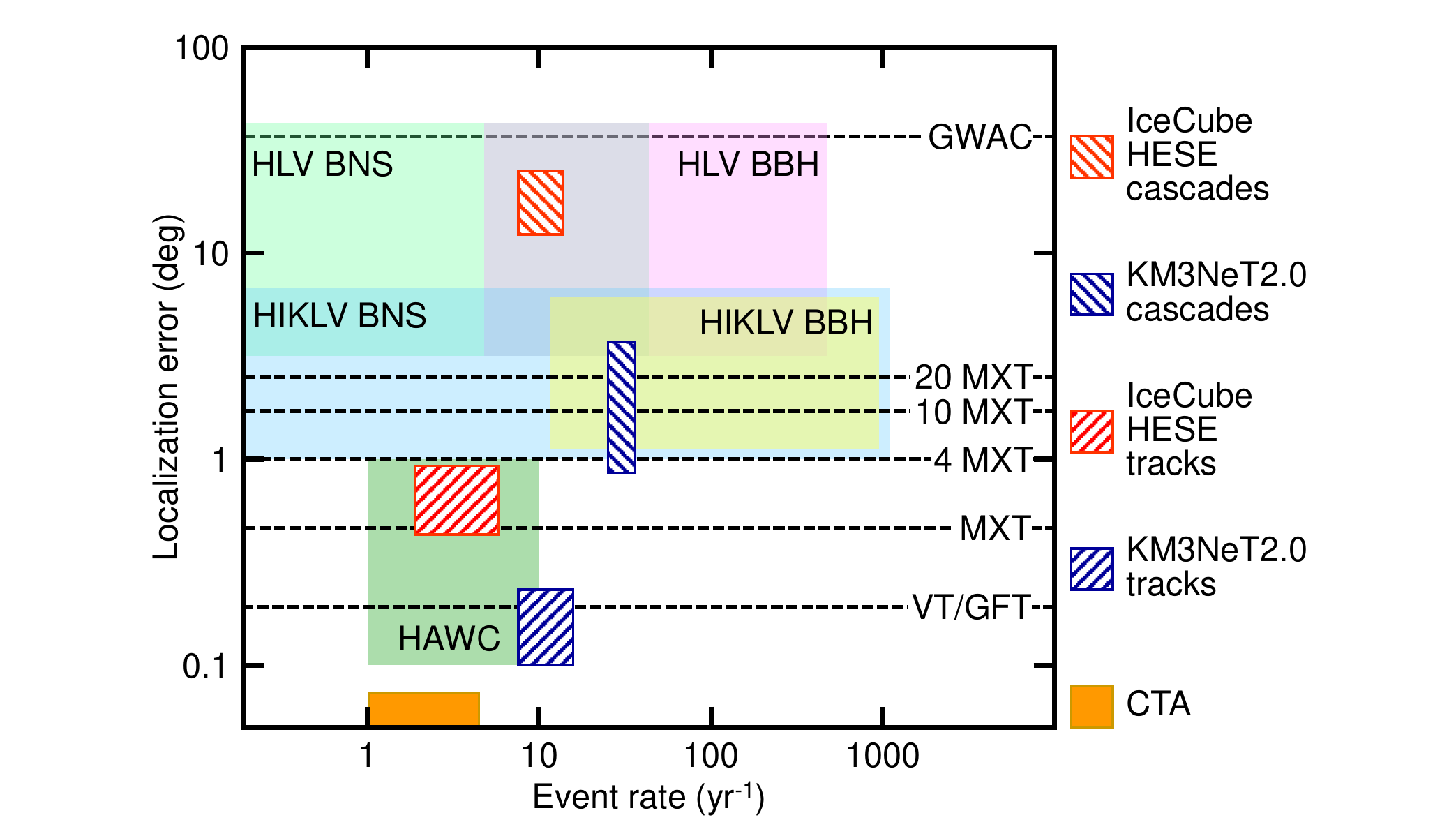}}
 \caption{\small Anticipated event rate and localization error for some forthcoming detectors operating in the field of multi-messenger astronomy. Predictions are shown for two configurations of gravitational-wave detectors: Hanford-Livingston-Virgo (HLV) and Hanford-India-Kagra-Livingston-Virgo (HIKLV) and two sources: binary neutron stars (BNS) and binary black-holes (BBH). Predictions are also shown for two phase 2 high-energy neutrinos detectors: IceCube for high-energy starting events (HESE) detected as cascades or tracks and KM3NeT, again for events detected as cascades or tracks. Predictions are finally shown for two high-energy gamma-ray detectors, HAWC and CTA. Horizontal dotted lines indicate the \textit{SVOM} follow-up instrument capabilities with 4 MXT, 10 MXT and 20 MXT corresponding to different MXT tilings with 4, 10 and 20 tiles.}
 \label{fig:Multi-Messengers}
\end{figure}

 The KM3NeT collaboration is building the second generation neutrino telescope in the Mediterranean Sea~\citep{Km3net2016}. Above 10 TeV, muon tracks have a typical angular resolution lower than 0.2 deg. while showers, the most promising events, have a 2 deg. localisation error. The expected event rates are indicated in Fig. \ref{fig:Multi-Messengers}. 
The existing multi-wavelength follow-up program of ANTARES (\citet{Antares2012}, \citet{Antares2016}) will be extended to KM3NeT.

Ultra-high energy earth-skimming neutrinos can be detected by the large cosmic-ray arrays such as the Pierre Auger Observatory in Argentina~\citep{Auger2015} and the large radio neutrino telescope GRAND in China~\citep{Grand2016}. Finding the source of these high energy neutrinos would have a huge impact on the astrophysical community since they are both related with the acceleration processes in the jets (section 4.5).  As seen in Fig. \ref{fig:Multi-Messengers}, the performances of \textit{SVOM} are perfectly tailored to follow neutrino alerts with the MXT and VT instruments on-board and the ground-based telescopes (GFTs and GWAC).


\subsubsection{High Energy Photons} 

The scientific return of the \textit{SVOM} mission will be further enhanced by its synergy with other space missions or ground-based telescopes providing external electromagnetic triggers.
Gamma-ray space detectors operating in the MeV range with poor localisation capabilities, e.g. similarly to the \textit{Fermi}/GBM \citep{Connaughton:2014xha}, might be operating at the time of \textit{SVOM}, providing numerous alerts with relatively small error boxes that can be scanned by the MXT or GFT.
At very high energies ($>$\,10$-$100 GeV), the main data acquisition system of the HAWC experiment will distribute GRB trigger times and positions to the world-wide GRB community, with an accuracy of ~1 deg or less \citep{Taboada:2013uza}.
In addition, a fraction of CTA observations might be performed in a wide-field mode, where several telescopes are offset to maximize the solid angle, with a corresponding rate of 2-3 GRBs observable per year \citep{Inoue:2013vy}. 

In both cases, the position accuracy will be small enough to initiate follow-up observations with the MXT in one single tile. The GFTs small field of view will be wide enough in some cases to cover the localisation error box in a single pointing.

\newpage

\section{\textit{SVOM} Advances on Observatory Science (\textit{SVOM} general program)}

The wide field surveys with ECLAIRs and the GRM  will permit to detect  transient activity from many different types of sources 
other than GRBs, while the narrow FoV instruments (MXT, VT, GFTs) will allow to extend the scientific analysis of dedicated sources 
to a wider spectral window. This ``general program" can cover and bring relevant information on a wide range of sources and scientific topics. 
With a minimum observing time of one orbit (effective time of $\sim 3000$~s with the MXT), the 0.3--6 keV $5\sigma$ detection 
limit of the MXT is $2$ $10^{-12}$~\ergcms ($\sim0.1$ mCrab). Extrapolation of the XMM-Newton high precision Log N--Log S indicates a 
source density of $\sim0.2$/deg$^2$ at high galactic latitudes \citep{mateos08} which shows the high potential of MXT for dedicated studies of specific sources.
For the wide field surveys, we estimated a total exposure time as low as 
$\sim200$~ks over one year on the galactic centre region. This leads to a 4--150 keV sensitivity limit of $\sim 7$~mCrab, roughly 2 $10^{-10}$\ergcms. 
 Such a limiting sensitivity permits to survey a large number of sources : the 17--60 keV \textit{INTEGRAL} Log N--Log S indicates a density of 0.02 
 (resp. 2 $10^{-4}$) source/deg$^2$ at $|b|$$<$5$^\circ$ (resp. $|b|>$5$^\circ$)] \citep{krivonos07}.  
 
\subsection{Active Galactic Nuclei}
\label{sec:agn}
\begin{figure}[hptb]
   \begin{center}
	\includegraphics[width=10cm,angle=0]{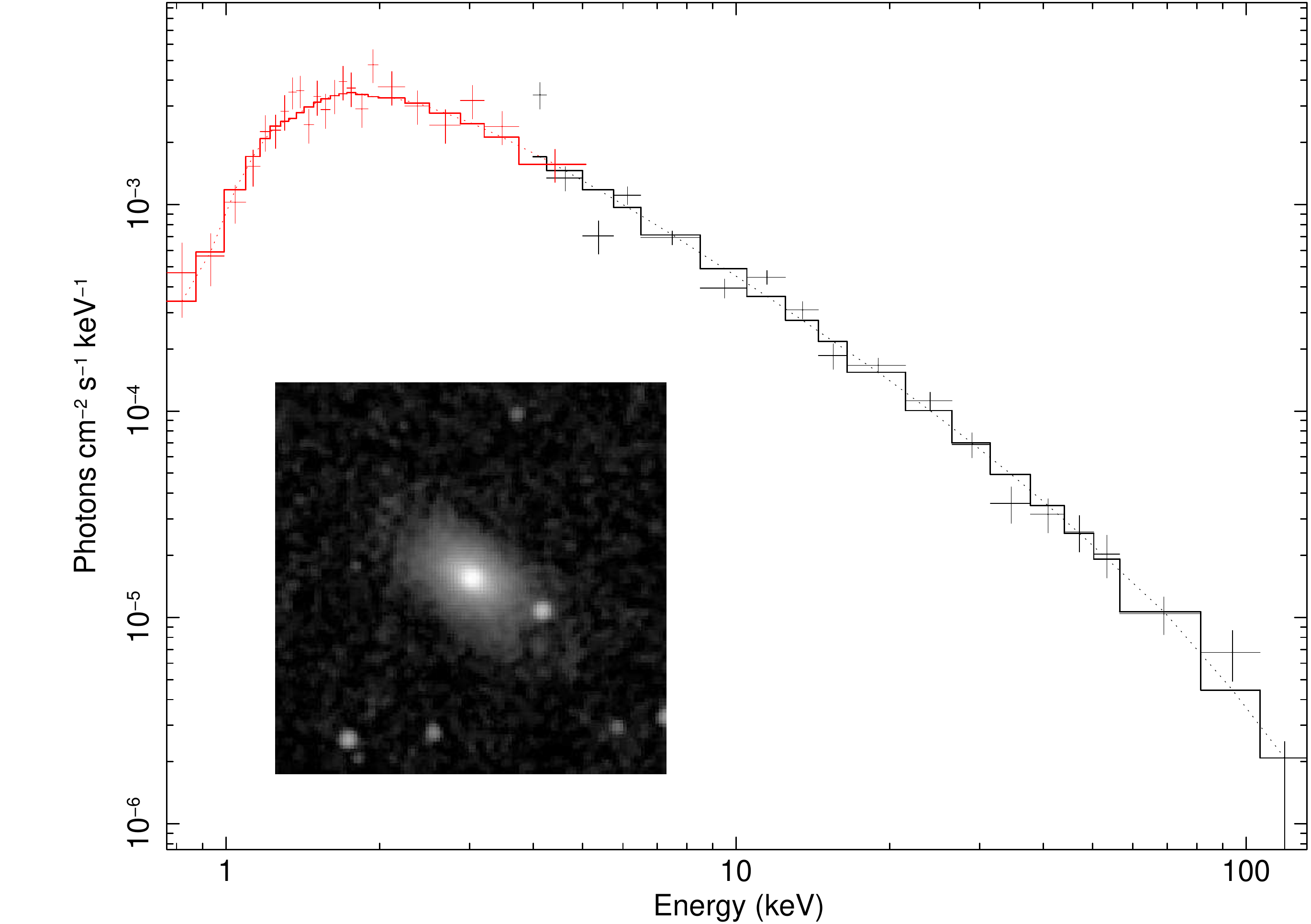}
	\caption{Simulated 10 ksec \textit{SVOM}/MXT observation of the absorbed Seyfert~1.9 galaxy MCG-5-23-16 ($f_{4-150 \rm keV} \simeq 8 \rm \, mCrab$), combined with the expected 1-year spectrum (effective exposure time of 300 ks) from ECLAIRs. The cut-off in the spectrum at $E_{\rm cut} = 72 \rm \, keV$ and the absorption in the line of sight can be determined with an accuracy of 10\%.  The optical image gives an impression of a \textit{SVOM}/VT observation of this source (based on a 2MASS image; \citealt{Skrutskie2006}).}
 \label{fig:AGN_MCG}
	\end{center}
\end{figure}

Active Galactic Nuclei (AGNs) are the most luminous persistent objects in the Universe. 
An important part of the observatory science of \textit{SVOM} will certainly include studies of persistent and transient AGNs at high energies with ECLAIRs, MXT, and, in 
the optical domain, with the VT, the GWACs, and the GFTs.
Photons emitted by the disk around the central super massive black hole are inverse Compton up-scattered on relativistic electron plasmas in the vicinity, e.g. in a corona partially covering the disk. Some AGNs develop powerful jets rising perpendicular to the accretion disk accelerating particles up to relativistic speeds. Thus they are divided into different classes, based on the angle of the line of sight with respect to the accretion disk and intrinsic absorption, and whether or not they exhibit a jet \citep{Beckmann2012}. 
\medbreak
ECLAIRs is going to open a new window on AGN large area surveys, as it will cover the 4--150~keV energy range with unprecedented sensitivity. Although there will be more sensitive all-sky 
surveys below 10~keV (e.g. eROSITA) and above 20~keV (\textit{Swift}/BAT, \textit{INTEGRAL}/IBIS), the ECLAIRs sensitivity between 10 and 20~keV (a gap not covered by any current and future 
missions)   will be crucial to study 
the inverse Compton scattering spectra of the accretion disk in AGNs. 
The capabilities of the narrow field instruments will be used to study AGN in outburst, in multi-wavelength (hereafter MWL) campaigns, or to clarify the nature of an object. 
\medbreak
Within a typical year of operations, \textit{SVOM}/ECLAIRs is going to detect 250~AGN. For the majority of these mainly Seyfert type galaxies, the GWACs are going to provide detailed lightcurves. After a 5~year mission life time, ECLAIRs will provide crucial information on about 700~AGN. 
For the brightest 30 of the persistent AGN, which are mainly absorbed Seyfert type galaxies, ECLAIRs data will allow to derive precise measurements of the continuum slope, determining the Compton reflection contribution to the flux. One example is the Seyfert~1.9 galaxy MCG-05-23-16 as shown in Fig. \ref{fig:AGN_MCG}. 
The whole survey sample will give us the average Compton reflection fraction and thus will be an important input to determine the sources of the cosmic X-ray background in this energy range \citep{Ueda2014}. It has to be noted that missions like NuSTAR are determining the contribution of AGN to the CXB in a similar energy range (6-70~keV), but only in pencil-beam surveys and thus at higher redshifts. The study of the CXB in the local Universe can only be done using wide field surveys like the one performed by ECLAIRs.
\medbreak
\textit{SVOM}'s all-sky survey will also allow to pick up transient AGNs, such as blazars, Seyfert galaxies in high state, and tidal disruption events (TDEs, section \ref{sec:extragalactic}). 
The on-board processing will allow to predefine a list of interesting AGN with individual trigger thresholds for automatic repointing of the satellite.
The narrow field X-ray instrument MXT on-board \textit{SVOM} will allow to follow-up on AGN outbursts and also to provide a means of identification of X-ray counterparts to optical AGN candidates as they will be seen in the VT. This will be vital in view of the deep all-sky survey telescopes operating during the \textit{SVOM} mission lifetime, such as the LSST and Euclid. Their multi-band photometric data will allow to pick out candidates for so far unknown transient X-ray AGNs. These can be either blazars in outburst or Seyfert galaxies, which show type changes \citep{Risaliti2011}, spectral variability \citep{Hernandez2015}, and/or significant brightening \citep{Soldi2014}, as we have seen for several Seyfert~2 galaxies, and also from radio galaxies like Pictor~A and Centaurus~A \citep{Steinle2010}.
The GFT network will be able to provide simultaneous optical spectra with information about the broad and narrow line regions and the underlying continuum from the accretion disk of the AGNs.  
\medbreak
Since the emission from blazars is often found to vary simultaneously across the spectrum, wide-field instruments like those provided by \textit{SVOM} are particularly useful in their study because they can be used to trigger MWL campaigns when a blazar enters a flaring state (e.g. for 3C~454.3; \citealt{Wehrle2012}). This is of particular interest to observers at very-high energies. Blazars are known to be extremely variable in the gamma-ray regime and most very high energy (VHE) observatories reserve significant fractions of their telescope time to follow-up blazar flares. In this way, MWL campaigns can be triggered when a known blazar enters into an active state. In addition to triggering such MWL campaigns, the \textit{SVOM} instruments and the GFT will be used to provide accurate measurements of the spectral energy distribution of high-state blazars.
Recent studies have shown that blazars at high redshifts ($z>4$) can be extremely bright, up to $L_{\rm X} > 10^{47} \rm \, erg \, s^{-1}$ in ECLAIRs' hard X-ray band \citep{Ajello2009}. This makes them ideal targets to study the early accretion (and emission) history of the Universe, and opens a window for \textit{SVOM} to study the early formation of relativistic jets \cite{Ghisellini2015}. 

%
\subsection{Ultra Luminous X-ray Sources and Tidal Disruption Events}
\label{sec:extragalactic}

Ultra-luminous X-ray sources (ULXs) are X-ray objects located outside the galaxy nucleus, with luminosities that exceed the Eddington luminosity (L$_{Edd}$) for a stellar mass black hole, where L$_{Edd}$ = 1.3 $\times$ 10$^{38}$ (M/M$_\odot$) erg s$^{-1}$. First discovered with the {\em Einstein} observatory \citep[1978-1981,][]{fabbiano:89} almost 40 years ago, many questions about these sources remain to be answered. Firstly, what is the nature and the mass of the compact object? Until recently the compact objects were assumed to be black holes, but \cite{bachetti:14} demonstrated that the ULX M82 X-2 (L$_{max} \sim$1.8 $\times$ 10$^{40}$ erg s$^{-1}$, 0.3-10.0 keV), hosts a neutron star accretor, with L$_x \sim$ 100L$_{Edd}$. This not only implies that the compact objects could be neutron stars, but also demonstrates that long term super-Eddington accretion appears possible. How a source can continuously accrete above the Eddington limit has yet to be discovered. Elucidating the accretion mechanism in ULXs requires observations at different luminosities/spectral regimes. Many ULXs are either transient \citep[e.g.][]{middleton:13} or variable, like M 82 X-2.

\medskip

A rather large number of known ULXs can be detected with the MXT (F$_{\mathrm{[0.3-10 keV]}} > 10^{-12}\ \mathrm{erg\ s^{-1} cm^{-2}}$) during a one orbit observation time. Several suitable targets are located in the following galaxies with good visibility throughout the year ($>$~100~days): NGC~55, NGC~253, NGC~5204, NGC~5408, NGC~5907, NGC~6946, NGC 7793. An X-ray monitoring of these sources will allow us to trigger observations with other X-ray and multi-wavelength observatories when they are observed to show spectral transitions at different flux levels.

For those ULXs with bright and isolated optical counterparts ($M_V < 23$), the VT will allow us to search for periodicities in the lightcurve of these objects. This in turn could reveal the orbital period of these systems, allowing us to gain a handle on the mass of the two stars (thanks to Kepler's laws) and provides further constraints on the accretion regime. An excellent example of such a study was done for the X-ray source P 13 in NGC 7793 \citep{motch:14}.

\medskip

Other events that \textit{SVOM} will lend itself well to discovering are tidal disruption events (TDEs). These occur when a star in a galaxy wanders too close to the central massive black hole. The star disrupts when the tidal forces exceed the self-gravity of the star and a previously undetected massive black hole will become extremely bright, allowing it to be studied.
The detection of three candidate relativistic tidal disruption flares (rTDFs) within 12 years of \textit{Swift} operations (e.g. \citealt{levan:11}) at initial X and hard X-ray fluxes (F$_{\mathrm{[0.3-10 keV]}} \sim $ F$_{\mathrm{[15-50 keV]}} \sim 10^{-9}\ \mathrm{erg\ s^{-1} cm^{-2}}$) suggest that up to one such event could be detectable with ECLAIRs within \textit{SVOM} initial lifetime.
Another possible prospect to find non-relativistic TDEs would be to survey galaxies with large globular cluster populations, those being in addition prime targets to hide the extremely rare intermediate mass black holes (IMBH, $\sim$10$^{2-5}$ M$_\odot$). This would be done with dedicated MXT observations. Without violating the B1 law, the Virgo cluster is an obvious candidate due to its close proximity and number density of galaxies. Regular monitoring of the cluster (i.e every one/two weeks for 10 ksec in a 3x3 tiling mode) should lead to the detection of 6 TDEs within \textit{SVOM} initial lifetime, with fluxes down to $10^{41-43}\ \mathrm{erg\ s^{-1}}$ that we can expect for a $10^5\ M_{\odot}$ BH.


\subsection{Galactic sources}
\label{sec:galactic}

The lower energy threshold and broader spectral coverage of \textit{SVOM} (compared to \textit{RXTE}, \textit{Swift} or \textit{INTEGRAL}) will permit to detect high energy 
galactic sources and unexpected events (outbursts, flares from binaries, thermonuclear explosion, ....) easily, thanks to the ECLAIRs sensitivity of 
50 mCrab in one orbit. \\

\medbreak
\subsubsection{Accreting systems}
\smallbreak
Cataclysmic variables (CVs) and X-ray binaries (XRBs) are systems that are powered by the accretion of matter onto the compact object (CO). The former systems 
contain a white dwarf (WD) and the latter either a neutron star (NS) or a black hole (BH). The compactness of the CO will affect the peak of the thermal emission of the disc, but 
also how accretion proceeds in the innermost regions when a significant magnetic field and/or a rigid surface enter into the game. By monitoring 
these sources from the early stages of their outbursts, one can better constrain models of  disk instabilities  and how they trigger outburst, but 
also how the different media (CO surface, accretion disk, ``corona'') interact and lead to the spectral state transitions. The \textit{SVOM} wide 0.1--5000 keV 
spectral coverage will permit to precisely measure the local absorption, probe the spectral shape of the high energy emission 
(black body  and/or bremsstrahlung temperature, power law photon index $\Gamma$, cut-off energy, reflection component, iron line, ...) 
while removing the degeneracy between $kT$, $N_H$ and $\Gamma$ thanks to the large ``lever arm'' brought by the 4--5000 keV band. 
The high temporal resolution of ECLAIRs and the GRM will also permit timing studies (pulsations, level of variability, quasi-periodic oscillations, ...) to 
be performed. Here we mention a few specific points for each systems that can be studied with \textit{SVOM}.\\

\indent The intermediate polars (IP), a sub-type of CVs containing intermediately magnetized white dwarf, are of particular relevance 
for the multi-wavelengths capabilities of \textit{SVOM}. These systems have been detected with both \textit{Swift} and \textit{INTEGRAL} \citep[e.g.][]{revnivtsev08}. 
Many lie at high galactic latitude, and are easily observable within the B1 law. 
Optical/IR and X-ray follow-up have shown levels above the limits of the MXT (F$_{\mathrm{IGR-CVs, 2-10 keV}}>10^{-12}$ erg/cm$^{2}$/s), 
the VT, and the GWAC  (M$_R\sim14-17$). Serendipitous activity (outburst, nova-like phenomenon) will easily be detected within 
the large ECLAIRs and GRM's FoV.\\

\indent Neutron star high mass XRBs are thought to be young systems hosting  pulsars with rather large  magnetic fields ($B\sim10^{12}$~G). 
Here the detection of resonant cyclotron scattering features can permit to measure the value of $B$, since
$\frac{B}{10^{12}~\mathrm{G}} = \frac{E_{cycl}}{11.6} \times(1 + z)$ where $E_{cycl}$ is the centroid of the line, and z the gravitational 
redshift near the NS. ECLAIRs has the capabilities  to detect these lines during the brightest phases of outburst in sources such as 
A0535$-$26, and also to search for coherent pulsations giving access to the pulsar spin, and eventually the spin derivative. 
GWAC and GFTs will, simultaneously, permit to study the companion, and/or the potential optical flaring 
activity and its possible relation with X-ray flares.\\
\indent The evolution of  BH and NS XRBs (and CVs in some respect) along their outbursts show a similar 
phenomenology of spectral transitions between states 
dominated by the accretion disk and those dominated by hard X-ray ``coronae".  
The mechanisms that drive the evolutions (smooth vs large flaring activity, and/or limit cycle oscillations), and the 
hysteresis (the fact that the  hard to soft transition is always brighter than the 
soft to hard one)  are, however, still not understood. In addition, many of these sources are also known as jet-emitting sources 
(as seen in the radio domain) and thus dubbed microquasars. While the association of jets with particular states of activity is now 
clear \citep[e.g.][to cite just a few]{corbel03,rodrigue08_1915a,mickael_1743}, the exact interplay between accretion and ejection, and how 
jets influe on the source evolution (self-regulation of outburst, energy and material feedback in the ISM or interaction with the local 
environment) is largely unknown.  Recent activity of some sources (e.g. IGR J17091$-$2624 or V404 Cyg) have also shown that some objects
largely depart from the ``standard" behavior (Fig. 21), with short (10--15 days) outburst and huge level of emission (up to 
50 Crab) on very short (hour) time scales. While these sources are mostly outside the B1 law, systematics (spectral and temporal) surveys of outbursts, during the 
10\% allowed time can thus  be used to better understand
the physics driving  the evolution of these sources, their impact on their surrounding and eventually the Galaxy. The right 
panels of Fig. 21 show spectral MXT+ECLAIRs simulations of a moderately bright state with a cold disk, and a faint hard state. \\

\begin{figure}
\centering
 \includegraphics[scale=0.7]{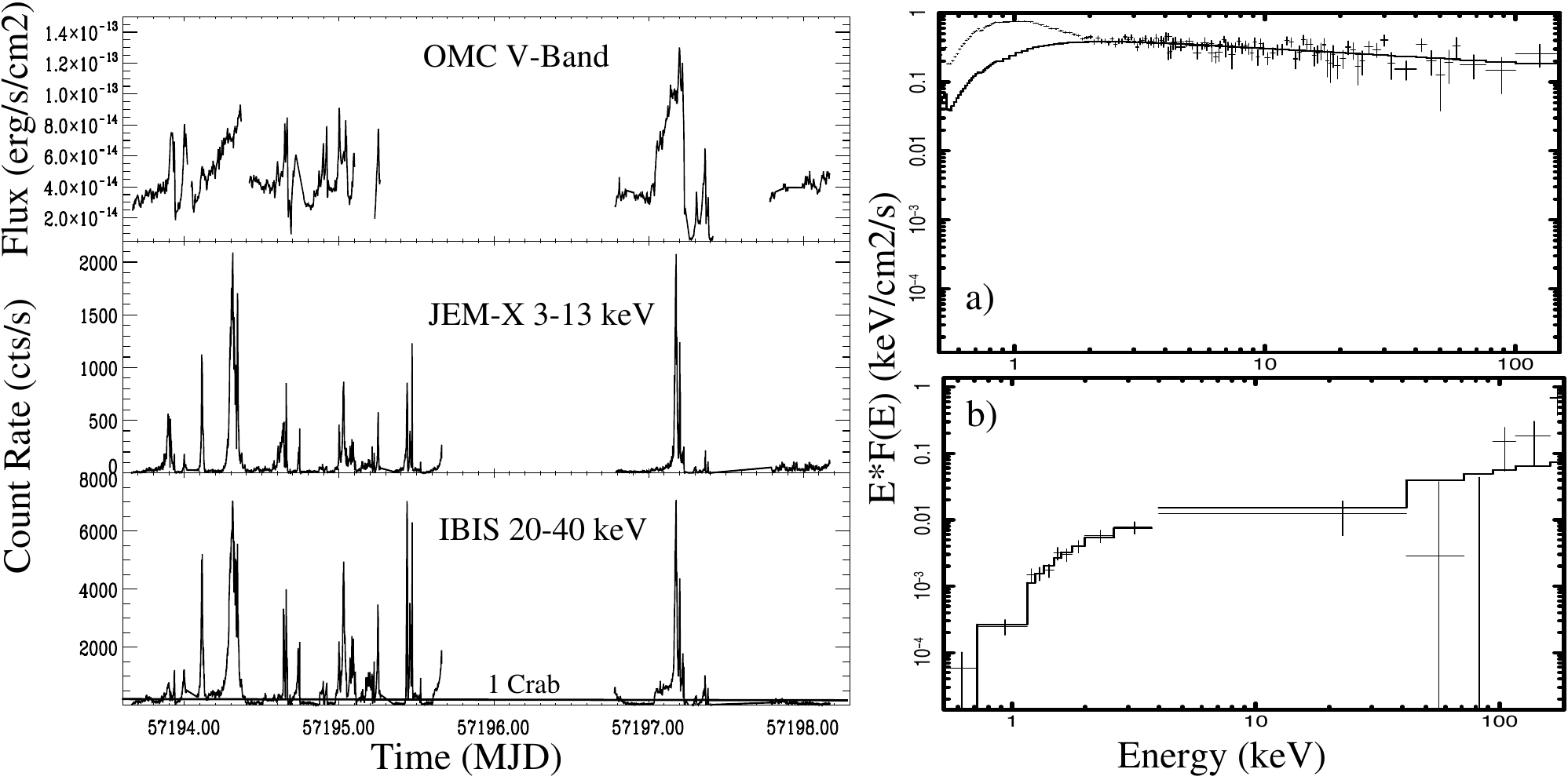}
\caption{{\bf{Left:}} \textit{INTEGRAL} light curves of V404 Cyg during part of 
its 2015-July outburst. From top to bottom optical V-band, soft (3--13 keV), and hard (20--40 keV ) X-rays.  Adapted from \citet{rodrigue15_V404}.  {\bf{Right:}} a)10 ks MXT-ECLAIRs simulated spectra of a moderately bright soft state (75 and 32 mCrab in the 1-10 and 20-200 keV ranges).  b) Same for a 1 mCrab (1--10 keV)/ 5 mCrab (20--200 keV) hard state. }
\label{fig:19}
\end{figure}

\medbreak
\subsubsection{Pulsars and magnetars}
\smallbreak  
Magnetars (either Anomalous X-ray Pulsars, AXPs, and Soft Gamma Repeaters, SGRs) are a small class of isolated NS that are, contrary to standard NS, powered by the decay of their huge magnetic field (B$\sim10^{14-15}$ G).
They can emit short ($< 1$ s) bursts, intermediate flares lasting for a few tens of seconds, and giant flares.  The latter peak at a luminosity of $10^{47}$ erg s$^{-1}$. 
A spectral measurement by ECLAIRs (at least before eventual instrumental saturation, or during the decaying tail) will permit to characterize these events. 
Giant flares in nearby ($< 100$ Mpc) galaxies could be discovered by ECLAIRs mimicking short-hard bursts, and followed with MXT. 
The $<10$ keV spectrum of the more common magnetar bursts is not well known. \citet{olive04} used a double black body to describe an intermediate flare of  SGR 1900+14, whereas  \citet{israel08}  showed that a Comptonized spectrum could represent the \textit{Swift}/XRT--BAT spectra of the same source well. Hence, the correct spectral modelling and consequent physical interpretation of short bursts at a few keV is not settled yet and 
requires further observations, and \textit{SVOM} with its unique multi-wavelength coverage can bring new diagnostics in this field. \\
\indent Recent results tend to show relatively low dipolar magnetic fields in at least two objects (SGR 0418+5729, and \textit{Swift} J1822$-$1606) with B close to standard
pulsar values. This may imply that the magnetar population includes many more X-ray dim (outside outbursts) members than previously thought, and that can be studied only using prompt follow-up observations by sensitive high-energy observatories like \textit{SVOM}. \\
\indent In addition, the first direct measurement of a magnetar magnetic field \citep{tiengo13} through a proton cyclotron 
feature at $\sim 2$ keV  in SGR 0418+5729 confirms a high magnetic field value  (B$\sim2-20 \times10^{14}$ G)  mainly stored in multi-pole components at the surface. This variable feature seen at  still a  high flux ($5 \times10^{-12}$ \ergcms) may be accessible to MXT thanks to its good spectral 
resolution and sensitivity with a 10 ks observation (sensitivity limit $7 \times10^{-13}$ \ergcms).
The frontier between pulsars and magnetars appears more tenuous than previously thought, objects belonging to one class being able to reproduce -- during some special 
periods of activity-- the behavior of the other family.   The timing and spectral capabilities of the instruments on board \textit{SVOM} will help to study the link between the 
two classes of objects, having interesting consequences on their formation.

\subsubsection{Flaring stars}
\smallbreak 
ECLAIRs will also be sensitive to flaring M and K stars that could produce huge flares. This could provide information on non-thermal processes happening in the stellar corona. This could help constraining particle acceleration mechanisms. Detection of X-ray lines in emission could also provide some clues on the shape/geometry of the magnetic loop reconnection leading to these events \citep{osten2010b}. Observations of super-flares from such small stars have also implications for astrobiology.

\subsection{Exoplanets and Solar System bodies}
\label{sec:planeto}
As of today only a handful of stars having soft X-ray luminosity between $10^{-13}$--$10^{-10}$ \ergcms\ are known to have planets in their habitable zones. 
In one case (HD187933), the transit of the hot Jupiter orbiting around the star has been observed in X-ray with \textit{Swift}/XRT 
\citep{lecavelier12}. The sensitivity of the MXT instrument onboard \textit{SVOM} would make the observation of exoplanets feasible only in the case 
of higher luminosity stars, by averaging several transits to obtain light curves where amplitude variations of a few percent could be detected. 
However, it can be expected that more stellar systems hosting planets will be discovered in the next years, therefore the search for exoplanets 
and the study of their atmospheres with \textit{SVOM} can be considered at this stage as a potential application.
\medbreak
Many solar system bodies emit X-rays like planets, moons, comets etc and simultaneous \textit{SVOM} X-ray and optical observations will be very 
promising for Solar System studies as shown by the \textit{Swift} mission that could study several comets.

\subsection{Additional science opportunities}
\label{sec:other}

The \textit{SVOM} mission will also bring new insights to a number of other important high-energy science topics
for which its specific operation strategy, with frequent Earth passes in the instrument FoVs, will favor accurate measurements. 
\medbreak

The  Cosmic X-ray Background (CXB) is a uniform, isotropic component of the sky X-ray emission, which is likely due to a population of 
unresolved weak point-sources. 
Indeed, the observed CXB surface brightness at more than few keV is likely provided by accreting super-massive black holes (SMBH) in galactic 
nuclei at various distances from us and it is therefore an overall measure of the growth of all SMBHs in the history of the Universe 
\citep{1982MNRAS.200..115S, 1999MNRAS.303L..34F}. Combining accurate CXB measurements  with detailed AGNs population synthesis models can provide the
 characteristics of this population and then trace the long term history of the SMBH formation, growth and evolution in the Universe \citep{Ueda2014}.
\medbreak

Unlike at lower energies,  above 7 keV most of the CXB is unresolved (only 2.5 \% in 20-60 keV) 
and will remain so in the near future (\textit{NuStar} is expected to resolve at most 30 \% of it).
This range traces the highly absorbed Compton thick AGN population, not visible at low energies, 
and an accurate CXB measurement here is needed for evaluating the role of this  population. 
A recent compilation of CXB measurements \citep{2014AstL...40..667R} shows that in the crucial range 
of the peak of the spectrum, the 10-50 keV band, the uncertainties in the absolute intensity 
are still of the order of 20-30 \%, which has a large impact on the computation of the Compton thick AGN demography.
\medbreak

In the determination of the CXB an important problem is the estimation of the intrinsic particle-induced background
 and of the other celestial components that contribute to the recorded flux. One of the most effective techniques to 
 evaluate the  background is to perform measurements during Earth occultation of the sky \citep{2007A&A...467..529C,2014AstL...40..667R}. 
\medbreak

Due to the \textit{SVOM} pointing strategy, ECLAIRs  will suffer 
from a large rate of partial Earth occultation  and it will therefore provide unprecedented
statistics for accurate CXB measurements in the range 4-150 keV. Given the high 
galactic latitudes of these observations, galactic point-sources and diffuse emission contamination  
will also be low, reducing the errors in the CXB estimations. 
\medbreak
 	 
Likewise, the Galactic plane is known to be an intense source of continuum high-energies emission. 
At gamma-ray energies, this emission is mainly of diffuse nature and is well understood, while
in the X-ray band it is referred as Galactic Ridge X-ray Emission (GRXE) and its origin
 is still debated.

\medbreak
At energies below 50 keV,
recent observations from the \textit{RXTE} and \textit{INTEGRAL} satellites showed that the global morphology of the GRXE reproduces the distribution of 
stars in the Milky Way, which points to a GRXE stellar origin  where accreting WDs are
the main contributors \citep{2007A&A...463..957K}. 

However, the various source populations responsible of the emission are not precisely known \citep{2013ApJ...766...14M}. 
In addition, the spectrum region around 50 keV is poorly constrained and the few recent measurements are dominated by systematics and modeling biases. 
Similarly, accurate measurements are needed to fix the characteristics of the power-law continuum below 100 keV, 
thought to be due to the interstellar radiation field interaction with the cosmic-ray electrons.

\medbreak
 Again, Earth occultation observations in which the Galactic plane is occulted by the Earth can be used.
Such a technique has been successfully applied to the data obtained with  \textit{INTEGRAL}/IBIS using few
 dedicated Earth occultation observations and obtaining a GRXE spectrum with a good accuracy in hard X-rays \citep{2010A&A...512A..49T}. 
 As mentioned above, the number of such Earth modulated observations is expected to be substantially higher for ECLAIRs, this will
  lead to a high signal-to-noise ratio measurement of the GRXE spectrum in the hard X-ray energy range 
  and will help to resolve the issues above mentioned.
\medbreak

Other topics that will be studied with \textit{SVOM} also thanks to observations pointed towards the Earth are solar and 
terrestrial events involving particle acceleration, confinement and interaction like in solar flares, south Atlantic anomaly,
aurorae and terrestrial gamma-ray flashes (TGF). The TGF are extremely rapid ($<$ 10 ms) bursts of photons 
in the range 20 keV -- 20 MeV associated with tropical thunderstorms in the atmosphere for which physical mechanisms 
are not yet understood. ECLAIRs will provide localisation of these events and help to resolve the open questions about TGFs 
in synergy with the future dedicated CNES mission \textit{Taranis} and the ESA experiment to be flown on the ISS.


\newpage

\begin{flushleft}
\Large\bf Acknowledgements
\end{flushleft}

Financial support for this work was provided by the CNES (French Space Agency), the PNHE  (French National High Energy Program), the European Research Group (Exploring the Dawn of the Universe with Gamma-Ray Bursts) and the UnivEarthS Labex program at Sorbonne Paris Cit\'{e} (ANR-10-LABX-0023 and ANR-11-IDEX-0005-02).



\newpage

\bibliographystyle{apj} 

 


\bibliography{biblio_wp_v20} 

\end{document}